\documentclass[acmsmall,natbib=false,nonacm=true]{acmart}

\pdfoutput=1

\usepackage[
datamodel=acmdatamodel,
style=acmauthoryear
]{biblatex}

\usepackage[colorinlistoftodos,prependcaption,textsize=tiny]{todonotes}
\NewDocumentCommand{\rknote}{s m}{
  \IfBooleanTF {#1}
    { \todo[backgroundcolor=green!30]{rak: #2} }
    { \todo[inline, backgroundcolor=green!30]{rak: #2} }
}

\newcommand\scalemath[2]{\scalebox{#1}{\mbox{\ensuremath{\displaystyle #2}}}}

\newcommand{\chase}{\textsc{Most}\xspace}

\usepackage{booktabs}   %
\usepackage{subcaption} %

\usepackage{macro}

\theoremstyle{remark}

\usepackage[createShortEnv, conf={restate, no link to proof}]{proof-at-the-end}

\defrule{WTu}{\rn{W\(\Tu\)}}{
  \jwftp
  { \Pi }
  { a : \Tu }
}{
}

\defrule{WTLbl}{\rn{W\({\Tplus}{\Tamp}\)}}{
  \jwftp
  { \Pi }
  { a : \odot_{l \in L} A_l }
}{
  \jwftp
  { \immred{\Pi}{\pLabel{a}{l}{a}} }
  { a : A_l }
  &
  (\forall l \in L)
  &
  {\odot} \in \Set{ {\Tplus}, {\Tamp} }
}

\defrule{WTChan}{\rn{W\({\Tot}{\Tlolly}\)}}{
  \jwftp
  { \Pi }
  { c : (a : A) \odot (b : B) }
}{
  \jwftp
  { \immred{\Pi}{\pi}, b : B }
  { a : A }
  &
  \jwftp
  { \immred{\Pi}{\pi}, a : A }
  { b : B }
  &
  \pi = \pChan{c}{a}{b}
  &
  {\odot} \in \Set{ {\Tot}, {\Tlolly} }
}

\defrule{WETu}{\rn{WE\(\Tu\)}}{
  \jwftp
  { \Pi, c : \Tu }
  { a : \TCaseU{c}{A} }
}{
  \jwftp
  { \immred{\Pi}{\pClose{c}} }
  { a : A }
}

\defrule{WETLbl}{\rn{WE\({\Tplus}{\Tamp}\)}}{
  \jwftp
  { \Pi, c : \odot_{l \in L} C_l }
  { a : \TCaseL{c}{\Set{l \Rightarrow A_l}_{l \in L \cup L'}} }
}{
  \jwftp
  { \immred{\Pi}{\pLabel{c}{l}{c}}, c : C_l }
  { a : A_l }
  &
  (\forall l \in L)
  &
  \jwftp
  { \Pi }
  { a : A_{l'} }
  &
  (\forall l' \in L')
  &
  {\odot} \in \Set{ {\Tplus}, {\Tamp} }
}

\defrule{WETChan}{\rn{WE\({\Tot}{\Tlolly}\)}}{
  \jwftp
  { \Pi, c : (b : B) \odot (c : C) }
  { a : \TCaseC{c}{b}{A} }
}{
  \jwftp
  { \immred{\Pi}{\pChan{c}{b}{c}}, b : B, c : C }
  { a : A }
  &
  {\odot} \in \Set{ {\Tot}, {\Tlolly} }
}

\defrule{WPiRed}{\rn{W\(\Pi{\Rightarrow}\)}}{
  \jwftp
  { \Pi }
  { a : A }
}{
  \jwftp
  { \Pi' }
  { a : A }
  &
  \smash[b]{\Pi \ecred{\pi} \Pi'} %
  &
  \freecn(\pi) \cap \freecn(A) = \emptyset
}

\defrule{WPi}{\rn{W\(\Pi\)}}{
  \jwftc{a_1 : A_1, \dotsc, a_n : A_n}
}{
  \jwftp{a_2 : A_2, \dotsc, a_n : A_n}{a_1 : A_1}
  &
  \cdots
  &
  \jwftp{a_1 : A_1, \dotsc, a_{n - 1} : A_{n - 1}}{a_n : A_n}
}

\defrule{WE}{\rn{WE}}{
  \jwfps
  { \Pi }
  { \cdot }
}{
  \jwftc
  { \Pi }
}

\defrule{WS}{\rn{WS}}{
  \jwfps
  { \Pi }
  { \proci
    { \Delta }
    { \Iota }
    { a : A } }
}{
  \jwftc
  { \Pi, \Delta, \Iota, a : A }
}

\defrule{WCons}{\rn{W\(\scons\)}}{
  \jwfps
  { \Pi }
  { \mc{G}_1 \scons \mc{G}_2 }
}{
  \jwfps
  { \Pi, \overline{\mc{G}_2} }
  { \mc{G}_1 }
  &
  \jwfps
  { \Pi, \overline{\mc{G}_1} }
  { \mc{G}_2 }
}

\defrule{STFoc-R}{\rn{\({\Vdash}\)R}}{
  \jtproc
  { \Pi }
  { \Iota }
  { P }
  { \Delta }
  { a : A }
  { \tfnt{T} }
}{
  \jtred
  { \Pi }
  { \Iota }
  { P }
  { \Delta }
  { \focus{a : A} }
  { \tfnt{T} }
  &
  \princ(P) = a
}

\defrule{STFoc-L}{\rn{\({\Vdash}\)L}}{
  \jtproc
  { \Pi }
  { \Iota }
  { P }
  { \Delta, a : A }
  { c : C }
  { \tfnt{T} }
}{
  \jtred
  { \Pi }
  { \Iota }
  { P }
  { \Delta, \focus{a : A} }
  { c : C }
  { \tfnt{T} }
  &
  \princ(P) = a
}

\defrule{Par}{\rn{Par}}{
  \jtproc
  { \Pi }
  { \Iota_1, a : A, \Iota_2 }
  { \tPar{a}{P}{Q} }
  { \Delta_1, \Delta_2 }
  { c : C }
  { \tfnt{T}_1 \tint \tfnt{T}_2 }
}{
  \jtproc
  { \Pi, \Delta_2, \Iota_2, c : C }
  { \Iota_1 }
  { P }
  { \Delta_1 }
  { a : A }
  { \tfnt{T}_1 }
  &
  \jtproc
  { \Pi, \Delta_1, \Iota_1 }
  { \Iota_2 }
  { Q }
  { \Delta_2, a : A }
  { c : C }
  { \tfnt{T}_2 }
  &
  \tfnt{T}_1 \therefore \tfnt{T}_2
}

\defrule{New}{\rn{New}}{
  \jtproc
  { \Pi }
  { \Iota }
  { \tNewV{(a_1 : A_1, \dotsc, a_n : A_n)}{P} }
  { \Delta }
  { c : C }
  { \tdel{\tfnt{T}}{\Set{a_1, \dotsc, a_n}} }
}{
  \jtproc
  { \Pi }
  { \Iota, a_1 : A_1, \dotsc, a_n : A_n }
  { P }
  { \Delta }
  { c : C }
  { \tfnt{T} }
}

\defrule{RTu-R}{\rn{\({\Rightarrow}\mblue{\Tu}\)R}}{
  \jtred
  { \Pi }
  { \cdot }
  { \tClose{a} }
  { \cdot }
  { \focus{a : \Tu} }
  { \Set{ \tcons{\pO}{\mClose{a}}{\tempt} } }
}{
}

\defrule{RTu-L}{\rn{\({\Rightarrow}\mblue{\Tu}\)L}}{
  \jtred
  { \Pi }
  { \Iota }
  { \tWait{a}{P} }
  { \Delta, \focus{a : \Tu} }
  { c : C }
  { \tcons{\pI}{\mClose{a}}{\tfnt{T}} }
}{
  \jtproc
  { \immred{\Pi}{\pi} }
  { \immred{\Iota}{\pi} }
  { P }
  { \immred{\Delta}{\pi} }
  { c : \immred{C}{\pi} }
  { \tfnt{T} }
  &
  \pi = \pClose{a}
}

\defrule{RTplus-R}{\rn{\({\Rightarrow}\mblue{\Tplus}\)R}}{
  \jtred
  { \Pi }
  { \Iota }
  { \tSendL{a}{k}{P} }
  { \Delta }
  { \focus{a : \Tplus_{l \in L \cup \Set{k}} A_l} }
  { \tcons{\pO}{\mLabel{a}{k}}{\tfnt{T}} }
}{
  \jtproc
  { \immred{\Pi}{\pi} }
  { \immred{\Iota}{\pi} }
  { P }
  { \immred{\Delta}{\pi} }
  { a : A_k }
  { \tfnt{T} }
  &
  \pi = \pLabel{a}{k}
}

\defrule{RTplus-L}{\rn{\({\Rightarrow}\mblue{\Tplus}\)L}}{
  \jtred
  { \Pi }
  { \Iota }
  { \tRecvL{a}{\Set{l \Rightarrow P_l}_{l \in L \cup L'}} }
  { \Delta , \focus{a : \Tplus_{l \in L} A_l} }
  { c : C }
  { \bigcup\nolimits_{l \in L} \tcons{\pI}{\mLabel{a}{k}}{\tfnt{T}_l} }
}{
  \jtproc
  { \immred{\Pi}{\pi} }
  { \immred{\Iota}{\pi} }
  { P_l }
  { \immred{\Delta}{\pi}, a : A_l }
  { c : \immred{C}{\pi} }
  { \tfnt{T}_l }
  &
  \pi = \pLabel{a}{l}
  &
  (\forall l \in L)
}

\defrule{RTamp-R}{\rn{\({\Rightarrow}\mblue{\Tamp}\)R}}{
  \jtred
  { \Pi }
  { \Iota }
  { \tRecvL{a}{\Set{l \Rightarrow P_l}_{l \in L \cup L'}} }
  { \Delta }
  { \focus{a : \Tamp_{l \in L} A_l} }
  { \bigcup\nolimits_{l \in L} \tcons{\pI}{\mLabel{a}{k}}{\tfnt{T}_l} }
}{
  \jtproc
  { \immred{\Pi}{\pi} }
  { \immred{\Iota}{\pi} }
  { P_l }
  { \immred{\Delta}{\pi} }
  { a : A_l }
  { \tfnt{T}_l }
  &
  \pi = \pLabel{a}{l}
  &
  (\forall l \in L)
}

\defrule{RTamp-L}{\rn{\({\Rightarrow}\mblue{\Tamp}\)L}}{
  \jtred
  { \Pi }
  { \Iota }
  { \tSendL{a}{k}{P} }
  { \Delta, \focus{a : \Tamp_{l \in L \cup \Set{k}} A_l} }
  { c : C }
  { \tcons{\pO}{\mLabel{a}{k}}{\tfnt{T}} }
}{
  \jtproc
  { \immred{\Pi}{\pi} }
  { \immred{\Iota}{\pi} }
  { P }
  { \immred{\Delta}{\pi}, a : A_k }
  { c : \immred{C}{\pi} }
  { \tfnt{T} }
  &
  \pi = \pLabel{a}{k}
}

\defrule{RTot-R}{\rn{\({\Rightarrow}{\mblue{\Tot}}\)R}}{
  \jtred
  { \Pi }
  { \Iota_1, \Iota_2 }
  { \tSendC{a}{b}{P}{Q} }
  { \Delta_1, \Delta_2 }
  { \focus{a : (b : B) \Tot (a : A)} }
  { \tcons{\pO}{\mChan{a}}[b]{(\tfnt{T}_1 \tint \tfnt{T}_2)} }
}{
  \jtproc
  { \immred{(\Pi, \Delta_2, \Iota_2)}{\pi}, a : A }
  { \immred{\Iota_1}{\pi} }
  { P }
  { \immred{\Delta_1}{\pi} }
  { b : B }
  { \tfnt{T}_1 }
  &
  \jtproc
  { \immred{(\Pi, \Delta_1, \Iota_1)}{\pi}, b : B }
  { \immred{\Iota_2}{\pi} }
  { Q }
  { \immred{\Delta_2}{\pi} }
  { a : A }
  { \tfnt{T}_2 }
  &
  \pi = \pChan{a}{b}
  &
  \tfnt{T}_1 \therefore \tfnt{T}_2
}

\defrule{RTot-L}{\rn{\({\Rightarrow}{\mblue{\Tot}}\)L}}{
  \jtred
  { \Pi }
  { \Iota }
  { \tRecvC{b}{a}{P} }
  { \Delta, \focus{ a : (b : B) \Tot (a : A) } }
  { c : C }
  { \tcons{\pI}{\mChan{a}}[b]{\tfnt{T}} }
}{
  \jtproc
  { \immred{\Pi}{\pi} }
  { \immred{\Iota}{\pi} }
  { P }
  { \immred{\Delta}{\pi}, b : B, a : A }
  { c : \immred{C}{\pi} }
  { \tfnt{T} }
  &
  \pi = \pChan{a}{b}
}

\defrule{RTlolly-R}{\rn{\({\Rightarrow}{\mblue{\Tlolly}}\)R}}{
  \jtred
  { \Pi }
  { \Iota }
  { \tRecvC{b}{a}{P} }
  { \Delta }
  { \focus{ a : (b : B) \Tlolly (a : A) } }
  { \tcons{\pI}{\mChan{a}}[b]{\tfnt{T}} }
}{
  \jtproc
  { \immred{\Pi}{\pi} }
  { \immred{\Iota}{\pi} }
  { P }
  { \immred{\Delta}{\pi}, b : B }
  { a : A }
  { \tfnt{T} }
  &
  \pi = \pChan{a}{b}
}

\defrule{RTlolly-L}{\rn{\({\Rightarrow}{\mblue{\Tlolly}}\)L}}{
  \jtred
  { \Pi }
  { \Iota }
  { \tSendC{a}{b}{P}{Q} }
  { \Delta_1, \Delta_2, \focus{ a : (b : B) \Tlolly (a : A) } }
  { c : C }
  { \tcons{\pO}{\mChan{a}}[b]{(\tfnt{T}_1 \tint \tfnt{T}_2)} }
}{
  \jtproc
  { \immred{(\Pi, \Delta_2, \Iota_2)}{\pi}, a : A }
  { \immred{\Iota_1}{\pi} }
  { P }
  { \immred{\Delta_1}{\pi} }
  { b : B }
  { \tfnt{T}_1 }
  &
  \jtproc
  { \immred{(\Pi, \Delta_1, \Iota_1)}{\pi}, b : B }
  { \immred{\Iota_2}{\pi} }
  { Q }
  { \immred{\Delta_2}{\pi}, a : A }
  { c : \immred{C}{\pi} }
  { \tfnt{T}_2 }
  &
  \pi = \pChan{a}{b}
  &
  \tfnt{T}_1 \therefore \tfnt{T}_2
}

\defrule{Red-RTu}{\rn{\({\Rightarrow}\Tu\)[R]}}{
  \jtred
  { \Pi, c : \Tu }
  { \Iota }
  { P }
  { \Delta }
  { \focus{a : \TCaseU{c}{A} } }
  { \tcons{\pC}{\mClose{c}}{\tfnt{T}} }
}{
  \jtred
  { \immred{\Pi}{\pi} }
  { \immred{\Iota}{\pi} }
  { P }
  { \immred{\Delta}{\pi} }
  { \focus{a : A} }
  { \tfnt{T} }
  &
  \pi = \pClose{c}
}

\defrule{Red-LTu}{\rn{\({\Rightarrow}\Tu\)[L]}}{
  \jtred
  { \Pi, c : \Tu }
  { \Iota }
  { P }
  { \Delta, \focus{a : \TCaseU{c}{A} } }
  { d : D }
  { \tcons{\pC}{\mClose{c}}{\tfnt{T}} }
}{
  \jtred
  { \immred{\Pi}{\pi} }
  { \immred{\Iota}{\pi} }
  { P }
  { \immred{\Delta}{\pi}, \focus{a : A} }
  { d : \immred{D}{\pi} }
  { \tfnt{T} }
  &
  \pi = \pClose{c}
}

\defrule{Red-RTLbl}{\rn{\({\Rightarrow}{\Tplus}{\Tamp}\)[R]}}{
  \jtred
  { \Pi, c : \odot_{l \in L'} C_l}
  { \Iota }
  { P }
  { \Delta }
  { \focus{a : \TCaseL{c}{ \Set{l \Rightarrow A_l}_{l \in L''} } } }
  { \bigcup\nolimits_{l \in L} \tcons{\pC}{\mLabel{c}{l}}{\tfnt{T}_l} }
}{
  \jtred
  { \immred{\Pi}{\pi_l}, c : C_l }
  { \immred{\Iota}{\pi_l} }
  { P }
  { \immred{\Delta}{\pi_l} }
  { \focus{a : A_l} }
  { \tfnt{T}_l }
  &
  \pi_l = \pLabel{c}{l}
  &
  (\forall l \in L)
  &
  L \subseteq L' \subseteq L''
  &
  {\odot} \in \Set{ {\Tplus}, {\Tamp} }
}

\defrule{Red-LTLbl}{\rn{\({\Rightarrow}{\Tplus}{\Tamp}\)[L]}}{
  \jtred
  { \Pi, c : \odot_{l \in L} C_l}
  { \Iota }
  { P }
  { \Delta, \focus{a : \TCaseL{c}{ \Set{l \Rightarrow A_l}_{l \in L \cup L'} } } }
  { d : D }
  { \bigcup\nolimits_{l \in L} \tcons{\pC}{\mLabel{c}{l}}{\tfnt{T}_l} }
}{
  \jtred
  { \immred{\Pi}{\pi_l}, c : C_l }
  { \immred{\Iota}{\pi_l} }
  { P }
  { \immred{\Delta}{\pi_l}, \focus{a : A_l}}
  { d : \immred{D}{\pi_l} }
  { \tfnt{T}_l }
  &
  \pi_l = \pLabel{c}{l}
  &
  (\forall l \in L)
  &
  {\odot} \in \Set{ {\Tplus}, {\Tamp} }
}

\defrule{Red-RTChan}{\rn{\({\Rightarrow}{\Tot}{\Tlolly}\)[R]}}{
  \jtred
  { \Pi, c : (b : B) \odot (c : C) }
  { \Iota }
  { P }
  { \Delta }
  { \focus{a : \TCaseC{c}{b}{A}} }
  { \tcons{\pC}{\mChan{c}}[b]{\tfnt{T}} }
}{
  \jtred
  { \immred{\Pi}{\pi}, b : B, c : C }
  { \immred{\Iota}{\pi} }
  { P }
  { \immred{\Delta}{\pi} }
  { \focus{a : A} }
  { \tfnt{T} }
  &
  \pi = \pChan{c}{b}
  &
  {\odot} \in \Set{ {\Tot}, {\Tlolly} }
}

\defrule{Red-LTChan}{\rn{\({\Rightarrow}{\Tot}{\Tlolly}\)[L]}}{
  \jtred
  { \Pi, c : (b : B) \odot (c : C) }
  { \Iota }
  { P }
  { \Delta, \focus{a : \TCaseC{c}{b}{A}} }
  { d : D }
  { \tcons{\pC}{\mChan{c}}[b]{\tfnt{T}} }
}{
  \jtred
  { \immred{\Pi}{\pi}, b : B, c : C }
  { \immred{\Iota}{\pi} }
  { P }
  { \immred{\Delta}{\pi}, \focus{a : A} }
  { d : \immred{D}{\pi} }
  { \tfnt{T} }
  &
  \pi = \pChan{c}{b}
  &
  {\odot} \in \Set{ {\Tot}, {\Tlolly} }
}

\defrule{tEmpt}{\rn{tEmpt}}{
  \jtt{\tempt}{\jwfts{\Pi}{\cdot}}
}{
}

\defrule{tRed}{\rn{tRed}}{
  \jtt{(\tcons{s}{p}[\vec y]{t})}{\jwfts{\Pi}{\mc{G}}}
}{
  \jwfts{\Pi}{\mc{G}} \psred{(s, \obsc{p}{\vec y})} \jwfts{\Pi'}{\mc{G}'}
  &
  \jtt{t'}{\jwfts{\Pi'}{\mc{G}'}}
}

\defrule{tE}{\rn{t\(\tempt\)}}{
  \jtcs[C]{{\cdot\,}}{{\cdot\,}}{\cdot}{\tempt}
  \mathstrut
}{
  \mathstrut
}

\defrule{tC}{\rn{tC}}{
  \jtcs[C, a]{I}{O}{S}{(\tcons{\pC}{p}[\vec x]{t})}
}{
  \jtcs[C, \vec x]{I}{O}{S}{t}
  &
  \carrcn(p) = a
}

\defrule{tS}{\rn{tS}}{
  \jtcs[C]{I}{O}{S, a}{(\tcons{\pS}{p}[\vec x]{t})}
}{
  \jtcs[C]{I}{O}{S, \vec x}{t}
  &
  \carrcn(p) = a
}

\defrule{tI}{\rn{tI}}{
  \jtcs[C]{I, a}{O}{S}{(\tcons{\pI}{p}[\vec x]{t})}
}{
  \jtcs[C]{I, \vec y}{O, \vec z}{S}{t}
  &
  \carrcn(p) = a
  &
  \vec x = \vec y \cup \vec z
}

\defrule{tO}{\rn{tO}}{
  \jtcs[C]{I}{O, a}{S}{(\tcons{\pO}{p}[\vec x]{t})}
}{
  \jtcs[C]{I, \vec y}{O, \vec z}{S}{t}
  &
  \carrcn(p) = a
  &
  \vec x = \vec y \cup \vec z
}

\title{Message-Observing Sessions}         %
\titlenote{This preprint is an extended version of the article:\\\fullcite{kavanagh_pientka_2024:_messag_obser_session}.}

\author{Ryan Kavanagh}
\orcid{0000-0001-9497-4276}             %
\affiliation{
  \position{Professeur}
  \department{Département d'informatique}              %
  \institution{Université du Québec à Montréal}            %
  \streetaddress{201 Du Président-Kennedy Avenue}
  \city{Montréal}
  \state{QC}
  \postcode{H2X 3Y7}
  \country{Canada}                    %
}
\email{kavanagh.ryan@uqam.ca}

\author{Brigitte Pientka}
\orcid{0000-0002-2549-4276}
\affiliation{
  \department{School of Computer Science}              %
  \institution{McGill University}            %
  \streetaddress{3480 University Street}
  \city{Montr{\'e}al}
  \state{QC}
  \postcode{H3A 0E9}
  \country{Canada}                    %
}
\email{bpientka@cs.mcgill.ca}

\begin{abstract}
  We present \chase, a process language with message-observing session types.
  Message-observing session types extend binary session types with type-level computation to specify communication protocols that vary based on messages observed on other channels.
  Hence, \chase allows us to express global invariants about processes, rather than just local invariants, in a bottom-up, compositional way.
  We give \chase a semantic foundation using \emph{traces with binding}, a semantic approach for compositionally reasoning about traces in the presence of name generation.
  We use this semantics to prove type soundness and compositionality for \chase processes.
 We see this as a significant step towards capturing
message-dependencies and providing more precise guarantees about
processes.
\end{abstract}


\newbool{appendix}
\booltrue{appendix}

\begin{document}

\maketitle

\section{Introduction}
\label{sec:introduction}

Session types~\cite{takeuchi_1994:_inter_based_languag,honda_1993:_types_dyadic_inter,honda_1998:_languag_primit_type} allow us to specify and statically verify that processes communicate according to prescribed protocols.
Hence, they rule out a wide class of communication-related bugs before executing a given protocol.

A \emph{binary} session type specifies the communication protocol
as seen from the point of view of one of the two participants. From a Curry-Howard perspective, it corresponds to the standard sequent calculus proof system for dual intuitionistic linear logic (see \cite{caires_pfenning_2010:_session_types_intuit_linear_propos,caires_2016:_linear_logic_propos}), thereby building a logical foundation for specifying and reasoning about concurrent communications.
Over the past decade, there have been different approaches to extend it and capture ever richer protocols:
value-dependent session types~\cite{toninho_2011:_depen_session_types, toninho_yoshida_2018:_depen_session_typed_proces} allow protocols to vary based on previously transmitted values;
label-dependent session types~\cite{thiemann_vasconcelos_2019:_label_depen_session_types} describe different communication behaviour depending on labels being observed on a given channel; manifest sharing~\cite{balzer_pfenning_2017:_manif_sharin_with_session_types} allows (binary) session types to capture shared communication channels.

In this paper, we describe \chase, a language for protocols using \emph{message-observing session types}.
A message-observing session type specifies how communication evolves taking into account messages that can be observed on other channels in a process's environment.
This is in contrast to both value and label-dependent session types, where a channel's type can only depend on messages that the channel previously carried, but not on messages on other channels in the process's environment. Hence, \chase allows us to express global invariants about processes rather than just local invariants.

To specify session types that vary based on messages on other channels, we extend binary session types with \emph{type-level processes}.
This allows us to capture a wider range of correctness guarantees than is currently possible.
To illustrate, we take a closer look at the identity process \(\ms{id}\) that provides an implementation for the session type \(A \coloneqq \Tplus \Set{ \ms{left} : \Tu, \ms{right} : \Tu }\).
This session type specifies that communication is a message containing either label \(\ms{left}\) or \(\ms{right}\), followed by a message signalling the end of communication.
We can implement \(\ms{id}\) between channels \(a\) and \(b\) of type \(A\) as follows:
\[
  \tRecvL{a}{\Set*{ \ms{left} \Rightarrow \tSendL{b}{\ms{left}}{\tWait{a}}{\tClose{b}} \mid \ms{right} \Rightarrow \tSendL{b}{\ms{right}}{\tWait{a}}{\tClose{b}}}}.
\]
Operationally, this process waits to receive a label on \(a\), selects the corresponding branch, and sends the label over \(b\).
It then waits for the channel \(a\) to close before closing \(b\). However, the session type for channel $a$ and channel $b$ does not capture this precisely.
In particular, the type of \(b\) does not rule out the following erroneous implementation that swaps labels:
\[
  \tRecvL{a}{\Set*{ \ms{left} \Rightarrow
{\color{red}{b.\ms{right}}} ;\ \tWait{a}{\tClose{b}} \mid \ms{right} \Rightarrow {\color{red}{b.\ms{left}}};\ \tWait{a}{\tClose{b}}}}
\]
The crux of the issue is that the messages we wish to allow on \(b\) depend on the messages that we observe on \(a\).
In \chase, we specify such a protocol for \(b\) using
type-level processes as follows:
\[
  b : \TCaseL{a}{\Set{ \ms{left} \Rightarrow {\Tplus \Set{ \ms{left} : \Tu }} \mid \ms{right} \Rightarrow {\Tplus \Set{ \ms{right} : \Tu}} }}.
\]
Intuitively, it specifies that the type of \(b\) is the unary choice \(\Tplus \Set{\ms{left} : \Tu}\) if a label \(\ms{left}\) is observed on \(a\), and symmetrically if \(\ms{right}\) is observed on \(a\).
A message-observing session type is necessarily open: it names the channel whose messages it observes.
This is analogous to processes, which contain the channel names on which they communicate.
The particular channel \(a\) observed by the type of \(b\) is determined by the process specification in which it appears.
For example, the following concrete syntax specifies \(\ms{id}\) by specifying the types of its channels, and it specifies that \(b\)'s type observes a channel \(a : A\) used by \(\ms{id}\):
\[
  \cproci{\ms{id}}
  {a : A}
  {b : \TCaseL{a}{\Set{ \ms{left} \Rightarrow {\Tplus \Set{ \ms{left} : \Tu }} \mid \ms{right} \Rightarrow {\Tplus \Set{ \ms{right} : \Tu}} }}}
  = \tRecvL{a}{\Set{\cdots}}.
\]
We will revisit this concrete syntax in \cref{sec:motivation}.

\chase provides a  ``bottom-up'' approach to specifying processes and their composition. We still specify the communication protocol
 from the point of view of one of the two participants, but take into account the messages observed on other channels in the environment.
An alternative to specify multi-channel communication patterns are \emph{multiparty} session types.
Multiparty session types~\cite{honda_2016:_multip_async_session_types} specify interactions between a static number participants using a \emph{global type}, and project these specifications onto individual participants for typechecking.
Despite recent work~\cite{denielou_yoshida_2011:_dynam_multir_session_types,stolze_2023:_compos_partial_multip}, multiparty session types are an inherently closed world or ``top-down'' approach, where the entire system must be designed before it can be implemented.
In contrast, we can compose protocols in \chase from the bottom-up, while still being able to capture some of the richer interactions of multiparty session~types.

Concretely, we make the following contributions:
\begin{enumerate}
\item We introduce a \textbf{message-observing binary session type system}, called \chase.
  A message-observing session type specifies how communication evolves taking into account communications that can be observed on other channels in our environment.
  This is achieved using \textbf{type-level processes} in session types.
  These type-level processes further restrict the eligible processes and their actions.
  We motivate this extension through a sequence of examples.

\item We give processes and their specifications a \textbf{semantics using \emph{traces with binding}}, a semantic foundation inspired by nominal sequences~\cite{gabbay_ghica_2012:_game_seman_nomin_model}.
  A frequent challenge when we define a trace semantics or ordered semantics for concurrency is that we must account for fresh channel name generation and propagate names involved in higher-order communications. Further, compositionality often requires infinite trace sets or infinite collections of semantic objects, which complicates implementation and obscures the computational content.
  Instead, we use name binding to quantify over all fresh channel in a manner reminiscent of higher-order abstract syntax, resulting in a more compact presentation. This abstraction allows us to retain compositionality.

\item We give a \textbf{semantically sound typechecking algorithm}.
  Process denotations characterize all possible communication behaviours, while specifications specify the processes traces they allow subject to constraints on ambient communications.
  Soundness ensures that well-typed processes only exhibit behaviours (traces) permitted by their specifications.
\end{enumerate}
We see this work as a significant step towards providing more precise guarantees about processes.

\section{Motivation}
\label{sec:motivation}

We introduce the main ideas of \chase through a sequence
of examples. This will allow us to showcase both the power of
message-observing session types and the design decisions.

\subsection{A Quick Guide to Session Types and Specifying Processes}
\label{sec:client-serv-arch}

Session types \(A, B, \dotsc\) specify the communication protocols on named channels \(a, b, \dotsc\).
Processes in our system are organized according to a client-server architecture.
This architecture is reflected in our concrete syntax for process specifications:
\[
  \cproci{\ms{proc}}[a_1 : A_1, \dotsc, a_n : A_n]{b_1 : B_1, \dotsc, b_n : B_n}{c_0 : C_0}
\]
It states that the process \(\ms{proc}\) is a \defin{server} for (or \defin{provides}) a distinguished service \(C_0\) over a channel \(c_0\).
Dually, it is a \defin{client} of (or \defin{uses}) zero or more services \(B_1, \dotsc, B_n\) on channels named \(b_1, \dotsc, b_n\), respectively.
Finally, its types may refer to zero or more ambient communication channels \(a_i : A_i\).
Ambient channels are instantiated either by composing \(\ms{proc}\) with processes that use or provide them, or by defining \(\ms{proc}\) as a composition of processes that communicate on these channels.
Ambient channels will permit compositional process specifications.
To keep our subsequent examples simple, most will feature empty internal and ambient contexts.
Each type can refer other channel names in the specification, allowing for mutual dependency between types.

For our first examples, we consider sending a sequence of \(n\) bits
on a channel \(a\). This is accomplished by sending a sequence of
$n$ labels drawn from set $\bit = \Set{\mt{0}, \mt{1}}$.
Such a sequence is specified by the session type \(\Tlist{\bit}{n}\) defined by induction on \(n\) using
the following pair of equations:%
\footnote{\chase omits language-level recursion and relies on meta-level induction to define inductive types.
  This simplifies \chase's presentation and avoids obscuring its key contributions.
  We nevertheless sketch how to extend \chase with recursive session types and processes in \cref{sec:recursion}.}
\begin{align*}
  \Tlist{\bit}{0} &= \Tu
  &
  \Tlist{\bit}{n + 1} &= \Tplus \Set{l : \Tlist{\bit}{n}}_{l \in \bit}.
\end{align*}
The empty sequence of bits is captured by the unit type \(\Tu\): it signals the end of communication on the channel.
To define a bit sequence of length $n + 1$, we specify the possible labels
(elements) that can be transmitted followed by the sequence of length
$n$.  To accomplish this, we use the internal choice type
$ \Tplus \Set{l : \Tlist{\bit}{n}}_{l \in \bit}$.\footnote{An internal choice \(\Tplus \Set{l : A_l}_{l \in L}\) specifies that a label \(l \in L\) will be transmitted, and then communication will satisfy~type~\(A_l\).}
Concretely, we can use the definition to generate the type for bit sequences of length $2$ as follows:
\[
  \Tlist{\bit}{2} = \Tplus \Set{ \mt{0} : \Tplus \Set{ \mt{0} : \Tu, \mt{1} : \Tu}, \mt{1} : \Tplus \Set{ \mt{0} : \Tu, \mt{1} : \Tu} }
\]
It specifies a sequence of two bits, \ie, a sequence of two labels,
$\mt{0}$ and $\mt{1}$ followed by termination.

We can then specify a family of identity processes, \(\idproc_n\), that use a channel $a :
\Tlist{\bit}{n}$ and provide an output channel $b$ of the same type by
the following pair of equations:
\[
  \begin{array}{ll@{}l@{}l@{}l@{}l@{~~=~~}l}
    \cproci{\idproc_0&}{&a : \Tlist{\bit}{0}&}{&b : \Tlist{\bit}{0}&} & \tWait{a}{\tClose{b}}\\
    \cproci{\idproc_{n + 1}&}{&a : \Tlist{\bit}{n + 1}&}{&b : \Tlist{\bit}{n + 1}&} & \tRecvL{a}{\Set{l \Rightarrow \tSendL{b}{l}{\idproc_{n}}}_{l \in \bit}}
  \end{array}
\]

In the base case, \(\idproc_0\), we wait to receive a close message on
\(a\) that signals the end of communication, and we immediately send
it on \(b\) before terminating.
In the recursive step, \(\idproc_{n + 1}\), we perform a case analysis on the label received over \(a\).
If we receive a label \(l\), then we take the corresponding branch and send the label \(l\) over \(b\).
At this point, \(a\) and \(b\) both have type \(\Tlist{\bit}{n}\), so
we continue as \(\idproc_{n}\).

\subsection{Specifying Identity Processes For Real Using
  Message-Observing Session Types}
\label{sec:spec-part-ident}

The session type specification of the identity process $\idproc_n$ in the previous
section fails to ensure that we copy the sequence of bits unchanged
from \(a\) to \(b\). In \chase, we can give a more precise type to the output channel by taking into account the message that we already have
received on a given channel using message-observing
sessions.
In particular, we can refine the type of \(b\) to the following more precise
family of types \(\Tidlist{a}{n}\) that observe messages on channel $a$. We give the definition
of the identity process next to it, to highlight the close
correspondence between the process and the message-observing session type:
\[
  \begin{array}{l@{~=~}l@{\quad}l@{~=~}l}
\multicolumn{2}{c}{\mbox{Process}} & \multicolumn{2}{c}{\mbox{Message-Observing
                                     Session Type}} \\ \midrule
 \idproc_0     & \tWait{a}{\tClose{b}}  & \Tidlist{a}{{\makebox[\widthof{\({}_{n + 1}\)}][l]{\(\scriptstyle 0\)}}} & \TCaseU{a}{\Tu}\\
 \idproc_{n+1} & \tRecvL{a}{\Set{l \Rightarrow
                \tSendL{b}{l}{\idproc_{n}}}_{l \in \bit}}  & \Tidlist{a}{n + 1} & \TCaseL{a}{\Set{l \Rightarrow \Tplus \Set{l : \Tidlist{a}{n}}}_{l \in \bit}}
  \end{array}
\]
In the base case, the type \(\TCaseU{a}{\Tu}\) specifies that we can
only signal termination on \(b\) after we have observed termination on
\(a\).
In the recursive step, \(\Tidlist{a}{n + 1}\) performs a case analysis on the label sent on \(a\).
If a label \(l\) is sent or received on \(a\), then the type reduces to the unary internal choice \(\Tplus \Set{l : \Tidlist{a}{n}}\) that only allows the label \(l\) to be sent, and continues as \(\Tidlist{a}{n}\).
We can then refine the specification of \(\ms{id}\) to use this more precise protocol:
\[
  \begin{array}{ll@{}l@{}l@{}l@{}l@{~~=~~}l}
    \cproci{\idproc_0&}[]{&a : \Tlist{\bit}{0}&}{&b : \Tidlist{a}{{\makebox[\widthof{\({}_{n + 1}\)}][l]{\(\scriptstyle 0\)}}}&} & \tWait{a}{\tClose{b}}\\
    \cproci{\idproc_{n + 1}&}[]{&a : \Tlist{\bit}{n + 1}&}{&b : \Tidlist{a}{n + 1}&} & \tRecvL{a}{\Set{l \Rightarrow \tSendL{b}{l}{\idproc_{n}}}_{l \in \bit}}
  \end{array}
\]
Note that the process and the type of the used channel \(a\) remain the
same as in the previous section: the only change is to the type of the channel \(b\),
which now observes messages on \(a\).
At the high-level, the type $\Tidlist{a}{n}$ ensures that \(\ms{id}_n\) can only output on \(b\) what it receives on \(a\).

We call the type \(\Tidlist{a}{n}\) \defin{message observing} because it observes communications on \(a\) to determine the range of permitted communications.
It can be thought of as a type-level process, as it mimics choices and behaviour specified in the process \(\ms{id}_n\) on the type-level.
This type reduces as messages are observed on the channel \(a\).
These observations and reductions exist only in the typechecking algorithm (types are not present at runtime), which tracks messages sent or received by the specified process and reduces types accordingly.
For example, to type check \(\idproc_{n + 1}\) against its specification, we check each of its branches ``\(\tSendL{b}{l}{\idproc_{n}}\)'' (to which it steps after receiving a label \(l\)) against the specification ``\(\cint{a : \Tlist{\bit}{n}}{b : \Tplus \Set{l : \Tidlist{a}{n}}}\)'', which is obtained by reducing all types by the label \(l\) received on \(a\).

This example illustrates two key points in \chase:
\begin{enumerate*}
\item message-observing session
types can express dependencies on messages on other
channels, and
\item message-observing types behave as type-level processes and incorporate
a notion of concurrent computation on the type-level.
\end{enumerate*}

On a practical level, this allows us to express more
precise safety guarantees that cannot be expressed by prior work.
On a more theoretical level, we see message-observing session types
as an essential step towards a type theory of session types.

\subsection{Expressing Mutual Observation %
 in Message-Observing Session Types}
\label{sec:mutu-depend-high}

Session-typed languages often feature a higher-order session type \(A \Tot B\),
which specifies a server that sends a channel of type \(A\) (call it \(a\)) and then communicates according to \(B\).

It is useful in applications for \(B\) to observe messages on \(a\).
To do so, we introduce the syntax \((a : A) \otimes B\), which binds a name \(a\) for the transmitted channel of type \(A\) in \(B\).
This bound name is eventually instantiated by the actual name of the transmitted channel.
This syntax then lets us specify, \eg, sending a pair of identical bit streams: \( \left(a : \Tlist{\bit}{n}\right) \Tot \left(\Tidlist{a}{n}\right) \).
A server of this type sends a channel \(a\) carrying \(n\) bits and then communicating according to the protocol \(\Tidlist{a}{n}\), which specifies transmitting a bit stream equal to the one carried by \(a\).

Communication on both channels occurs independently, so it is natural to ask whether \(A\) can also observe messages on the channel of type \(B\) (call it \(b\)).
From the dependent-types perspective in the sequential setting, such mutual dependency may seem unusual, but it captures the concurrent behaviour where the processes providing \(A\) and \(B\) run concurrently.
To allow for mutual observation, we use the syntax \((a : A) \Tot (b : B)\) for higher-order sessions, where name \(a\) is bound in \(B\) and name \(b\) is bound in \(A\).
It allows the type \(B\) to observe messages on the transmitted channel \(a : A\),
while also allowing the type \(A\) to observe messages on the channel \(b : B\).
In practice, \(b\) will always be instantiated with the name of the channel of type \((a : A) \Tot (b : B)\); we treat it as bound in the type only to ensure that the type makes sense independently of the channel that it types.
As a result, we can specify mutually observing higher-order protocols.

To illustrate mutual observation, we consider a load-balanced list service.
This service provides a pair of lists, and then forces the client to alternate between lists when receiving elements.
It is given by \((a : \Tlblist{b}{\bit}{n}) \Tot (b : \Tlblist{a}{\bit}{n}) \), where \(\Tlblist{c}{\bit}{n}\) is inductively defined by:
\begin{align*}
  \Tlblist{c}{\bit}{0} &= \Tu &
  \Tlblist{c}{\bit}{n + 1} &= \Tplus \Set{x : \TCaseL{c}{\Set{y \Rightarrow \Tlblist{c}{\bit}{n}}_{y \in \bit}}}_{x \in \bit}.
\end{align*}
Intuitively, a channel \(a : \Tlblist{c}{\bit}{n + 1}\) provides a list element \(x \in \bit\), and then requires the client to observe some element \(y\) on \(c\) before it can observe another list element from \(a\).

To see how the protocol \((a : \Tlblist{b}{\bit}{n + 1}) \Tot (b : \Tlblist{a}{\bit}{n + 1}) \) enforces alternation, consider one of its clients.
After receiving the channel provided by this service, the client has two channels: channels \(a : \Tlblist{b}{\bit}{n + 1}\) and \(b : \Tlblist{a}{\bit}{n + 1}\).
Assume, without loss of generality, that it first receives a label on \(a\).
The internal choice type specifies that communication on \(a\) then satisfies \(\TCaseL{b}{\Set{y \Rightarrow \Tlblist{b}{\bit}{n}}_{y \in \bit}}\).
In particular, the client cannot communicate on \(a\) until it reduces the \(\ms{CASE}\) analysis.
To do so, the client must receive a label on \(b\).
It can do so, for the type of \(b\) is an internal choice \(\Tplus \Set{x : \Tlblist{a}{\bit}{n}}_{x \in \bit}\) (it was reduced by the label received on \(a\)).
After receiving a label on \(b\), the type of \(a\) reduces to \(\Tlblist{b}{\bit}{n}\), while the type of \(b\) becomes \(\Tlblist{a}{\bit}{n}\).

For convenience, we write \(A \Tot (b : B)\) for \((a : A) \Tot (b : B)\) when \(a\) does not appear in \(B\).
The abbreviations \((a : A) \Tot B\) and \(A \Tot B\) are analogous.

\subsection{Depending on Higher-Order Sessions}
\label{sec:depend-high-order}

Protocols can depend on higher-order sessions using an elimination form reminiscent of positive product elimination in functional languages.
We illustrate this by specifying an \textsc{and} logic gate, given by the process \(\ms{AND}\):
\begin{alignat*}{2}
  \cproci{\ms{AND}}{b : \Tlist{\bit}{1} \Tot \Tlist{\bit}{1}}{c : \Tlist{\bit}{1}} = {}\\
  \tRecvC{a}{b}{\tRecvL{a}{
  \{\,&\mt{1} &&\Rightarrow \tRecvL{b}{\Set{\mt{1} \Rightarrow \tSendL{c}{\mt{1}}{\ms{end}} \mid \mt{0} \Rightarrow \tSendL{c}{\mt{0}}{\ms{end}}} }\\
  {}\mid {}&\mt{0} &&\Rightarrow \tRecvL{b}{\Set{ x \Rightarrow \tSendL{c}{0}{\ms{end}}}_{x \in \bit}}}\,\}}
\end{alignat*}
where \(\cproci{\ms{end}}{a : \Tu, b : \Tu}{g : \Tu} = (\tWait{a}{\tWait{b}{\tClose{g}}})\).
From its specification, we see that \(\ms{AND}\) receives a pair of bits on \(b\) and sends a bit on \(c\).
Its implementation starts by receiving a channel of type \(\Tlist{\bit}{1}\) over \(b\) and binding it to the name \(a\) (syntax: ``\(\tRecvC{a}{b}{\cdots}\)'').
At this point, the process is a client of \(a : \Tlist{\bit}{1}\) and \(b : \Tlist{\bit}{1}\) and provides \(c : \Tlist{\bit}{1}\).
The process then observes the bits on \(a\) and \(b\), sends their \textsc{and} over \(c\), and terminates.

Observe that the above specification does not specify that \(\ms{AND}\) correctly implements an \textsc{and} gate.
However, we can use message observation to specify this functionality at the type level.
To do so, we refine the above specification to
\[
  \cproci{\ms{AND}}{b : \Tlist{\bit}{1} \Tot \Tlist{\bit}{1}}{c : \ms{andOf}\ b} = {\cdots}
\]
where the following type specifies an \textsc{and} gate:
\begin{align*}
  \ms{andOf}\ b =
  \TCaseC{b}{a}{\TCaseL{a}{
  \{\,\mt{1} &\Rightarrow \TCaseL{b}{\Set{\mt{1} \Rightarrow \Tplus \Set{\mt{1} : \Tu} \mid \mt{0} \Rightarrow \Tplus \Set{\mt{0} : \Tu} }\\
  {} \mid \mt{0} &\Rightarrow \Tplus \Set{\mt{0} : \Tu}} }\,\}}.
\end{align*}
This type-level process uses the new construct \(\TCaseC{b}{a}{C}\), which binds the name of a channel transmitted over \(b\) to \(a\) in \(C\).
This construct reduces to \(\subst{\alpha}{a}{C}\) when a channel \(\alpha\) is transmitted on \(b\).
This reduction closely mimics the behaviour of the channel-receiving process construct, \(\tRecvC{a}{b}{P}\), which becomes \(\subst{\alpha}{a}{P}\) after receiving \(\alpha\) on \(b\).
In the case of ``\(\ms{andOf}\ b\)'', it reduces to the internal choice \(\Tplus \Set{ \mt{1} : \Tu }\) if the transmitted channel and its carrier both carry a \(\mt{1}\) bit, and otherwise it reduces to \(\Tplus \Set{ \mt{0} : \Tu }\).

\subsection{A Fair Auction}
\label{sec:an-auction}

In our examples so far, there has been a close correspondence between
processes and the message-observing types in their specifications. In fact, this correspondence
is partially by design as we take into account the process
communication in defining a message-observing type. The next example
shows that this correspondence between processes and
message-observing types is not always trivial, and that processes can be
more complex than the types specifying their communications.

Assume two bidders participating in an auction, where each may bid 0
(described by the label $\mt{0}$) or 1 (described by the label $\mt{1}$).
We want to ensure that the bidder with the highest bid wins.
With traditional session types, we can specify the protocols for a bidder service as follows, where the external choice \(\Tamp \{ \cdots \}\) means that the bidder \emph{receives} the label:
\begin{align*}
  \ms{bidder} &= \Tplus \Set{ \mt{0} : \ms{result}, \mt{1} : \ms{result} }\\
  \ms{result} &= \Tamp \Set{ \mt{lost} : \Tu, \mt{tie} : \Tu, \mt{win} : \Tu }
\end{align*}
A bidder is a process that provides a channel \(b\) of type \(\ms{bidder}\); the name \(b\) can freely be renamed, provided that it is kept distinct from other names in the process and its specification:
\[
  \cproci{\ms{Bidder}}{}{b : \ms{bidder}} = {\cdots}
\]
An \(\ms{Auctioneer}\) is a client of two bidders and signals the end of the auction on its provided channel \(c\).
We can implement it as follows, where  \(\cproci{\ms{end}}{b_1 : \Tu, b_2 : \Tu}{c : \Tu} = (\tWait{b_1}{\tWait{b_2}{\tClose{c}}})\):
\begin{alignat*}{6}
  \mathclap{\cproci{\ms{Auctioneer}}[]{b_1 : \ms{bidder}, b_2 : \ms{bidder}}{c : \Tu} = {}}\\
  \tRecvL{b_1}{&\{\, \mt{0} &&\Rightarrow \tRecvL{b_2}{&&\{\, \mt{0} &&\Rightarrow \tSendL{b_1}{\mt{tie}}{&&\tSendL{b_2}{\mt{tie}}{&&\ms{end}}}\\
             &&&&&\mid \mt{1} &&\Rightarrow \tSendL{b_1}{\mt{lost}}{&&\tSendL{b_2}{\mt{win}}{&&\ms{end}}} \,\}}\\
  &\mid \mt{1} &&\Rightarrow \tRecvL{b_2}{&&\{\, \mt{0} &&\Rightarrow \tSendL{b_1}{\mt{win}}{&&\tSendL{b_2}{\mt{lost}}{&&\ms{end}}}\\
             &&&&&\mid \mt{1} &&\Rightarrow \tSendL{b_1}{\mt{tie}}{&&\tSendL{b_2}{\mt{tie}}{&&\ms{end}}} \,\}\,\}}}
\end{alignat*}
Observe that the specification does not rule out an unfair auctioneer that privileges bidder \(b_1\) and always give them the win, \ie, an auctioneer where each branch is \(\tSendL{b_1}{\mt{win}}{\tSendL{b_2}{\mt{lost}}{\ms{end}}}\).

\newcommand{\fairAuct}[1]{\ms{fairAuct}~#1}

To prohibit unfair auctioneers, we use a message-observing protocol that specifies the interactions for one bidder in terms of the actions of its opponent \(o\):
\begin{alignat*}{2}
  \fairAuct o &= \Tplus \{\, &&\mt{0} : \TCaseL{o}{\Set{ \mt{0} \Rightarrow \Tamp \Set{ \mt{tie} : \Tu } \mid  \mt{1} \Rightarrow \Tamp \Set{ \mt{lost} : \Tu } }},\\
  &&&\mt{1} : \TCaseL{o}{\Set{ \mt{0} \Rightarrow \Tamp \Set{ \mt{win} : \Tu } \mid  \mt{1} \Rightarrow \Tamp \Set{ \mt{tie} : \Tu } }} \,\}
\end{alignat*}
The following specification then ensures that \(\ms{Auctioneer}\) is fair:
\[
  \cproci{\ms{Auctioneer}}[]{b_1 : \fairAuct{b_2}, b_2 : \fairAuct{b_1}}{c : \Tu} = {\cdots}.
\]
Indeed, the fair auction protocol \(\fairAuct{b_2}\) specifying communications on \(b_1\) ensures that if \(b_1\) sends a label \(\mt{0}\), then the auctioneer can only send \(\mt{tie}\) or \(\mt{lost}\) to \(b_1\), and then only after a label \(\mt{0}\) or \(\mt{1}\) has been observed on \(b_2\), respectively.
The case when \(b_1\) bids 1 is analogous.

This example illustrates that specifications do not simply lift
process implementations to the type level. In particular, processes can specify more
complex interactions than their types.
This illustration is even more striking when we consider the following implementation of a bidder.
It bids 0, and then terminates after having received the result of the auction:
\begin{align*}
  &\cproci{\ms{BidsZero}}[b_2 : \fairAuct{b_1}]{}{b_1 : \fairAuct{b_2}} ={}\\
  &\quad\tSendL{b_1}{\mt{0}}{\tRecvL{b_1}{\Set{ \ms{result} \Rightarrow \tClose{a} }_{\ms{result} \in \Set{ \mt{lost}, \mt{tie}, \mt{win} }}}}.
\end{align*}
This example illustrates how we track the ambient channel \(b_2 : \fairAuct{b_1}\) in the specification to ensure that the type of \(b_1\) remains well-scoped.
When typechecking \(\ms{BidsZero}\) against its specification, our typechecking algorithm generates constraints on messages that can appear on \(b_2\) that are then checked when \(\ms{BidsZero}\) is composed with a process implementing \(b_2\).

The ambient context helps ensure that we know the types of all channels observed by types in process interfaces.
As a result, we can ensure that types are well-formed relative to each other.
It also means that we can modularly specify processes: we specify the types of local channels, but these can observe channels used by other processes in our execution environment.

\subsection{Process composition}
\label{sec:process-composition}

\newcommand{\negBit}[1]{\ms{negBit}\ #1}

To illustrate how \chase typechecks compositions, we consider the composition of two bit-flipping processes.
A bit-flipping process is a client of a bit that provides its negation:
\[
  \cproci{\ms{neg}}{i : \Tlist{\bit}{1}}{o : \Tlist{\bit}{1}}
  =
  \tRecvL{i}{\Set{ \mt{0} \Rightarrow \tSendL{o}{\mt{1}}{\ms{end}} \mid \mt{1} \Rightarrow \tSendL{o}{\mt{0}}{\ms{end}} }}
\]
where  \(\cproci{\ms{end}}{i : \Tu}{o : \Tu} = (\tWait{i}{\tClose{o}})\).
We can give \(\ms{neg}\) more precise specifications.
For example, we can specify that
\begin{alignat*}{2}
  &\quad&&\cproci{\ms{neg}}{i : \Tlist{\bit}{1}}{o : \negBit i} = {\cdots}\\
  \text{or}&\quad&&\cproci{\ms{neg}}[c : \Tlist{\bit}{1}]{i : \negBit c}{o : \Tidlist{c}{1}} = {\cdots},\\
  \text{where}&\quad&&\negBit a = \TCaseL{a}{\Set{ \mt{0} \Rightarrow \Tplus \Set{ \mt{1} : \Tu} \mid \mt{1} \Rightarrow \Tplus \Set{ \mt{0} : \Tu} }}.
\end{alignat*}
The first specification ensures that \(\ms{neg}\) sends the negation of the bit it receives on \(i\) over \(o\).
The second is more interesting: it specifies that if \(\ms{neg}\) receives on \(i\) the negation of a bit sent on some ambient channel \(c\), then it outputs that original bit on \(o\).

Using these two specifications, we can ensure that flipping a bit twice behaves as the identity.
To do so, we compose an implementation of \(\ms{neg}\) with itself:
\[
  \cproci{\ms{doubleNeg}}[c : \negBit i]{i : \Tlist{\bit}{1}}{o : \Tidlist{i}{1}} =
  \tPar{c}
  {(i \leftrightarrow \ms{neg} \leftrightarrow c)}
  {(c \leftrightarrow \ms{neg} \leftrightarrow o)}.
\]
The syntax \((i \leftrightarrow \ms{neg} \leftrightarrow c)\) instantiates the implementation of \(\ms{neg}\) with the channel names \(i\) and \(c\), while the syntax \(P \mathbin{\|_c} Q\) composes a server \(P\) with its client \(Q\) on \(c\).
The type of the composition channel \(c\) is determined by the ambient context \(\Set{c : \negBit i}\).
The type checking algorithm checks each composed process against the types specified by the outer specification, \ie, that
\begin{alignat*}{4}
  \cproci{\ms{neg}}[&&]{&i : \Tlist{\bit}{1}&}{&c : \negBit i&}&{} = {\cdots}\\
  \cproci{\ms{neg}}[&i : \Tlist{\bit}{1}&]{&c : \negBit i&}{&o : \Tidlist{i}{1}&}&{} = {\cdots}.
\end{alignat*}
We treat the channel \(i\) as ambient in the second specification (it is not accessible to its process).
Typechecking the first process poses no difficulties; checking second process is interesting because it is a client of the channel \(c\) whose type immediately observes the ambient channel \(i\).
To check the second process, we do a case analysis on the messages that could appear on \(i\), reduce the types of \(c\) and \(o\) accordingly in each case, and check that \(\ms{neg}\) is well typed in the reduced specification.
For instance, in the case where \(\mt{0}\) appears on \(i\), we check that:
\[
  \cproci{\ms{neg}}[i : \Tu]{c : \Tplus \Set{\mt{1} : \Tu}}{o : \Tplus \Set{\mt{0} : \Tu}} = {\cdots}.
\]
As a side effect, we generate a constraint relating the label \(\mt{0}\) on \(i\) to the behaviour of \(\ms{neg}\) on \(c\) and \(o\).
After checking both processes independently, our algorithm checks that they impose mutually consistent constraints, and that they satisfy each other's constraints where applicable.
In this example, the second process produces a constraint that \(\ms{neg}\) receives \(\mt{1}\) on \(c\) only if \(\mt{0}\) appears on \(i\); this constraint is clearly satisfied by the first negation process.
Assuming that all constraints are satisfied, we conclude that the composition is well-typed.

The types of composition channels can be locally specified using the program syntax \(\tNew{a}{A}{P}\), which binds a private channel \(a : A\) in \(P\).
This privacy also ensures that \(a\) cannot externally be observed.
For example, we can hide the fact that \(\ms{doubleNeg}\) is implemented as a composition and ensure that no other processes can observe its composition channel \(c\) by hiding \(c : \negBit i\):
\[
  \cproci{\ms{doubleNeg'}}{i : \Tlist{\bit}{1}}{o : \Tidlist{i}{1}} =
  \tNew{c}{\negBit i}{\ms{doubleNeg}}.
\]

More generally, the syntax \(\tNewV{(a_1 : A_1, \dotsc, a_n : A_n)}{P}\) binds channels \(a_1, \dotsc, a_n\)  in \(P\).
To check \(\tNewV{\overrightarrow{a : A}}{P}\) against a specification, we check \(P\) against the same specification extended with the ambient channels \(\overrightarrow{a : A}\).
For example, checking \(\ms{doubleNeg'}\) entails checking the specification~\(\ms{doubleNeg}\).

Our treatment of composition departs from many simply session typed systems~\cite{caires_pfenning_2010:_session_types_intuit_linear_propos, wadler_2014:_propos_as_session} that combine parallel composition with hiding, \ie, that define the composition of \(P\) and \(Q\) along \(a\) as \(\tNew{a}{A}{\tPar{a}{P}{Q}}\).
We distinguish these operations to more flexibly specify process compositions and to ensure that process composition is associative and partially commutative.
Indeed, suppose we wished to chain together three processes:
\begin{align*}
  &\cproci{P_1}[a_3 : A_3(a_2)]{a_1 : \Tu}{a_2 : A_2(a_3)}\\
  &\cproci{P_2}{a_2 : A_2(a_3)}{a_3 : A_3(a_2)}\\
  &\cproci{P_3}[a_2 : A_2(a_3)]{a_3 : A_3(a_2)}{a_4 : \Tu},
\end{align*}
where we write \(A_i(a_j)\) to mean \(A_i\) observes \(a_j\), to form a process \(\cproci{P}{a_1 : \Tu}{a_4 : \Tu}\).
The specification of \(P\) is clearly well-defined.
However, if we always combine parallel composition with hiding, then neither of the following composition specifications is well-defined:
\begin{align*}
  &\cproci{P_{12}}{a_1 : \Tu}{a_3 : A_2(a_3)} = \tNew{a_2}{A_2(a_3)}{\left(\tPar{a_2}{P_1}{P_2}\right)}\\
  &\cproci{P_{23}}{a_2 : A_2(a_3)}{a_4 : \Tu} = \tNew{a_3}{A_3(a_2)}{\left(\tPar{a_3}{P_2}{P_3}\right)}
\end{align*}
This is because the types \(A_2(a_3)\) and \(A_3(a_2)\) are ill-scoped after hiding the channels \(a_2\) and \(a_3\).
In contrast, if we hide \(a_2\) and \(a_3\) \emph{after} the parallel composition, we can successfully define \(P\) as:
\[
  \cproci{P}{a_1 : \Tu}{a_4 : \Tu} = \tNewV{\left(a_2 : A_2(a_3), a_3 : A_3(a_2)\right)}{\left(\tPar{a_2}{P_1}{\left(\tPar{a_3}{P_2}{P_3}\right)}\right)}.
\]

Our examples show how \chase modularly specifies processes while guaranteeing rich invariants.

\section{A Higher-Order Process Language}
\label{sec:higher-order-process}

Processes \(P\) are generated by the following grammar, where \(a\) and \(b\) range over channel names, \(L\) over sets of choice labels, \(l\) and \(k\) over labels, and \(A\) over types:
\begin{align*}
\text{Process}~ P,Q &\Coloneqq \tClose{a} && \text{End communication on \(a\) and terminate}\\
    &\grammid \tWait{a}{P} && \text{Wait for communication to end on \(a\); continue as \(P\)}\\
    &\grammid \tSendL{a}{k}{P} && \text{Send label \(k\) on \(a\); continue as \(P\)}\\
    &\grammid \tRecvL{a}{\Set{l \Rightarrow P_l}_{l \in L}} && \text{Continue as \(P_l\) after receiving label \(l \in L\) on \(a\)}\\
    &\grammid \tSendC{a}{b}{P}{Q} && \text{Send a channel provided by \(P\) on \(a\); continue as \(Q\)}\\
    &\grammid \tRecvC{b}{a}{P} && \text{Receive channel \(b\) on \(a\); continue as \(P\)}\\
    &\grammid \tPar{a}{P}{Q} && \text{Compose \(P\) and \(Q\) along channel \(a\)}\\
    &\grammid \tNewV{(a_1 : A_1, \dotsc, a_n : A_n)}{P} && \text{Introduce private channels \(a_i\) of type \(A_i\) in \(P\)}
\end{align*}
The channel name \(b\) is bound in \(P\) in \(\tRecvC{b}{a}{P}\) and in \(\tSendC{a}{b}{P}{Q}\).
In \(\tSendC{a}{b}{P}{Q}\), it represents the sent channel provided by \(P\).
Intuitively, this process forks \(P\) with \(b\) instantiated by a fresh name, and sends this name over \(a\).
The names \(a_1, \dotsc, a_n\) are bound in \(P\) in \(\tNewV{(a_1 : A_1, \dotsc, a_n : A_n)}{P}\), and the type annotations exist only to simplify type checking.
For the semantics to be reasonable, we require that \(a\) does not appear free in \(P\) in \(\tWait{a}{P}\), that \(P\) and \(Q\) have disjoint sets of free names in \(\tSendC{a}{b}{P}{Q}\), and that their free names intersect only in \(a\) in \(\tPar{a}{P}{Q}\).
As is usual for session-typed processes, we consider \emph{open} processes: the free names in a process name its used and provided channels.

\subsection{Trace Semantics}
\label{sec:trace-semantics}

A process denotes a set of traces that describe its possible executions.
Traces are sequences of elements \((s; m)\), where \(m\) is a message on a channel and \(s\) is a tag indicating if \(m\) was sent or received, or if it appeared on an unhidden internal channel.
To model sending and receiving channels (process constructs \(\tSendC{a}{b}{P}{Q}\) and \(\tRecvC{b}{a}{P}\)), we extend traces with a binding structure similar to the coasbraction operator for nominal sequences~\cite{gabbay_ghica_2012:_game_seman_nomin_model}.
To help ensure that our semantics is compositional, we bind \(b\) in the tail of a trace after a channel transmission message.
This avoids free name clashes when interleaving traces in the denotation of process composition.
Advantageously, our approach keeps the denotations of processes finite, a key property in ensuring that typechecking terminates.
Explicitly, traces \(t\), messages \(m\), and signs \(s\) are given by the grammar:
\begin{align*}
\text{Observable Signs}~ \poS &\Coloneqq \pO \grammid \pI \grammid \pS &
\text{Messages}~ m &\Coloneqq \mClose{a}   &
\text{Traces}~ t &\Coloneqq \tempt\\
\text{Signs}~ s &\Coloneqq \poS \grammid \pC  &  &\grammid \mLabel{a}{l} &   &\grammid \tcons{s}{m}[\vec a]{t}\\
    &&    &\grammid \mChan{a}  &   &
\end{align*}
Observable \defin{signs} specify actions performed by processes.
They are \(\pI\) (input), \(\pO\) (output), and \(\pS\)
(internal synchronization).
The constraint sign \(\pC\) is not used in this section, but will be used in the semantics of message-observing types.
It specifies a constraint on an ambient channel, \ie, that a message must be observed in the ambient environment before the trace can continue.
\defin{Messages} \(m\) describe possible messages on a given channel \(a\).
The message ``\(\mClose{a}\)'' means the end of communication, ``\(\mLabel{a}{l}\)'' describes sending a label \(l\),
and ``\(\mChan{a}\)'' captures sending a channel.
We write \(\carrcn(m)\) for the name of the channel carrying \(m\); in
the above cases, \(\carrcn(m) = a\).
Finally, \defin{traces} follow a list-like structure.
The empty trace is \(\tempt\).
The trace \(\tcons{s}{m}[\vec a]{t}\) prefixes a message \(m\) with sign \(s\) onto trace \(t\), with zero or more names \(\vec a\) bound in \(t\).
We identify traces up to \(\alpha\)-equivalence.
The \defin{free channel names} \(\freecn(t)\) in a trace \(t\) are defined by induction on the structure of \(t\): \(\freecn(\tempt) = \emptyset\) and \(\freecn(\tcons{s}{m}[\vec a]{t}) = \Set{ \carrcn(m) } \cup (\freecn(t) \setminus \vec a) \).

Processes denote sets of traces, where each trace denotes a possible interleaving of its actions.
The denotation \(\ptraces{P}\) of a process \(P\) is given by induction on its syntax.
We explain the semantic clauses and the trace operators they use below:
\begin{align}
  \ptraces{\tClose{a}} &= \Set{ \tcons{\pO}{\mClose{a}}{\tempt} } \label{eq:process-language:1}\\
  \ptraces{\tWait{a}{P}} &= \tcons{\pI}{\mClose{a}}{ \ptraces{P} } \label{eq:process-language:2}\\
  \ptraces{\tSendL{a}{k}{P}} &= \tcons{\pO}{\mLabel{a}{k}}{ \ptraces{P} } \label{eq:process-language:3}\\
  \ptraces{\tRecvL{a}{\Set{ l \Rightarrow P_l }_{l \in L}}} &= \bigcup\nolimits_{l \in L} \tcons{\pI}{\mLabel{a}{l}}{ \ptraces{P_l} } \label{eq:process-language:4}\\
  \ptraces{\tSendC{a}{b}{P}{Q}} &= \tcons{\pO}{\mChan{a}}[b]{ (\sembr{P} \pint \sembr{Q}) } \label{eq:process-language:5}\\
  \ptraces{\tRecvC{b}{a}{P}} &= \tcons{\pI}{\mChan{a}}[b]{ \ptraces{P} } \label{eq:process-language:6}\\
  \ptraces{\tPar{a}{P}{Q}} &= \sembr{P} \pint \sembr{Q} \label{eq:process-language:7}\\
  \ptraces{\tNewV{(a_1 : A_1, \dotsc, a_n : A_n)}{P}} &= \tdel{\sembr{P}}{\Set{a_1, \dotsc, a_n}} \label{eq:process-language:8}\\
  \text{where}\quad\tcons{s}{m}[\vec{a}]{T} &= \Set{ \tcons{s}{m}[\vec{a}]{t} \given t \in T}\nonumber\\
  T_1 \pint T_2 &= \Set{ t \given t_1 \in T_1,\, t_2 \in T_2,\, t \in (t_1 \pint t_2) }\nonumber\\
  \tdel{T}{\Set{a_1, \dotsc, a_n}} &= \Set{ \tdel{t}{\Set{a_1, \dotsc, a_n}} \given t \in T }\nonumber
\end{align}
\Cref{eq:process-language:1} specifies that the only action performed
by the process \((\tClose{a})\) is sending a close message on \(a\).
Dually, \cref{eq:process-language:2} specifies that every execution of
the process \((\tWait{a}{P})\) receives a close message on \(a\) before continuing as \(P\).

\Cref{eq:process-language:3} similarly captures sending a label \(k\) on \(a\).
\Cref{eq:process-language:4} specifies that after receiving a label \(l\), the process \(P = ({\tRecvL{a}{\Set{ l \Rightarrow P_l }_{l \in L}}})\) continues executing as \(P_l\).
The meaning of the process \(P\) is then the union of all the traces for $P_l$ prefixed by the received label $l$.

\Cref{eq:process-language:5} specifies sending a channel \(b\) provided by \(P\) over \(a\) and then continuing as \(Q\).
Processes \(P\) and \(Q\) execute independently, so the traces that follows sending \(b\) are interleavings of the executions of \(P\) and \(Q\), \ie, elements of \(\sembr{P} \pint \sembr{Q}\).
The interleaving operator \(\pint\) is defined in \cref{def:process-language:1}.
We assume without loss of generality that the bound name \(b\) is chosen distinct from any free name in \(Q\).
Because \(P\) and \(Q\) are assumed to have disjoint sets of free channel names, so will their traces.
Therefore, \(\pint\) will effectively compute arbitrary interleavings of these two traces.
\Cref{eq:process-language:6} specifies receiving a channel over \(a\) and binding it to the name \(b\) in \(P\).

\Cref{eq:process-language:7} states that the parallel composition of processes is given by the synchronized interleavings of their traces.
The synchronized interleavings of a pair of traces are defined in such a way that each receive action synchronizes with a send action when available.
In particular, it captures a synchronous communication semantics.
The process syntax \(\tCut{a}{P}{Q}\) specifies the channel name \(a\) along which \(P\) and \(Q\) are composed to specify the only name shared between \(P\) and \(Q\) and to guide typechecking.
In contrast, the interleaving operator does not mention \(a\).
Our operator is sufficiently general to synchronize traces on multiple channels.
This is required, \eg, to handle synchronization for channel transmission, where we must synchronize processes traces on two channels: the transmitted channel and the channel carrying it.

Finally, \cref{eq:process-language:8} hides the bound channels \(a_i\) by deleting all messages appearing on \(a_i\) from the traces in \(\sembr{P}\).
It uses the deletion operation \(\tdel{t}{\Set{a_1, \dotsc, a_n}}\) which is given in \cref{def:process-language:2}.
This deletion operator is defined in such a manner that it deletes not only actions on the \(a_i\), but also actions appearing on the channels they carried, and so on and so forth.

To avoid proliferating definitions for trace operators, we postpone their formal definitions until we have discussed the slightly more general case of traces with constraints; readers wishing to skip ahead are invited to see \cref{sec:operation-traces}.
Meanwhile, we consider a few intuition-building examples:

\begin{example}
  \label{ex:process-language:1}
  We compute the denotation of the double negation process \(\ms{doubleNeg'}\) of \cref{sec:process-composition}.
  We start by computing the denotation of the bit flipping process \(\ms{neg}\) composing it:
  \begin{align*}
    \sembr{i \leftrightarrow \ms{neg} \leftrightarrow c} = \{\,
    &\tcons{\pI}{\mLabel{i}{\mt{0}}}{
      \tcons{\pO}{\mLabel{c}{\mt{1}}}{
      \tcons{\pI}{\mClose{i}}{
      \tcons{\pO}{\mClose{c}}{\tempt}}}},\\
    &\tcons{\pI}{\mLabel{i}{\mt{1}}}{
      \tcons{\pO}{\mLabel{c}{\mt{0}}}{
      \tcons{\pI}{\mClose{i}}{
      \tcons{\pO}{\mClose{c}}{\tempt}}}} \,\},\\
    \sembr{c \leftrightarrow \ms{neg} \leftrightarrow o} = \{\, & \text{analogous} \,\}.
  \end{align*}
  The denotation of \(\ms{doubleNeg}\) uses the synchronized interleaving operator \(\pint\) to interleave traces from each denotation.
  Intuitively, it matches \((\pO; m)\) and \((\pI; m)\) elements and replaces them by synchronization elements \((\pS; m)\); the remaining elements are freely interleaved.
  For example, it interleaves the following traces, respectively from \(\sembr{i \leftrightarrow \ms{neg} \leftrightarrow c}\) and \(\sembr{c \leftrightarrow \ms{neg} \leftrightarrow o}\),
  \begin{align*}
    &\tcons{\pI}{\mLabel{i}{\mt{0}}}{
      \tcons{\pO}{\mLabel{c}{\mt{1}}}{
      \tcons{\pI}{\mClose{i}}{
      \tcons{\pO}{\mClose{c}}{\tempt}}}},\\
    &\tcons{\pI}{\mLabel{c}{\mt{1}}}{
      \tcons{\pO}{\mLabel{o}{\mt{0}}}{
      \tcons{\pI}{\mClose{c}}{
      \tcons{\pO}{\mClose{o}}{\tempt}}}},
  \end{align*}
  to produce, among others, the following traces in \(\sembr{\tPar{c}
    {(i \leftrightarrow \ms{neg} \leftrightarrow c)}
    {(c \leftrightarrow \ms{neg} \leftrightarrow o)}}\):
  \begin{align*}
    &\quad\tcons{\pI}{\mLabel{i}{\mt{0}}}{
      \tcons{\pS}{\mLabel{c}{\mt{1}}}{
      \tcons{\pO}{\mLabel{o}{\mt{0}}}{
      \tcons{\pI}{\mClose{i}}{
      \tcons{\pS}{\mClose{c}}{\cdots}}}}},\\
    &\quad\tcons{\pI}{\mLabel{i}{\mt{0}}}{
      \tcons{\pS}{\mLabel{c}{\mt{1}}}{
      \tcons{\pI}{\mClose{i}}{
      \tcons{\pO}{\mLabel{o}{\mt{0}}}{
      \tcons{\pS}{\mClose{c}}{
      \cdots}}}}}.
  \end{align*}
  Contrasting the two given traces illustrates the concurrent nature of message-passing processes.
  In the first execution, the left \(\ms{neg}\) process sends \(\mt{0}\) before the right receives its close message on \(i\); this order is reversed in the second trace.
  Finally, the denotation of \(\ms{doubleNeg'}\) hides all messages on \(c\):
  \begin{align*}
    \sembr{\tNew{c}{\negBit i}{\ms{doubleNeg}}} = \{\,
    &\tcons{\pI}{\mLabel{i}{\mt{0}}}{
      \tcons{\pO}{\mLabel{o}{\mt{0}}}{
      \tcons{\pI}{\mClose{i}}{
      \cdots}}},\\
    &\tcons{\pI}{\mLabel{i}{\mt{0}}}{
      \tcons{\pI}{\mClose{i}}{
      \tcons{\pO}{\mLabel{o}{\mt{0}}}{
      \cdots}}}, \dotsc \,\}.
  \end{align*}
\end{example}

\begin{example}
  \label{ex:process-language:2}
  To illustrate how traces with binding help model channel transmission, consider the parallel composition \(\cproci{\ms{comp}}{}{c : \Tu} = \tPar{a}{P}{Q}\), where
  \begin{alignat*}{2}
    &\cproci{P}{}{a : \Tu \Tot \Tu} &&= \tSendC{a}{c}{\tClose{c}}{\tClose{a}}\\
    &\cproci{Q}{a : \Tu \Tot \Tu}{c : \Tu} &&= \tRecvC{b}{a}{\tWait{a}{\tWait{b}{\tClose{c}}}}
  \end{alignat*}
  Process \(P\) denotes two traces, capturing the independent executions of processes \(\tClose{a}\) and \(\tClose{c}\):
  \begin{align*}
    \{\,
      &\tcons{\pO}{\mChan{a}}[c]{
        \tcons{\pO}{\mClose{c}}{
          \tcons{\pO}{\mClose{a}}{
            \tempt}}},
    \\
      &\tcons{\pO}{\mChan{a}}[c]{
        \tcons{\pO}{\mClose{a}}{
          \tcons{\pO}{\mClose{c}}{
            \tempt}}}
    \,\}
  \end{align*}
  The process \(Q\) exhibits no non-determinism and denotes a single trace:
  \[
    \sembr{Q} = \Set{
      \tcons{\pI}{\mChan{a}}[b]{
        \tcons{\pI}{\mClose{a}}{
          \tcons{\pI}{\mClose{b}}{
            \tcons{\pO}{\mClose{c}}{
              \tempt}}}}
    }
  \]
  Their parallel composition denotes a synchronized interleaving of these two trace sets.
  It is here that binding pays off: we can \(\alpha\)-vary the bound name \(c\) identifying the transmitted channel in the traces of \(P\) to a fresh name \(b\) to avoid clashing with the free name \(c\) in the trace of \(Q\).
  The interleaved trace set captures synchronization on both \(a\) and the transmitted channel:
  \[
    \sembr{\ms{comp}} = \Set{
       \tcons{\pS}{\mChan{a}}[b]{
        \tcons{\pS}{\mClose{a}}{
          \tcons{\pS}{\mClose{b}}{
            \tcons{\pO}{\mClose{c}}{
              \tempt}}}}
    }
  \]
  It contains a single trace, because only the first of \(P\)'s executions successfully synchronizes with~\(Q\).
\end{example}

There are no non-terminating executions, \ie, every process performs a finite number of actions:

\begin{propositionE}[Termination]
  \label{prop:process-language:2}
  For all processes \(P\) and traces \(t \in \ptraces{P}\), \(t\) is finite.
\end{propositionE}

\begin{proofE}
  By induction on the structure of \(P\), observing that each semantic clause extends a finite trace by a finite number of actions.
\end{proofE}

\section{Message-Observing Session Types and Process Specifications}
\label{sec:sess-types-proc}

Our type soundness result will show that if a process typechecks against a specification, then its range of behaviours (the traces it denotes) is among those allowed by its specification.
We specify which behaviours a process specification allows by giving specifications a trace semantics in \cref{sec:type-reduction}.
First, we define message-observing session types and process specifications.

Recall that each process is a client of some services and the server of a distinguished service.
From the perspective of a server of type \(A\), session types are generated by the grammar:
\begin{align*}
\text{Session Types}~~  A, B &\Coloneqq \Tu && \text{End communication and terminate}\\
       &\grammid \Tplus_{l \in L} A_l && \text{Internal choice}\\
       &\grammid \Tamp_{l \in L} A_l && \text{External choice}\\
       &\grammid (a : A) \Tot (b : B) && \text{Channel transmission}\\
       &\grammid (a : A) \Tlolly (b : B) && \text{Channel reception}\\
       &\grammid \TCaseU{a}{A} && \text{Termination observation}\\
       &\grammid \TCaseL{a}{\Set{l \Rightarrow A_l}_{l \in L}} && \text{Label observation}\\
       &\grammid \TCaseC{a}{b}{A} && \text{Higher-order observation}
\\[0.5em]
\text{Session Type Context}~~\Pi, \Delta, \Iota & \Coloneqq a_1 : A_1,  \dotsc, a_n :A_n
\end{align*}

We explain session types from the perspective of a server.
A server of \(\Tu\) sends a close message signalling the end of communication on a channel.
The internal choice type \(\Tplus_{l \in L} A_l\) specifies sending a label \(l \in L\), and then communicating according to \(A_l\).
Dually, the external choice type \(\Tamp_{l \in L} A_l\) specifies receiving a label \(l \in L\), and then communicating according to \(A_l\).
Many session-typed languages provide a session type \(A \Tot B\).
A server for a channel \(c\) of type \(A \Tot B\) sends a channel \(a\) of type \(A\) on \(c\) and then communicates on \(c\) according to \(B\).
Our syntax \((a : A) \Tot (b : B)\) extends the usual syntax to support mutual observation between the types \(A\) and \(B\).
It binds the name \(a\) in \(B\) to allow \(B\) to observe communications on the transmitted channel, and it binds \(b\) in \(A\) to allow \(A\) to observe subsequent communications on the carrier channel.
A server of a channel \(c\) of type \((a : A) \Tot (b : B)\) sends a channel \(a\) of type \(\subst{c}{b}{A}\), and then communicates on \(c\) according to \(B\).
Symmetrically, a server of \(c : (a : A) \Tlolly (b : B)\) specifies receiving a channel \(a\) of type \(\subst{c}{b}{A}\) and then communicating on \(c\) according to \(B\).
This type has an identical binding structure.
We call these types \defin{weak head normal} because they do not involve computation at the outermost level.

\defin{Observing types} restrict communication based on communication on ambient channels.
The type \(\TCaseU{a}{A}\) observes an ambient channel \(a\) and reduces to \(A\) when a close message is sent on \(a\).
The type \(\TCaseL{a}{\Set{l \Rightarrow A_l}_{l \in L}}\) reduces to the type \(A_l\) when a label \(l\) is sent on \(a\).
Finally, \(\TCaseC{a}{b}{A}\) reduces to \(A\) if a channel named \(b\) is transmitted over \(a\).
Because a priori, we do not know the name of the transmitted channel, \(\TCaseC{a}{b}{A}\) binds the name \(b\) in \(A\).

The free channels in a type are inductively defined on the structure in the obvious manner.

Our motivating examples used a concrete syntax ``\( \cproci{\ms{name}}[\Sigma]{\Delta}{c : C} = P \)'' to specify a process \(\ms{name}\), implemented by \(P\), that used channels \(\Delta\) to provide a service \(c : C\), where these channels were free to observe ambient channels \(\Sigma\).
These ambient channels were either (unhidden) internal channels, or channels in other processes whose meaning would be given by process composition.
By inspecting the syntax of \(P\), we can partition \(\Sigma\) into a context \(\Iota\) of internal channels and a context \(\Pi\) containing the remaining ambient channels.
We can then translate the concrete specification syntax \(\cint[\Pi, \Iota]{\Delta}{c : C}\) into the abstract syntax \(\jwfps{\Pi}{\proci{\Delta}{\Iota}{c : C}}\).
Our abstract syntax distinguishes \(\Pi\) and \(\Iota\) because they play different roles in typechecking: channels in \(\Iota\) can be hidden by \(P\), while those in \(\Pi\) cannot appear in \(P\).
\chase's typing judgment \(\jtproc{\Pi}{\Iota}{P}{\Delta}{c : C}{\tfnt{T}}\) checks \(P\) against its specification, generating constraints \(\tfnt{T}\).
Our concrete syntax names processes only to simplify examples; process names do not appear in our formal system.

To specify channel transmission, we need to specify multiple independent processes that execute concurrently.
Indeed, \(\tSendC{a}{b}{P}{Q}\) spawns process \(P\) providing a fresh \(b\), sends \(b\) over \(a\), and runs \(P\) concurrently with \(Q\) providing \(a\).
Checking \(\tSendC{a}{b}{P}{Q}\) against its specification requires decomposing its specification into specifications for \(P\) and \(Q\), and checking \(P\) and \(Q\).
Accordingly, we extend our abstract syntax for specifications to simultaneously specify multiple independent processes.
We write \(\jwfps{\Pi}{(\proci{\Delta_1}{\Iota_1}{c_1 : C_1} \scons \dotsb \scons \proci{\Delta_n}{\Iota_n}{c_n : C_n})}\) to simultaneously specify \(n\) processes, where channels \(\Pi\) and \(\Delta_i, \Iota_i, c_i : C_i\) are ambient in \(\Delta_j, \Iota_j, c_j : C_j\) for \(i \neq j\).
We call \(\proci{\Delta}{\Iota}{c : C}\) a process \defin{interface}: it specifies the communication interface for a process.
We range over interface sequences \((\proci{\Delta_1}{\Iota_1}{c_1 : C_1} \scons \dotsb \scons \proci{\Delta_n}{\Iota_n}{c_n : C_n})\) using the meta-variable \(\mc{G}\).

A specification \(\jwfps{\Pi}{\mc{G}}\) is well-defined only if for each \(a : A\) in \(\Pi\) or \(\mc{G}\), the free channels in \(A\) are contained in the domain of \(\jwfps{\Pi}{\mc{G}}\), \ie, are assigned types by the specification.
This precludes, \eg, \(\jwfps{\cdot}{\proci{a : A}{\cdot}{c : \TCaseU{b}{C}}}\) because the channel \(b\) is not typed by the specification.

\subsection{Type reductions and the semantics of specifications}
\label{sec:type-reduction}

In our motivating examples, we informally reduced types in specifications after observing messages on channel.
To make this explicit, we first define reduction on types, and then extend it to entire specifications.
Type reduction is given by a syntactic operation \(\immred{A}{\pi}\) that reduces a type \(A\) by an observed communication \(\pi\).
Observed communication, \(\pClose{a}\), \(\pLabel{a}{l}\), and \(\pChan{a}{b}\), closely resembles messages from the trace semantics.
The only difference is in the case of channel transmissions, where a free channel name \(b\) identifies the transmitted channel.
This free name will be used by the process typing judgment to update types with the name of the received channel.
We write \(\freecn(\pi)\) for the free channels in \(\pi\), and \(\carrcn(\pi) = a\) for the \defin{carrier channel}.

\defin{Type reduction} is partial, capture-avoiding, and inductively defined by:
\begin{align*}
  \immred{\Tu}{\pi} &= \Tu\\
  \immred{\left(\odot_{l \in L} A_l\right)}{\pi} &= \odot_{l \in L} \left(\immred{A_l}{\pi}\right) \qquad\qquad\quad\;\;\, ({\odot} \in \Set{ {\Tplus}, {\Tamp} })\\
  \immred{\left((a : A) \odot (b : B)\right)}{\pi} &= (a : \immred{A}{\pi}) \odot (b : \immred{B}{\pi}) \qquad ({\odot} \in \Set{ {\Tot}, {\Tlolly} })\\
  \immred{\left(\TCaseU{c}{A}\right)}{\pi} &= \begin{cases}
                                                A &\text{if }\pi = \pClose{c}\\
                                                \TCaseU{c}{\immred{A}{\pi}} &\text{if }\carrcn(\pi) \neq c
                                              \end{cases}\\
  \immred{\left(\TCaseL{c}{\Set{l \Rightarrow A_l}_{l \in L}}\right)}{\pi} &= \begin{cases}
                                                                                A_l &\text{if }\pi = \pLabel{c}{l}\text{ and }l \in L\\
                                                                                \TCaseL{c}{\Set{l \Rightarrow \immred{A_l}{\pi}}_{l \in L}} &\text{if }\carrcn(\pi) \neq c
                                                                              \end{cases}\\
  \immred{\left(\TCaseC{c}{b}{A}\right)}{\pi} &= \begin{cases}
                                                   A &\text{if }\pi = \pChan{c}{b}\\
                                                   \TCaseC{c}{b}{\immred{A}{\pi}} &\text{if }\carrcn(\pi) \neq c
                                                 \end{cases}
\end{align*}
All other cases are undefined.

When a message is observed on a channel in a process specification, we simultaneously reduce all types in the specification.
This eliminates the need to track past observations and it simplifies the presentation.
The syntax \(\immred{(a_1 : A_1,  \dotsc, a_n : A_n)}{\pi} = {a_1 : (\immred{A_1}{\pi})}, \dotsc, {a_n : (\immred{A_n}{\pi})}\) captures the simultaneous reduction of types in a context; it is defined whenever all \(\immred{A_i}{\pi}\) are defined.
It lifts to process specifications in context-by-context.

A specification \(\jwfps{\Pi}{\mc{G}}\) specifies a collection of permitted behaviours, \ie, a trace set \(\sembr{\jwfps{\Pi}{\mc{G}}}\).
To make this discussion concrete, consider a specification \(\jwfps{\Pi}{\proci{\Delta}{\Iota}{c : C}}\).
The traces in \(\sembr{\jwfps{\Pi}{\proci{\Delta}{\Iota}{c : C}}}\) are process traces that involve sending or receiving messages on \(\Delta, c : C\), with synchronizations on \(\Iota\).
However, these traces may also contain \defin{constraint elements} \((\pC; m)\).
An element \((\pC; \mClose{a})\) means that the execution described by the remaining tail of the trace is permitted only if a close message appears on the ambient channel \(a \in \dom(\Pi)\).
Constraint elements \((\pC; \mLabel{a}{l})\) and \((\pC; \mChan{a})\) are analogous.
We use constraint elements to semantically specify that a process is well-typed, provided that some other processes perform some action.

A trace set \(\sembr{\jwfps{\Pi}{\mc{G}}}\) is the least set satisfying the following collection of inequalities.
It is well-defined because type reduction reduces the number of type operators in a specification, so the right-hand specification always contains fewer type operators than the left-hand specification.
The right-hand side may be undefined; in this case, we treat it as an empty set.

We first specify label transmission.
Where \(\pi = \pLabel{a}{k}\), \(m = \mLabel{a}{k}\), and \(k \in L\):
\begin{align}
  \sembr{
  \jwfts
  { \Pi }
  { \mc{G}
  \scons
  (\jwfpi
  { \Delta }
  { \Iota }
    { a : \Tplus_{l \in L} A_l })}
    }
  &\supseteq
    \tcons{\pO}{m}{
    \sembr{
    \jwfts
    { \immred{\Pi}{\pi} }
    { \immred{\mc{G}}{\pi}
    \scons
    (\jwfpi
    { \immred{\Delta}{\pi} }
    { \immred{\Iota}{\pi} }
    { a : A_k })}
    }}
    \label{psred:Tplus-R}
      \\
  \sembr{
  \jwfts
  { \Pi }
  { \mc{G}
  \scons
  (\jwfpi
  { \Delta, a : \Tplus_{l \in L} A_l }
  { \Iota }
  { c : C })}
  }
  &\supseteq
      \scalemath{0.99}{
    \tcons{\pI}{m}{
    \sembr{
    \jwfts
    { \immred{\Pi}{\pi} }
    { \immred{\mc{G}}{\pi}
    \scons
    (\jwfpi
    { \immred{\Delta}{\pi}, a : A_k }
    { \immred{\Iota}{\pi} }
    { c : \immred{C}{\pi} })}
    }}}
    \label{psred:Tplus-L}
  \\
  \sembr{
  \jwfts
  { \Pi }
  { \mc{G}
  \scons
  (\jwfpi
  { \Delta }
  { \Iota, a : \Tplus_{l \in L} A_l }
  { c : C })}
  }
  &\supseteq
    \scalemath{0.99}{
    \tcons{\pS}{m}{
    \sembr{
    \jwfts
    { \immred{\Pi}{\pi} }
    { \immred{\mc{G}}{\pi}
    \scons
    (\jwfpi
    { \immred{\Delta}{\pi} }
    { \immred{\Iota}{\pi}, a : A_k }
    { c : \immred{C}{\pi} })}
    }}}
\end{align}
The first clause specifies that a provider of type \(\Tplus_{l \in L} A_l\) can send any label \(l\) in \(L\).
The provider must obey its original specification, reduced by the just-observed message \(\pLabel{a}{l}\).
Receiving a label is dually specified.
The last clause specifies that a process formed of a composition along \(a : \Tplus_{l \in L} A_l\) can perform a synchronization on that channel.
We define clauses with inequalities to avoid imposing an order on communication on different channels.
For example, given a specification \(\jwfps{\cdot}{\proci{a : \Tplus (\mt{0} : \Tu)}{\cdot}{b : \Tplus (\mt{0} : \Tu)}}\), the above clauses ensure that \(a\) and \(b\) can be used in either order because its denotation includes traces starting with either channel.

A provider of type \(\Tu\) can terminate only if it has no other channels.
This ensures that channels do not accidentally get discarded and it is captured by the first clause below.
Receiving and synchronizing on channels of type \(\Tu\) ends communication on that channel.
The last clause captures termination when there are no processes left.
Where \(\pi = \pClose{a}\):
\begin{align}
   \sembr{
    \jwfts{\Pi}
    { \mc{G}
      \scons
      (\jwfpi
      { \cdot }
      { \cdot }
      { a : \Tu })}
  }
  &\supseteq
  \tcons{\pO}{\mClose{a}}{
    \sembr{
      \jwfts{\immred{\Pi}{\pi}}{ \immred{\mc{G}}{\pi} }
    }}
    \label{psred:Tu-R}
  \\
  \sembr{
  \jwfts
  { \Pi }
  { \mc{G}
  \scons
  (\jwfpi
  { \Delta, a : \Tu }
  { \Iota }
  { c : C })}
  }
  &\supseteq
    \tcons{\pI}{\mClose{a}}{
    \sembr{
    \jwfts
    { \immred{\Pi}{\pi} }
    { \immred{\mc{G}}{\pi}
    \scons
    (\jwfpi
    { \immred{\Delta}{\pi} }
    { \immred{\Iota}{\pi} }
    { c : \immred{C}{\pi} })}
    }}
        \label{psred:Tu-L}
  \\
    \sembr{
  \jwfts
  { \Pi }
  { \mc{G}
  \scons
  (\jwfpi
  { \Delta }
  { \Iota, a : \Tu }
  { c : C })}
  }
  &\supseteq
    \tcons{\pS}{\mClose{a}}{
    \sembr{
    \jwfts
    { \immred{\Pi}{\pi} }
    { \immred{\mc{G}}{\pi}
    \scons
    (\jwfpi
    { \immred{\Delta}{\pi} }
    { \immred{\Iota}{\pi} }
    { c : \immred{C}{\pi} })}
    }}
  \\
  \sembr{
    \jwfts{\Pi}
    { \cdot }
  }
  &\supseteq
    \Set{ \tempt }
\end{align}

Channel transmission changes the shape of specifications to capture spawning processes.
Where \(\pi = (\pChan{a}{b})\) with \(b\) fresh:
\begin{align}
  &\sembr{\jwfts
    { \Pi }
    { \mc{G}
    \scons
    (\jwfpi
    { \Delta_1, \Delta_2 }
    { \Iota_1, \Iota_2 }
    { a : (b : B) \Tot (a : A) })}}
    \nonumber\\
  &\supseteq
    \tcons{\pO}{\mChan{a}}[b]{
    \sembr{
    \jwfts
    { \immred{\Pi}{\pi} }
    { \immred{\mc{G}}{\pi}
    \scons
    (\jwfpi
    { \immred{\Delta_1}{\pi} }
    { \immred{\Iota_1}{\pi} }
    { b : B })
    \scons
    (\jwfpi
    { \immred{\Delta_2}{\pi}, a : A }
    { \immred{\Iota_2}{\pi} }
    { c : \immred{C}{\pi} })
    }}}
    \label{psred:Tot-R}
  \\
  &\sembr{\jwfts
    { \Pi }
    { \mc{G}
    \scons
    (\jwfpi
    { \Delta, a : (b : B) \Tot (a : A) }
    { \Iota }
    { c : C })}}
    \nonumber\\
  &\supseteq
    \tcons{\pI}{\mChan{a}}[b]{
    \sembr{
    \jwfts
    { \immred{\Pi}{\pi} }
    { \immred{\mc{G}}{\pi}
    \scons
    (\jwfpi
    { \immred{\Delta}{\pi}, b : B, a : A }
    { \immred{\Iota}{\pi} }
    { c : \immred{C}{\pi} })}
    }}
    \label{psred:Tot-L}
\end{align}%
The first clause specifies the permitted behaviours of a provider \(\tSendC{a}{b}{P}{Q}\).
Operationally, this process forks \(P\) and \(Q\), respectively providing \(b\) and \(a\), and sends \(b\) over \(a\).
The used and internal channels of \(\tSendC{a}{b}{P}{Q}\) are then divided between \(P\) and \(Q\).
Accordingly, the first clause splits its interface \(\jwfpi { \Delta_1, \Delta_2 } { \Iota_1, \Iota_2 } { a : (b : B) \Tot (a : A) }\) in two.
The channels assigned to \(P\) and \(Q\) depend on the particular implementation, so the clause allows for arbitrary partitions \(\Delta_1, \Delta_2\) and \(\Iota_1, \Iota_2\).
This clause also illustrates why the observed message \(\pChan{a}{b}\) (used in type reductions) specifies a free name \(b\), in contrast to the message element \(\mChan{a}\) used in traces.
When we reduce the remaining types in the specification to account for the channel transmission, we must specify the free name \(b\) that carries the communications of \(P\).
However, this name \(b\) is not externally meaningful, and we bind it in the trace to avoid name clashes in process compositions.
The second clause captures receiving a channel over \(b\); it updates the context to contain the received channel.
The elided clause for synchronization is analogous to the second clause.

Finally, we specify the constraints generated by ambient channels in \(\Pi\):
\begin{align}
  \sembr{
  \jwfts
  { \Pi, a : \Tu }
  { \mc{G} }
  }
  &\supseteq
    \tcons{\pC}{\pClose{a}}{
    \sembr{\jwfts
    { \immred{\Pi}{\pi} }
    { \immred{\mc{G}}{\pi} }
    }},
    \label{eq:session-types:1}
  \\
  \text{where }\pi &= \pClose{a};\nonumber
  \\
    \sembr{
  \jwfts
  { \Pi, a : \odot_{l \in L} A_l }
  { \mc{G} }
  }
  &\supseteq
    \tcons{\pC}{\pLabel{a}{l}}{
    \sembr{\jwfts
    { \immred{\Pi}{\pi}, a : A_l }
    { \immred{\mc{G}}{\pi} }
    }},
    \label{eq:session-types:2}
  \\
  \text{where } \pi &= \pLabel{a}{l}, l \in L, \text{ and } {\odot} \in \Set{ {\Tplus}, {\Tamp} };\nonumber
  \\
    \sembr{
  \jwfts
  { \Pi, a : (b : B) \odot (a : A) }
  { \mc{G} }
  }
  &\supseteq
    \tcons{\pC}{\mChan{a}}[b]{
    \sembr{\jwfts
    { \immred{\Pi}{\pi}, a : A, b : B }
    { \immred{\mc{G}}{\pi} }
    }},
    \label{eq:session-types:3}
  \\
  \text{where } \pi &= \pChan{a}{b}, \text{ and } {\odot} \in \Set{ {\Tot}, {\Tlolly} }.\nonumber
\end{align}
The first clause specifies that if we reduce a specification by observing a close message on an ambient channel \(a\), then the traces permitted by the reduced specification are valid provided that some process in the environment closes \(a\).
The second clause is analogous.
The third clause captures the fact that sending a channel results in a pair of channels in a context: the transmitted channel and its carrier.
We assume without loss of generality that \(b\) is fresh in the third clause.

The omitted clauses for \(\Tlolly\) and \(\Tamp\) are analogous.
There are no clauses for observing types: they allow communication only after reducing to a weak-head normal type.
In particular, type reductions are silent and they occur as a result of communication on channels with weak-head normal types.

\begin{example}
  \label{ex:session-types:2}
  The concrete specification \( \cproci{\ms{neg}}{i : \Tlist{\bit}{1}}{c : \negBit{i}} \) of \cref{sec:process-composition} corresponds to the specification \( \jwfps{\cdot}{\proci{i : \Tlist{\bit}{1}}{\cdot}{c : \negBit{i}}}\).
  It only allows the executions
  \begin{align*}
    &\tcons{\pI}{\mLabel{i}{\mt{0}}}{
      \tcons{\pO}{\mLabel{c}{\mt{1}}}{
      \tcons{\pI}{\mClose{i}}{
      \tcons{\pO}{\mClose{c}}{
      \tempt
      }}}},\\
    &\tcons{\pI}{\mLabel{i}{\mt{0}}}{
      \tcons{\pI}{\mClose{i}}{
      \tcons{\pO}{\mLabel{c}{\mt{1}}}{
      \tcons{\pO}{\mClose{c}}{
      \tempt
      }}}},
  \end{align*}
  plus the two traces obtained by exchanging \(\mt{0}\) and \(\mt{1}\).
  The first two traces illustrate the concurrent nature of communication on different channels.
  Indeed, \(\negBit{i}\) only allows communication on \(c\) after communication on \(i\), but no other ordering is imposed between messages on \(i\) and \(c\).
\end{example}

\begin{example}
  \label{ex:session-types:3}
  The concrete specification \( \cproci{\ms{neg}}[i : \Tlist{\bit}{1}]{c : \negBit i}{o : \Tidlist{i}{1}} \) of \cref{sec:process-composition} corresponds to the specification \( \jwfps{i : \Tlist{\bit}{1}}{\proci{c : \negBit i}{\cdot}{o : \Tidlist{i}{1}}}\).
  It allows:
  \begin{align*}
    &\tcons{\pC}{\mLabel{i}{\mt{0}}}{
      \tcons{\pI}{\mLabel{c}{\mt{1}}}{
      \tcons{\pO}{\mLabel{o}{\mt{0}}}{{}\\
    &\qquad\qquad
      \tcons{\pC}{\mClose{i}}{
      \tcons{\pI}{\mClose{c}}{
      \tcons{\pO}{\mClose{o}}{
      \tempt
      }}}}}},\\
    &\tcons{\pC}{\mLabel{i}{\mt{0}}}{
      \tcons{\pC}{\mClose{i}}{
      \tcons{\pI}{\mLabel{c}{\mt{1}}}{
      \tcons{\pO}{\mLabel{o}{\mt{0}}}{{}\\
    &\qquad\qquad
      \tcons{\pI}{\mClose{i}}{
      \tcons{\pO}{\mClose{o}}{
      \tempt
      }}}}}},\ \text{\etc}
  \end{align*}
  Interpreting each trace as an execution or sequence of actions permitted to \(\ms{neg}\), the first trace specifies that if a label \(\mt{0}\) is sent on an ambient channel \(i\), then \(\ms{neg}\) can receive a bit \(\mt{1}\) on \(c\), send \(\mt{0}\) on \(o\), and so on and so forth.
  The second trace is analogous, but illustrates that we impose no extraneous orderings on messages on different channels.
\end{example}

\begin{example}
  \label{ex:session-types:4}
  The specification \(\jwfps{a : \Tu, b : \Tu}{\proci{\cdot}{\cdot}{c : \TCaseU{a}{\TCaseU{b}{\Tu}}}}\) permits two traces:
  \begin{align*}
    &\tcons{\pC}{\mClose{a}}{
    \tcons{\pC}{\mClose{b}}{
    \tcons{\pO}{\mClose{c}}{
      \tempt}}},\\
    &\tcons{\pC}{\mClose{b}}{
    \tcons{\pC}{\mClose{a}}{
    \tcons{\pO}{\mClose{c}}{
    \tempt}}}.
  \end{align*}
  These two traces illustrate that the type of \(c\) does not impose an order on ambient communications: because we can reduce under \(\ms{CASE}\), communication on \(a\) and \(b\) can occur in any order.
  Indeed, the first trace corresponds to reducing the specification to \(\jwfps{b : \Tu}{\proci{\cdot}{\cdot}{c : \TCaseU{b}\Tu}}\) (after observing \(\pClose{a}\)) and then to \(\jwfps{\cdot}{\proci{\cdot}{\cdot}{c : \Tu}}\) (after observing \(\pClose{b}\)); the second captures first reducing to \(\jwfps{a : \Tu}{\proci{\cdot}{\cdot}{c : \TCaseU{a}{{\Tu}}}}\), then to \(\jwfps{\cdot}{\proci{\cdot}{\cdot}{c : \Tu}}\).
\end{example}

\begin{example}
  \label{ex:session-types:5}
  The specification \(\jwfps{\cdot}{\proci{\cdot}{\cdot}{a : (b : \Tu) \Tot \TCaseU{b}{\Tu}}}\) specifies sending a channel of type \(\Tu\), and with the carrier closing after the sent channel.
  Its trace set is:
  \begin{align*}
    &\tcons{\pO}{\mChan{a}}[b]{\sembr{\jwfps{\cdot}{\proci{\cdot}{\cdot}{b : \Tu} \scons \proci{\cdot}{\cdot}{a : \TCaseU{b}{\Tu}}}}}\\
    &= \tcons{\pO}{\mChan{a}}[b]{\tcons{\pO}{\mClose{b}}{\sembr{\jwfps{\cdot}{\proci{\cdot}{\cdot}{a : \Tu}}}}}\\
    &= \Set{ \tcons{\pO}{\mChan{a}}[b]{\tcons{\pO}{\mClose{b}}{\tcons{\pO}{\mClose{a}}{\tempt}}} }.
  \end{align*}
  This example shows how multiple interfaces in a specification are used to specify relationships between messages from different processes that are spawned in the course of channel transmission.
\end{example}

\section{Operations on Traces}
\label{sec:operation-traces}

Our trace semantics for processes uses two operations on traces: deletion and synchronized interleaving.
Name deletion deletes all messages in a trace whose names appear in a given set, while synchronized interleavings interleave two traces while ensuring that input and output actions match up.
We also specify how to delete constraints elements from traces.
All three operations will be used by our typechecking algorithm when generating traces with constraints.

\subsection{Deleting channel names from traces}
\label{def:process-language:2}

The construct \(\tNewV{(\overrightarrow{a_i : A_i})}{P}\) binds the channels \(a_i\) to hide them from external view.
Semantically, hiding is captured by deleting all messages on channels \(a_i\) from the traces of \(P\).
To account for channel transmission, we may need to delete messages on other channel names.
In particular, if \(a_i\) carries some channel \(b\), then we must also delete all messages on \(b\) from the trace.

Given a set \(X\) of names and a trace \(t\), the deletion of \(X\) in \(t\) is the trace \(\tdel{t}{X}\) inductively defined~by:
\[
  \tdel{\tempt}{X} = \tempt
  \quad
  \text{and}
  \quad
  \tdel{(\tcons{s}{m}[c_1, \dotsc, c_n]{t})}{X} =
                                                \begin{cases}
                                                  \tdel{t}{(X \cup \Set{c_1, \dotsc, c_n})} &\text{if } \carrcn(m) \in X\\
                                                  \tcons{s}{m}[c_1, \dotsc, c_n]{(\tdel{t}{X})} &\text{otherwise}
                                                \end{cases}
\]
We assume without loss of generality that the names \(c_i\) are chosen distinct from those already in \(X\).

\begin{example}
  \label{ex:trace-operations:1}
  If \(t = \tcons{\pI}{\mChan{a}}[b]{\tcons{\pI}{\mClose{a}}{\tcons{\pI}{\mClose{b}}{\tcons{\pO}{\mClose{c}}{\tempt}}}}\), then \(\tdel{t}{\Set{a}} = \tcons{\pO}{\mClose{c}}{\tempt}\).
\end{example}

\subsection{Deleting constraints from traces}

The operation \(\tquoc{t}\) deletes all constraints from \(t\).
We will use it when checking that processes have consistent constraint sets when typing process compositions.
It is inductively defined by
\[
  \tquoc{\tempt} = \tempt
  \qquad
  \text{and}
  \qquad
  \tquoc{\tcons{s}{m}[\vec x]{t}}
  =
  \begin{cases}
    \tquoc{t} &\text{if } s = \pC\\
    \tcons{s}{m}[\vec x]{\tquoc{t}} &\text{otherwise}
  \end{cases}
\]

\subsection{Trace reduction}
\label{sec:immediate-reductions}

When interleaving two traces, we need to delete constraints in one trace satisfied by messages in the other trace.
We do so using a partial operation \(\immred{t}{\pi}\) that deletes a constraint \((\pC; m)\) in \(t\) if it is satisfied by the observed communication \(\pi\).
It is inductively defined on the structure of the trace:
\begin{align*}
  \immred{\tempt}{\pi} &= \tempt &  \immred{(\tcons{\pC}{\mClose{a}}{t})}{(\pClose{a})} &= t\\
  \immred{(\tcons{\pC}{\mLabel{a}{l}}{t})}{(\pLabel{a}{l})} &= t &
  \immred{(\tcons{\pC}{\mChan{a}}[c]{t})}{(\pChan{a}{b})} &= \subst{b}{c}{t}\\
  \immred{(\tcons{s}{m}[\vec x]{t})}{\pi} &= \mathrlap{\tcons{s}{m}[\vec x]{(\immred{t}{\pi})} \quad \text{if } \carrcn(m) \neq \carrcn(\pi),} &&
\end{align*}
and it is undefined in all other cases.
The fourth clause instantiates a bound name \(c\) for a transmitted channel with the free name \(b\) actually observed in the course of channel transmission.
The last clause handles an observation on a channel different from the head of the trace.
It is defined only if \(\immred{t}{\pi}\) is defined.
We always \(\alpha\)-vary \(\vec x\) to be distinct from the names in \(\pi\).

\begin{example}
  \label{ex:trace-operations:2}
  The reduction \(\immred{(\tcons{\pC}{\mLabel{a}{k}}{\tempt})}{(\pLabel{a}{l})}\) is undefined because the label \(l\) observed on \(a\) does not match the label \(k\) expected by the constraint.
  Similarly, the reduction \(\immred{{(\tcons{\pO}{\mClose{a}}{\tempt})}}{(\pLabel{a}{l})}\) is undefined because the trace we are reducing sends on \(a\), while the observation ``\(\pLabel{a}{l}\)'' is meant to be an observation on an ambient channel \(a\).
\end{example}

\begin{textAtEnd}
  To help characterize the behaviour of reduction on traces with the behaviour of type reduction in \cref{prop:session-types:36}, we introduce a reduction relation on contexts and specifications, and a few technical lemmas (\cref{lemma:session-types:10,lemma:session-types:14,prop:session-types:14}).
  We assume that bound names are unique and disjoint from free names.
  The relation is defined by the following schemata, which hold only if both sides of the arrow are~defined:
  \begin{alignat}{2}
    \Pi, a : \Tu &\ecred{\pClose{a}} \immred{\Pi}{(\pClose{a})} &&\label{ecred:tu}\\
    \Pi, a : \odot_{l \in L} A_l &\ecred{\pLabel{a}{l}} \immred{\Pi}{(\pLabel{a}{l})}, a : A_l & (l \in L, {\odot} \in \Set{ {\Tplus}, {\Tamp} })& \label{ecred:lbl}\\
    \Pi, a : (b : B) \odot (a : A) &\ecred{\pChan{a}{b}} \immred{\Pi}{(\pChan{a}{b})}, b : B, a : A & ({\odot} \in \Set{ {\Tot}, {\Tlolly} })& \label{ecred:chan}
  \end{alignat}
  The relation \(\psred{(\pC, \pi)}\) on pairs of well-formed specifications is defined as:
  \begin{align}
    \jwfts
    { \Pi }
    { \mc{G} }
    &\psred{(\pC, \pi)}
      \jwfts
      { \Pi' }
      { \immred{\mc{G}}{\pi} }
      \qquad
      \text{if }\Pi \ecred{\pi} \Pi'
      \label{psred:env}
  \end{align}
  We invite the reader to compare reduction \eqref{psred:env} with the clauses generating traces of the form \(\tcons{\pC}{m}[\vec x]{t}\) in \cref{sec:type-reduction}.
\end{textAtEnd}

\begin{lemmaE}[Partial Commutativity of Reduction][all end]
  \label{lemma:session-types:10}
  Let \(\pi_1\) and \(\pi_2\) be observed communications such that \(\freecn(\pi_1)\) and \(\freecn(\pi_2)\) are disjoint.
  \begin{enumerate}
  \item Both \(\immred{A}{\pi_1}\) and \(\immred{A}{\pi_2}\) are defined if and only if both \(\immred{(\immred{A}{\pi_1})}{\pi_2}\) and \(\immred{(\immred{A}{\pi_2})}{\pi_1}\) are defined.
  \item If \(\immred{(\immred{A}{\pi_1})}{\pi_2}\) and \(\immred{(\immred{A}{\pi_2})}{\pi_1}\) are both defined, then they are equal.
  \end{enumerate}
\end{lemmaE}

\begin{proofE} %
  It is immediate that if \(\immred{(\immred{A}{\pi_1})}{\pi_2}\) and \(\immred{(\immred{A}{\pi_2})}{\pi_1}\) are both defined, then so are \(\immred{A}{\pi_1}\) and \(\immred{A}{\pi_2}\).
  Assume that \(\immred{A}{\pi_1}\) and \(\immred{A}{\pi_2}\) are both defined.
  We show by induction on the structure of \(A\) that \(\immred{(\immred{A}{\pi_1})}{\pi_2}\) and \(\immred{(\immred{A}{\pi_2})}{\pi_1}\) are both defined and equal.

  The weak-head normal cases follow readily by the induction hypotheses.

  Assume next that \(A = \TCaseU{c}{B}\).
  We proceed by case analysis on \(\pi_1\).
  Assume first that \(\pi_1 = \pClose{c}\).
  In this case, \(\carrcn(\pi_2) \neq c\), so \(\immred{A}{\pi_2} = \TCaseU{c}{\immred{B}{\pi_2}}\) is defined, so \(\immred{B}{\pi_2}\) is defined.
  Then \( \immred{(\immred{A}{\pi_1})}{\pi_2} = \immred{B}{\pi_2} = \immred{(\immred{A}{\pi_2})}{\pi_1} \) are both defined and equal.
  Otherwise, if neither \(\pi_1\) or \(\pi_2\) is \(\pClose{c}\), then \(\carrcn(\pi_1) \neq c\) and \(\carrcn(\pi_2) \neq c\).
  It follows that \(\immred{B}{\pi_i}\) is defined for \(i = 1, 2\).
  We then deduce by the induction hypothesis:
  \[
    \immred{(\immred{A}{\pi_1})}{\pi_2}
    = \TCaseU{c}{\immred{(\immred{B}{\pi_1})}{\pi_2}}
    = \TCaseU{c}{\immred{(\immred{B}{\pi_2})}{\pi_1}}
    = \immred{(\immred{A}{\pi_2})}{\pi_1}.
  \]

  Assume next that \(A = \TCaseL{c}{\Set{ l \Rightarrow A_l }_{l \in L}}\).
  We proceed by case analysis on \(\pi_1\).
  Assume first that \(\pi_1 = \pLabel{c}{l}\), where \(l \in L\).
  In this case, \( \immred{(\immred{A}{\pi_1})}{\pi_2} = \immred{A_l}{\pi_2} = \immred{(\immred{A}{\pi_2})}{\pi_1} \) are defined and equal.
  Otherwise, \(\carrcn(\pi_1) \neq c\) and \(\carrcn(\pi_2) \neq c\).
  In this case, \(\immred{A_l}{\pi_i}\) is defined for \(l \in L\) and \(i = 1, 2\).
  We deduce by the induction hypothesis that the following are defined and equal:
  \[
    \immred{(\immred{A}{\pi_1})}{\pi_2}
    = \TCaseL{c}{\Set{ l \Rightarrow \immred{(\immred{A_l}{\pi_1})}{\pi_2} }_{l \in L} }
    = \TCaseL{c}{\Set{ l \Rightarrow \immred{(\immred{A_l}{\pi_2})}{\pi_1} }_{l \in L} }
    = \immred{(\immred{A}{\pi_2})}{\pi_1}.
  \]

  The case \(A = \TCaseC{c}{b}{B}\) is analogous.
\end{proofE}

\begin{textAtEnd}
  The following \namecref{lemma:session-types:14} is used to relate two different reductions in the proof of the diamond property (\cref{prop:session-types:14}):
\end{textAtEnd}

\begin{lemmaE}[][all end]
  \label{lemma:session-types:14}
  If \(\immred{\Pi}{\pi_1}\) is defined, \(\Pi \ecred{\pi_2} \Gamma\), and \(\freecn(\pi_1) \cap \freecn(\pi_2) = \emptyset\), then \(\immred{\Pi}{\pi_1} \ecred{\pi_2} \immred{\Gamma}{\pi_1}\).
\end{lemmaE}

\begin{proofE} %
  By case analysis on the transition \(\Pi \ecred{\pi_2} \Gamma\).
  \begin{proofcases}
  \item[\eqref{ecred:tu}] Assume the transition is given by \(\Pi = \Delta, a : \Tu \ecred{\pi_2} \immred{\Delta}{\pi_2} = \Gamma\) where \(\pi_2 = \pClose{a}\).
    Then \(\immred{\Delta}{\pi_1}\) and \(\immred{\Delta}{\pi_2}\) are both defined, so \(\immred{(\immred{\Delta}{\pi_1})}{\pi_2} = \immred{(\immred{\Delta}{\pi_2})}{\pi_1}\) are both defined by \cref{lemma:session-types:10}.
    It follows from \eqref{ecred:tu} that:
    \[
      \immred{\Pi}{\pi_1} = \immred{\Delta}{\pi_1}, a : \Tu \ecred{\pi_2} \immred{(\immred{\Delta}{\pi_1})}{\pi_2} = \immred{(\immred{\Delta}{\pi_2})}{\pi_1} = \immred{\Gamma}{\pi_1}.
    \]
  \item[\eqref{ecred:lbl}] Assume the transition is given by \(\Pi = \Delta, a : \odot_{l \in L} A_l \ecred{\pi_2} \immred{\Delta}{\pi_2}, a : A_l = \Gamma\) with \(\pi_2 = \pLabel{a}{l}\), and \({\odot} \in \Set{ {\Tplus}, {\Tamp} }\).
    Then \(\immred{\Delta}{\pi_1}\) and \(\immred{\Delta}{\pi_2}\) are both defined, so \(\immred{(\immred{\Delta}{\pi_1})}{\pi_2} = \immred{(\immred{\Delta}{\pi_2})}{\pi_1}\) are both defined by \cref{lemma:session-types:10}.
    It follows from \eqref{ecred:lbl} that:
    \[
      \immred{\Pi}{\pi_1} = \immred{\Delta}{\pi_1}, a : \odot_{l \in L} \immred{A_l}{\pi_1} \ecred{\pi_2} \immred{(\immred{\Delta}{\pi_1})}{\pi_2}, a : \immred{A_l}{\pi_1} = \immred{(\immred{\Delta}{\pi_2}, a : A_l)}{\pi_1} = \immred{\Gamma}{\pi_1}.
    \]
  \item[\eqref{ecred:chan}] Assume the transition is given by \(\Pi = \Delta, a : ({b : B}) \odot ({a : A}) \ecred{\pi_2} \immred{\Delta}{\pi_2}, {b : B}, {a : A} = \Gamma\) with \(\pi_2 = \pChan{a}{b}\), and \({\odot} \in \Set{ {\Tot}, {\Tlolly} }\).
    Then \(\immred{\Delta}{\pi_1}\) and \(\immred{\Delta}{\pi_2}\) are both defined, so \(\immred{(\immred{\Delta}{\pi_1})}{\pi_2} = \immred{(\immred{\Delta}{\pi_2})}{\pi_1}\) are both defined by \cref{lemma:session-types:10}.
    It follows from \eqref{ecred:chan} that:
    \begin{multline*}
      \immred{\Pi}{\pi_1} = \immred{\Delta}{\pi_1}, a : (b : \immred{B}{\pi_1}) \odot (a : \immred{A}{\pi_1})\\
      \ecred{\pi_2} \immred{(\immred{\Delta}{\pi_1})}{\pi_2}, b : \immred{B}{\pi_1}, a : \immred{A}{\pi_1} = \immred{(\immred{\Delta}{\pi_2}, b : B, a : A)}{\pi_1} = \immred{\Gamma}{\pi_1}.\qedhere
    \end{multline*}
  \end{proofcases}
\end{proofE}

\begin{propositionE}[Diamond Property][all end]
  \label{prop:session-types:14}
  If \(\Pi \ecred{\pi_1} \Pi_1\) and \(\Pi \ecred{\pi_2} \Pi_2\) where \(\freecn(\pi_1) \cap \freecn(\pi_2) = \emptyset\), then there exists a \(\Pi_3\) such that \(\Pi_1 \ecred{\pi_2} \Pi_3\) and \(\Pi_2 \ecred{\pi_1} \Pi_3\).
\end{propositionE}

\begin{proofE} %
  If \(\Pi \ecred{\pi_1} \Pi_1\) and \(\Pi \ecred{\pi_2} \Pi_2\) where \(\freecn(\pi_1) \cap \freecn(\pi_2) = \emptyset\), then \({\Pi = \Delta, a_1 : A_1, a_2 : A_2}\) for some \(\Delta\) and \(a_i : A_i\) where \(a_i : A_i \ecred{\pi_i} \Delta_i\) for \(i = 1, 2\).
  Then
  \begin{align*}
    \Pi_1 &= \immred{(\Delta, a_2 : A_2)}{\pi_1}, \Delta_1,\\
    \Pi_2 &= \immred{(\Delta, a_1 : A_1)}{\pi_2}, \Delta_2.
  \end{align*}
  By \cref{lemma:session-types:14},
  \begin{align*}
    &\immred{(\Delta, a_2 : A_2)}{\pi_1} \ecred{\pi_2} \immred{(\immred{\Delta}{\pi_1})}{\pi_2}, \immred{\Delta_2}{\pi_1}\\
    &\immred{(\Delta, a_1 : A_1)}{\pi_2} \ecred{\pi_1} \immred{(\immred{\Delta}{\pi_2})}{\pi_1}, \immred{\Delta_1}{\pi_2}.
  \end{align*}
  In particular, \(\immred{\Delta_1}{\pi_2}\) and \(\immred{\Delta_2}{\pi_1}\) are defined.
  This implies that the following transitions are defined:
  \begin{align*}
    &\Pi_1 \ecred{\pi_2} \immred{(\immred{\Delta}{\pi_1})}{\pi_2}, \immred{\Delta_2}{\pi_1}, \immred{\Delta_1}{\pi_2}\\
    &\Pi_2 \ecred{\pi_1} \immred{(\immred{\Delta}{\pi_2})}{\pi_1}, \immred{\Delta_2}{\pi_1}, \immred{\Delta_1}{\pi_2}.
  \end{align*}
  The two right sides are equal by \cref{lemma:session-types:10}, and we take them to be \(\Pi_3\).
\end{proofE}

\begin{textAtEnd}
  The following \namecref{prop:session-types:36} relates trace reductions and type reductions:
\end{textAtEnd}

\begin{propositionE}[][all end]
  \label{prop:session-types:36}
  If \(\jtt{t}{\jwfps{\Pi}{\mc{G}}}\), \(\jwfps{\Pi}{\mc{G}} \psred{(\pC, \pi)} \jwfps{\Pi'}{\immred{\mc{G}}{\pi}}\), and \(\immred{t}{\pi}\) is defined,\\
  then \({\jtt{\immred{t}{\pi}}{\jwfps{\Pi'}{\immred{\mc{G}}{\pi}}}}\).
\end{propositionE}

\begin{proofE}
  By induction on the derivation \(\jtt{t}{\jwfps{\Pi}{\mc{G}}}\).

  The base case occurs when \(t = \tempt\), in which case the result is immediate.

  The inductive steps occur when \(t = \tcons{s}{m}[\vec x]{t'}\) for some \(s\), \(m\), \(\vec x\) and \(t'\).
  Each inductive case corresponds to a different possible choice of \((s, m)\).
  We give a few illustrative cases.

  Assume first that \(s = \pC\) and \(\pi = \obsc{m}{\vec x}\).\footnote{Recall that \(\obsc{m}{\vec x}\) is defined on \cpageref{def:obsc}.}
  By inversion, \(\jtt{t}{\jwfps{\Pi}{\mc{G}}}\) because \(\jtt{t'}{\jwfps{\Pi'}{\immred{\mc{G}}{\pi}}}\).
  Observe that \(\immred{t}{\pi} = t'\).
  This gives the result.

  In all other cases, the fact that \(\immred{t}{\pi}\) is defined implies that \(\carrcn(\pi) \neq \carrcn(m)\).
  This in turn implies that \(\immred{t'}{\pi}\) is defined.
  Moreover, inversion on \(\jtt{t}{\jwfps{\Pi}{\mc{G}}}\) implies that \(\jtt{t'}{\jwfps{\Pi_0}{\mc{G}_0}}\) for some \(\Pi_0\) and \(\mc{G}_0\) such that \(\Pi, \overline{\mc{G}} \psred{\obsc{m}{\vec x}} \Pi_0, \overline{\mc{G}_0}\), where we write \(\overline{\mc{G}}\) for the union of the contexts appearing in \(\mc{G}\).
  By the diamond property (\cref{prop:session-types:14}), it follows that there exists a \(\Pi_0'\) such that \(\jwfps{\Pi_0}{\mc{G}_0} \psred{(\pC, \pi)} \jwfps{\Pi_0'}{\immred{\mc{G}_0}{\pi}}\), and such that the following two reductions are defined:
  \begin{align*}
    &\Pi, \overline{\mc{G}} \psred{\obsc{m}{\vec x}} \Pi_0, \overline{\mc{G}_0} \psred{\pi} \Pi_0', \immred{\overline{\mc{G}_0}}{\pi}\\
    &\Pi, \overline{\mc{G}} \psred{\pi} \Pi', \immred{\overline{\mc{G}}}{\pi} \psred{\obsc{m}{\vec x}}  \Pi_0', \immred{\overline{\mc{G}_0}}{\pi}.
  \end{align*}
  Observe that \(\Pi_0, \overline{\mc{G}_0} \psred{\pi} \Pi_0', \immred{\overline{\mc{G}_0}}{\pi}\) induces a reduction  \(\jwfps{\Pi_0}{\mc{G}_0} \psred{(\pC, \pi)} \jwfps{\Pi_0'}{\immred{\mc{G}_0}{\pi}}\).
  By the induction hypothesis on \(\jtt{t'}{\jwfps{\Pi_0}{\mc{G}_0}}\) and this reduction, we deduce \(\jtt{\immred{t'}{\pi}}{\jwfps{\Pi_0'}{\immred{\mc{G}_0}{\pi}}}\).
  Case analysis on \(s\) and \(m\) with the reduction \(\Pi', \immred{\overline{\mc{G}}}{\pi} \psred{\obsc{m}{\vec x}}  \Pi_0', \immred{\overline{\mc{G}_0}}{\pi}\) implies \(\jtt{(\tcons{s}{m}[\vec x]{(\immred{t'}{\pi})})}{\jwfps{\Pi'}{\immred{\mc{G}}{\pi}}}\).
    But \(\immred{t}{\pi} = (\tcons{s}{p}[\vec x]{(\immred{\tau}{\pi})})\), so this gives the result.
\end{proofE}

\subsection{Synchronized trace interleavings}
\label{def:process-language:1}

Synchronized interleavings interleave traces according to a synchronous communication semantics.
It marks matching input-output elements as synchronized, and ensures that constraints imposed by one trace are not violated by the other trace.
This commutative operator is defined by lexicographic induction on the length of the traces.
We group its clauses into four categories:
\begin{enumerate}
\item the base cases: interleaving a trace with an empty trace;
\item synchronizing cases: the heads of each trace perform actions on the same channel;
\item commuting cases: the heads of each trace perform actions on different channels;
\item ill-defined cases: unsynchronizeable sends and receives, \etc
\end{enumerate}

The base cases specify that the empty trace is the unit for interleaving:
\begin{equation*}
  \tempt \pint t = \Set{ t }, \qquad t \pint \tempt = \Set{ t }.
\end{equation*}

The first synchronizing case holds whenever two traces send and receive the same message:
\begin{equation*}
  (\tcons{\pO}{m}[\vec x]{t_1}) \pint (\tcons{\pI}{m}[\vec x]{t_2})
  = \tcons{\pS}{m}[\vec x]{\left(t_1 \pint t_2\right)}.
\end{equation*}
It specifies that these two traces synchronize (captured by the \(\pS\) sign), and that the tail is given by interleaving their tails.
To ensure that the resulting trace is well-formed, we require that both traces bind the same names.
It is this implicit \(\alpha\)-variation that matches sent and received channel names \(b\) when interleaving traces of the form \(\tcons{\pO}{\mChan{a}}[b]{t_1}\) and \(\tcons{\pI}{\mChan{a}}[b]{t_2}\).

The other principal case manages constraints.
It states that a constraint in one trace can be dropped whenever it is satisfied by a corresponding action in the other trace, or whenever both traces have the same constraint.
If a constraint is matched against a different message on that channel, then the constraint cannot be satisfied and there are no interleavings.
\begin{equation*}
  (\tcons{s}{m}[\vec x]{t_1}) \tint (\tcons{\pC}{m}[\vec x]{t_2})
  = \tcons{s}{m}[\vec x]{\left(t_1 \tint t_2\right)}
\end{equation*}

The commuting cases handle heads that act on different channels.
Let \(\obsc{m}{\vec x}\)\label{def:obsc} be the observed communication induced by a message \(m\) and channel names \(\vec x\): \(\obsc{\mClose{a}}{\emptyset} = \pClose{a}\); \(\obsc{\mLabel{a}{l}}{\emptyset} = \pLabel{a}{l}\); \(\obsc{\mChan{a}}{b} = \pChan{a}{b}\); and \(\obsc{m}{\vec x}\) is undefined otherwise.
Given a partial map \(f\), let \(f(x) \kleeneq y\) define \(y\) whenever \(f(x)\) is defined.
Where \(T_i = \tcons{s_i}{m_i}[\vec x_i]{t_i}\),
\begin{equation*}
  \left.
    \begin{aligned}
      &\tcons{s_1}{m_1}[\vec x_1]{t_1} \tint \tcons{s_2}{m_2}[\vec x_2]{t_2}\\
      &= \Set{ \tcons{s_1}{m_1}[\vec x_1]{t} \given \immred{T_2}{\obsc{m_1}{\vec x_1}} \kleeneq T_2' \land t \in t_1 \tint T_2' } \cup {}\\
      &\cup \Set{ \tcons{s_2}{m_2}[\vec x_2]{t} \given \immred{T_1}{\obsc{m_2}{\vec x_2}} \kleeneq T_1' \land t \in T_1' \tint t_2 }
    \end{aligned}
    \quad
  \right\}
  \quad \text{if }\carrcn(m_1) \neq \carrcn(m_2).
\end{equation*}
This case is well-defined because the lengths of \(T_1'\) and \(T_2'\) are respectively at most the lengths of \(T_1\) and \(T_2\).
Our trace reduction operator hides considerable complexity, so we unpack the definition.

First, observe that the commuting case ensures that if a synchronization between \((s_1; m_1)\) and an element of \(T_2\) is eventually possible, then it occurs.
Indeed, if \(\carrcn(m_1)\) appears free in \(T_2\) with an observable sign, then \(\immred{T_2}{\obsc{m_1}{\vec x_1}}\) is undefined (see, \eg, \cref{ex:trace-operations:2}).
As a result, the first set (prefixing \((s_1; m_1)\)) is empty, and we interleave \(T_1\) with the tail of \(T_2\) in the second set to eventually synchronize \((s_1; m_1)\) with the corresponding element of \(T_2\).

\begin{example}
  \label{ex:trace-operations:3}
  The interleaving \(\tcons{\pO}{\mClose{a}}{\tempt} \tint \tcons{\pI}{\mClose{b}}{\tcons{\pI}{\mClose{a}}{\tempt}}\) is determined by the commuting case.
  The reduction in the first set of the clause is undefined, so the first set is empty.
  The second, \( \Set{ \tcons{\pI}{\mClose{b}}{t} \given t \in (\tcons{\pO}{\mClose{a}}{\tempt}) \tint (\tcons{\pI}{\mClose{a}}{\tempt}) } \), is defined in terms of a principal case.
  This principal case captures the synchronization on \(a\), and the interleavings are given by \( \Set{ \tcons{\pI}{\mClose{b}}{\tcons{\pS}{\mClose{a}}{\tempt}} }\).
\end{example}

Second, the commuting case detects deadlock.
Indeed, if two traces attempt to synchronize on different channels in opposite orders, then the carrier of the head of one trace is free in the other trace.
This implies that the reduction in each set is undefined, resulting in no interleavings:
\[
  (\tcons{\pI}{\mClose{a}}{\tcons{\pO}{\mClose{b}}{\tempt}})
  \pint
  (\tcons{\pI}{\mClose{b}}{\tcons{\pO}{\mClose{a}}{\tempt}})
  =
  \emptyset.
\]

Third, the commuting case uses reduction to handle constraint satisfaction.
Indeed, if \((s_1; m_1)\) satisfies a constraint in \(T_2\), then the reduction \(\immred{T_2}{\obsc{m_1}{\vec x_1}}\) deletes that constraint before interleaving the result with the tail \(t_1\).

\begin{example}
  \label{ex:trace-operations:4}
  Interleaving \((\tcons{\pO}{\mClose{a}}{\tempt}) \tint (\tcons{\pI}{\mClose{b}}{\tcons{\pC}{\mClose{a}}{\tempt}})\) produces the interleavings \( \Set{ \tcons{\pO}{\mClose{a}}{\tcons{\pI}{\mClose{b}}{\tempt}},\;\; \tcons{\pI}{\mClose{b}}{\tcons{\pO}{\mClose{a}}{\tempt}} } \), where the send on \(a\) in the first trace satisfies the constraint in the second trace.
\end{example}

Finally, the commuting case ensures that constraint satisfaction obeys synchronization boundaries.
For example, consider interleaving the traces \(\tcons{\pC}{\mClose{a}}{\tcons{\pI}{\mClose{b}}{\tempt}}\) and \({\tcons{\pO}{\mClose{b}}{\tcons{\pO}{\mClose{a}}{\tempt}}}\).
Intuitively, they synchronize on \(b\).
However, the constraint on \(a\) cannot be satisfied by the send on \(a\), because the constraint is imposed before the synchronization, while the send occurs after the synchronization on \(b\).
This unsatisfiability implies that the traces should have no synchronizing interleavings.
This is captured by our interleaving operator: the only applicable clause is the commuting case, and both sets defining it are empty.

All other cases are taken to be ill-defined, so we define their set of interleavings to be empty.
These include, \eg, cases of the form \((\tcons{\pO}{\mClose{a}}{t}) \tint (\tcons{\pO}{\mLabel{a}{l}}{t'})\) (sending on the same channel) and mismatched sends and receives: \((\tcons{\pO}{\mClose{a}}{t}) \tint (\tcons{\pI}{\mLabel{a}{l}}{t'})\).

A trace is \defin{safely constrained}\label{page:safely-constrained} if any carrier channel associated with a constraint sign is not also associated with input, output, or synchronization.
By \cref{prop:session-types:29}, constraints can only limit the interleavings of observable actions.
In other words, checking constraints when interleaving traces does not introduce new observable behaviours.
The opposite inclusion will play a key role when typechecking process~composition:

\begin{textAtEnd}
  Recall from \cpageref{page:safely-constrained} that a trace is safely constrained if any carrier channel associated with a constraint sign is not also associated with input, output, or synchronization.
\end{textAtEnd}

\begin{lemmaE}[][all end]
  \label{lemma:trace-operations:1}
  If \(\tcons{o}{m}[\vec x]{t_1}\) and \(\tcons{\pC}{m}[\vec x]{t_2}\) are safely constrained,
  then
  \[
    \tcons{o}{m}[\vec x]{\left((\tquoc{t_1}) \tint (\tquoc{t_2})\right)} \subseteq \left(\tquoc{\left(\tcons{o}{m}[\vec x]{t_1}\right)}\right) \tint \left(\tquoc{\left(\tcons{\pC}{m}[\vec x]{t_2}\right)}\right).
  \]
\end{lemmaE}

\begin{proofE}
  We remark that the hypothesis that both traces are safely constrained is essential.
  Indeed, the following traces are not safely constrained (they send/receive on \(a\) while also imposing a constraint on \(a)\)
  \begin{align*}
    &\tcons{\pO}{\mLabel{a}{l}}{\tcons{\pC}{\mClose{a}}{\tempt}}\\
    &\tcons{\pC}{\mLabel{a}{l}}{\tcons{\pI}{\mClose{a}}{\tempt}},
  \end{align*}
  and they do not satisfy the desired identity:
  \begin{align*}
    & \tcons{\pO}{\mLabel{a}{l}}{\left(\tquoc{\left(\tcons{\pC}{\mClose{a}}{\tempt}\right)}\right) \tint \left(\tquoc{\left(\tcons{\pI}{\mClose{a}}{\tempt}\right)}\right)}\\
    &= \tcons{\pO}{\mLabel{a}{l}}{\left(\tempt \tint \left(\tcons{\pI}{\mClose{a}}{\tempt}\right)\right)} \\
    &= \Set{ \tcons{\pO}{\mLabel{a}{l}}{\tcons{\pI}{\mClose{a}}{\tempt}} }\\
    &\nsubseteq \emptyset\\
    &= \left(\tcons{\pO}{\mLabel{a}{l}}{\tempt}\right)
      \tint
      \left(\tcons{\pI}{\mClose{a}}{\tempt}\right)\\
    &= \left(\tquoc{\left(\tcons{\pO}{\mLabel{a}{l}}{\tcons{\pC}{\mClose{a}}{\tempt}}\right)}\right)
      \tint
      \left(\tquoc{\left(\tcons{\pC}{\mLabel{a}{l}}{\tcons{\pI}{\mClose{a}}{\tempt}}\right)}\right).
  \end{align*}

  Assume that both traces are safely constrained.
  We examine the traces on the right side of the inclusion and observe:
  \begin{align*}
    \tquoc{\left(\tcons{o}{m}[\vec x]{t_1}\right)} &= \tcons{o}{m}[\vec x]{(\tquoc{t_1})}\\
    \tquoc{\left(\tcons{\pC}{m}[\vec x]{t_2}\right)} &= \tquoc{t_2}.
  \end{align*}
  The hypothesis that the traces are safely constrained implies that \(\carrcn(m)\) does not appear free in \(\tcons{\pC}{m}[\vec x]{t_2}\).
  It follows that the interleaving on the right side of the inclusion, which by the last two equalities is
  \(
    \left(\tcons{o}{m}[\vec x]{(\tquoc{t_1})}\right) \tint \left(\tquoc{t_2}\right)
  \),
  is either undefined or given by either a base case (when \(\tquoc{t_2} = \tempt\)) or a commuting case.

  If it is given by a base case, then the result follows easily:
  \begin{align*}
    &\tcons{o}{m}[\vec x]{\left((\tquoc{t_1}) \tint (\tquoc{t_2})\right)}\\
    &= \tcons{o}{m}[\vec x]{\left((\tquoc{t_1}) \tint \tempt\right)}\\
    &= \Set{ \tcons{o}{m}[\vec x]{(\tquoc{t_1})} }\\
    &= \tcons{o}{m}[\vec x]{(\tquoc{t_1})} \tint \tempt\\
    &= \left(\tquoc{\left(\tcons{o}{m}[\vec x]{t_1}\right)}\right) \tint \left(\tquoc{\left(\tcons{\pC}{m}[\vec x]{t_2}\right)}\right).
  \end{align*}

  If it is given by a commuting case, then observe that \(\immred{(\tquoc{t_2})}{\obsc{m}{\vec x}} = \tquoc{t_2}\) because \(\carrcn(m)\) does not appear free in \(t_2\).
  Using the definition of \(\tint\) in the commuting case, we then compute:
  \begin{align*}
    & \tcons{o}{m}[\vec x]{\left((\tquoc{t_1}) \tint (\tquoc{t_2})\right)}\\
    &= \Set{ \tcons{o}{m}[\vec x]{t} \given t \in \left((\tquoc{t_1}) \tint (\tquoc{t_2})\right)} \\
    &= \Set{ \tcons{o}{m}[\vec x]{t} \given \immred{(\tquoc{t_2})}{\obsc{m}{\vec x}} \kleeneq t_2' \land t \in \tquoc{t_1} \tint t_2' }\\
    &\subseteq \left(\tcons{o}{m}[\vec x]{\left(\tquoc{t_1}\right)}\right) \tint \tquoc{t_2}\\
    &= \left(\tquoc{\left(\tcons{o}{m}[\vec x]{t_1}\right)}\right) \tint \left(\tquoc{\left(\tcons{\pC}{m}[\vec x]{t_2}\right)}\right).
  \end{align*}

  This case analysis is exhaustive, so we conclude the result.
\end{proofE}

\begin{lemmaE}[][all end]
  \label{lemma:trace-operations:2}
  If \(t\) is safely constrained, then \({\tquoc{(\immred{t}{\obsc{m}{\vec x}})}}\), \({\immred{(\tquoc{t})}{\obsc{m}{\vec x}}}\), and \(\tquoc{t}\) are all defined and equal whenever \(\immred{t}{\obsc{m}{\vec x}}\) is defined.
\end{lemmaE}

\begin{proofE}
  The hypothesis that \(t\) is safely constrained is essential.
  Indeed, consider the unsafely constrained trace \( t = \tcons{\pC}{\mClose{a}}{\tcons{\pI}{\mClose{a}}{\tempt}} \).
  Then \(\immred{t}{\pClose{a}}\) is defined but \( \immred{(\tquoc{t})}{\pClose{a}} = \immred{(\tcons{\pI}{\mClose{a}}{\tempt})}{\pClose{a}} \) is undefined.

  Assume that \(t\) is safely constrained and that \(\immred{t}{\obsc{m}{\vec x}}\) is defined.
  It is immediate that traces \({\tquoc{(\immred{t}{\obsc{m}{\vec x}})}}\) and \(\tquoc{t}\) are defined.
  We proceed by induction on the length of \(t\) to show that \({\tquoc{(\immred{t}{\obsc{m}{\vec x}})}} = {\immred{(\tquoc{t})}{\obsc{m}{\vec x}}} = \tquoc{t}\).
  The base case occurs when \(t = \tempt\), in which case the result is immediate.
  Assume next that \(t = \tcons{s'}{m'}[\vec{y}]{t'}\).
  We consider two subcases, based on whether or not \(\carrcn(m) = \carrcn(m')\).

  Assume first that \(\carrcn(m) = \carrcn(m')\).
  In this case, the fact that \(\immred{t}{\obsc{m}{\vec x}}\) is defined implies \(s' = \pC\).
  Then, where the middle equality is given by the induction hypothesis:
  \[
    \tquoc{(\immred{t}{\obsc{m}{\vec x}})} = \tquoc{t'} = \immred{(\tquoc{t'})}{\obsc{m}{\vec x}} = \immred{(\tquoc{t})}{\obsc{m}{\vec x}},
  \]
  and \(\tquoc{(\immred{t}{\obsc{m}{\vec x}})} = \tquoc{t'} = \tquoc{t}\).
  These transitively give the equalities we wished to show.

  Assume otherwise that \(\carrcn(m) \neq \carrcn(m')\).
  We consider two subcases: if \(s' = \pC\) or \(s' \neq \pC\).
  If \(s' = \pC\), then:
  \begin{align*}
    &\tquoc{(\immred{t}{\obsc{m}{\vec x}})}\\
    \shortintertext{by definition:}
    &= \tquoc{(\tcons{s'}{m'}[\vec{y}]{(\immred{t'}{\obsc{m}{\vec x}})})}\\
    &= \tquoc{(\immred{t'}{\obsc{m}{\vec x}})}\\
    \shortintertext{by the induction hypothesis:}
    &= \immred{(\tquoc{t'})}{\obsc{m}{\vec x}}\\
    &= \immred{(\tquoc{t})}{\obsc{m}{\vec x}},
  \end{align*}
  which establishes the first equality.
  The second equality is similarly established by appealing to the induction hypothesis:
  \[
    \tquoc{(\immred{t}{\obsc{m}{\vec x}})}
    = \tquoc{(\immred{t'}{\obsc{m}{\vec x}})}
    = \tquoc{t'}
    = \tquoc{t}.
  \]
  If \(s' \neq \pC\), then the result follows by a similarly straightforward analysis.
\end{proofE}

\begin{propositionE}
  \label{prop:session-types:29}
  \label{prop:session-types:24}
  Given safely constrained traces \(t_1\) and \(t_2\), \(\tquoc{(t_1 \tint t_2)} \subseteq (\tquoc{t_1}) \tint (\tquoc{t_2})\).
\end{propositionE}

\begin{proofE}
  By lexicographic induction on the lengths of traces \(t_1\) and \(t_2\).

  The base cases are clear.
  If \(t_1 = \tempt\), then
  \[
    \tquoc{(\tempt \tint t_2)} = \tquoc{\Set{ t_2 }} = \Set{\tquoc{t_2}} = \tempt \pint (\tquoc{t_2}) = (\tquoc{\tempt}) \pint (\tquoc{t_2}).
  \]
  The analysis for \(t_2 = \tempt\) is symmetric.

  Assume next that both traces are non-empty.
  We proceed by case analysis on which clause in the definition determines \(t_1 \tint t_2\).

  The first clause is the synchronizing case in which the two traces send and receive the same message:
  \begin{equation*}
    t_1 \tint t_2
    = (\tcons{\pO}{m}[\vec x]{t_1'}) \pint (\tcons{\pI}{m}[\vec x]{t_2'})
    = \tcons{\pS}{m}[\vec x]{\left(t_1' \pint t_2'\right)}
  \end{equation*}
  By the induction hypothesis, \( \tquoc{\left(t_1' \pint t_2'\right)} \subseteq (\tquoc{t_1'}) \tint (\tquoc{t_2}') \).
  By definition of \(\tquoc{\cdot}\) and monotonicity:
  \begin{align*}
    &\tquoc{(t_1 \tint t_2)}\\
    &= \tquoc{\left(\tcons{\pS}{m}[\vec x]{\left(t_1' \pint t_2'\right)}\right)}\\
    &= \tcons{\pS}{m}[\vec x]{\left(\tquoc{\left(t_1' \pint t_2'\right)}\right)}\\
    &\subseteq \tcons{\pS}{m}[\vec x]{\left((\tquoc{t_1'}) \tint (\tquoc{t_2'})\right)}\\
    &= (\tquoc{(\tcons{\pO}{m}[\vec x]{t_1'})}) \pint (\tquoc{(\tcons{\pI}{m}[\vec x]{t_2'})})\\
    &= (\tquoc{t_1}) \tint (\tquoc{t_2}).
  \end{align*}
  This is what we wanted to show.

  The second clause is the principal case in which, without loss of generality, the head of \(t_1\) is \((s, m_1)\) and the head of \(t_2\) is \((\pC, m_2)\), where both \(m_1\) and \(m_2\) have the same carrier channel.
  It has two subcases.
  In the first subcase, \(m_1 \neq m_2\) or the two traces bind different numbers of variables, and \(t_1 \tint t_2\) is empty.
  It follows that \(\tquoc{(t_1 \tint t_2)}\) is empty, so the result holds trivially.
  The second subcase is when the interleaving is defined:
  \[
    (\tcons{s}{m}[\vec x]{t_1'}) \tint (\tcons{\pC}{m}[\vec x]{t_2'})
    = \tcons{s}{m}[\vec x]{\left(t_1' \tint t_2'\right)}.
  \]
  Assume first that \(s = \pC\), then:
  \begin{align*}
    &\tquoc{\left((\tcons{s}{m}[\vec x]{t_1'}) \tint (\tcons{\pC}{m}[\vec x]{t_2'})\right)}\\
    &= \tquoc{\left(\tcons{s}{m}[\vec x]{\left(t_1' \tint t_2'\right)}\right)}\\
    &= \tquoc{\left(t_1' \tint t_2'\right)}\\
    \shortintertext{which by the induction hypothesis,}
    &\subseteq \tquoc{t_1'} \tint \tquoc{t_2'}\\
    &= (\tquoc{t_1}) \tint (\tquoc{t_2}).
  \end{align*}
  Assume otherwise that \(s \neq \pC\):
  We then compute, using the induction hypothesis, monotonicity, and the definition of \(\tint\):
  \begin{align*}
    &\tquoc{\left((\tcons{s}{m}[\vec x]{t_1'}) \tint (\tcons{\pC}{m}[\vec x]{t_2'})\right)}\\
    &= \tquoc{\left(\tcons{s}{m}[\vec x]{\left(t_1' \tint t_2'\right)}\right)}\\
    &= \tcons{s}{m}[\vec x]{\left(\tquoc{\left(t_1' \tint t_2'\right)}\right)}\\
    &\subseteq \tcons{s}{m}[\vec x]{\left((\tquoc{t_1'}) \tint (\tquoc{t_2'})\right)}\\
    \shortintertext{which by \cref{lemma:trace-operations:1}:}
    &\subseteq (\tquoc{t_1}) \tint (\tquoc{t_2}).
  \end{align*}
  This completes the analysis of the synchronizing cases.

  Next, we show the result for the commuting cases, where the heads of each trace perform actions on different channels.
  In this case, \(t_i = \tcons{s_i}{m_i}[\vec x_i]{t_i'}\) for \(1 \leq i \leq 2\) with \(\carrcn(m_1) \neq \carrcn(m_2)\).
  We compute:
  \begin{align*}
    &\tquoc{(t_1 \tint t_2)}\\
    &= \bigcup_{1 \leq i \neq j \leq 2} \tquoc{\Set{ \tcons{s_i}{m_i}[\vec x_i]{t} \given \immred{t_j}{\obsc{m_i}{\vec x_i}} \kleeneq T_j' \land t \in t_i' \tint T_j' }}\\
    \shortintertext{using the fact the operation \(\tquoc{\cdot}\) is idempotent:}
    &= \bigcup_{1 \leq i \neq j \leq 2} \tquoc{\Set{ \tcons{s_i}{m_i}[\vec x_i]{(\tquoc{t})} \given \immred{t_j}{\obsc{m_i}{\vec x_i}} \kleeneq T_j' \land t \in t_i' \tint T_j' }}\\
    &= \bigcup_{1 \leq i \neq j \leq 2} \tquoc{\Set{ \tcons{s_i}{m_i}[\vec x_i]{t} \given \immred{t_j}{\obsc{m_i}{\vec x_i}} \kleeneq T_j' \land t \in \tquoc{(t_i' \tint T_j')} }}\\
    \shortintertext{by the induction hypothesis:}
    &= \bigcup_{1 \leq i \neq j \leq 2} \tquoc{\Set{ \tcons{s_i}{m_i}[\vec x_i]{t} \given \immred{t_j}{\obsc{m_i}{\vec x_i}} \kleeneq T_j' \land t \in (\tquoc{t_i'}) \tint (\tquoc{T_j'}) }}\\
    &= \bigcup_{1 \leq i \neq j \leq 2} \tquoc{\Set{ \tcons{s_i}{m_i}[\vec x_i]{t} \given \tquoc{(\immred{t_j}{\obsc{m_i}{\vec x_i}})} \kleeneq T_j' \land t \in (\tquoc{t_i'}) \tint T_j' }}\\
  \end{align*}
  Let \(X\) be the last set in this sequence.
  We proceed by case analysis on \(s_1\) and \(s_2\).

  If \(s_1 = s_2 = \pC\), then by \cref{lemma:trace-operations:2}, \(X\) is:
  \begin{align*}
    X &\subseteq \bigcup_{1 \leq i \neq j \leq 2} \tquoc{\Set{ \tcons{s_i}{m_i}[\vec x_i]{t} \given \immred{(\tquoc{t_j})}{\obsc{m_i}{\vec x_i}} \kleeneq T_j' \land t \in (\tquoc{t_i'}) \tint T_j' }}\\
    &\subseteq \bigcup_{1 \leq i \neq j \leq 2} \tquoc{\Set{ \tcons{s_i}{m_i}[\vec x_i]{t} \given \tquoc{t_j} \kleeneq T_j' \land t \in (\tquoc{t_i'}) \tint T_j' }}\\
      &= \bigcup_{1 \leq i \neq j \leq 2} \Set{ t \given \tquoc{t_j} \kleeneq T_j' \land t \in (\tquoc{t_i'}) \tint T_j' }\\
      &= \bigcup_{1 \leq i \neq j \leq 2} \Set{ t \given \in (\tquoc{t_i'}) \tint (\tquoc{t_j'}) }\\
      &= (\tquoc{t_1'}) \tint (\tquoc{t_2'})\\
      &= (\tquoc{t_1}) \tint (\tquoc{t_2}),
  \end{align*}
  which is what we wanted to show.

  If neither of \(s_1\) or \(s_2\) is \(\pC\), then by \cref{lemma:trace-operations:2}, \(X\) is:
  \begin{align*}
    X &\subseteq \bigcup_{1 \leq i \neq j \leq 2} \tquoc{\Set{ \tcons{s_i}{m_i}[\vec x_i]{t} \given \immred{(\tquoc{t_j})}{\obsc{m_i}{\vec x_i}} \kleeneq T_j' \land t \in (\tquoc{t_i'}) \tint T_j' }}\\
      &= \bigcup_{1 \leq i \neq j \leq 2} \Set{ \tcons{s_i}{m_i}[\vec x_i]{t} \given \immred{(\tquoc{t_j})}{\obsc{m_i}{\vec x_i}} \kleeneq T_j' \land t \in (\tquoc{t_i'}) \tint T_j' }\\
      &= \bigcup_{1 \leq i \neq j \leq 2} \Set{ \tcons{s_i}{m_i}[\vec x_i]{t} \given \immred{(\tcons{s_j}{m_j}[\vec x_j]{(\tquoc{t_j'})})}{\obsc{m_i}{\vec x_i}} \kleeneq T_j' \land t \in (\tquoc{t_i'}) \tint T_j' }\\
      &= (\tcons{s_1}{m_1}[\vec x_1]{(\tquoc{t_1'})}) \tint (\tcons{s_2}{m_2}[\vec x_2]{(\tquoc{t_2'})})\\
      &= (\tquoc{t_1}) \tint (\tquoc{t_2}).
  \end{align*}

  If, without loss of generality, \(s_1 = \pC\) and \(s_2 \neq \pC\), then \(X\) is:
  \begin{align*}
    X &= \tquoc{\Set{ \tcons{\pC}{m_1}[\vec x_1]{t} \given \tquoc{(\immred{t_2}{\obsc{m_1}{\vec x_1}})} \kleeneq T_2' \land t \in (\tquoc{t_1'}) \tint T_2' }} \cup {}\\
      &\phantom{{}={}} \tquoc{\Set{ \tcons{s_2}{m_2}[\vec x_2]{t} \given \tquoc{(\immred{t_1}{\obsc{m_2}{\vec x_2}})} \kleeneq T_1' \land t \in (\tquoc{t_2'}) \tint T_1' }},\\
    \shortintertext{which by \cref{lemma:trace-operations:2}:}
      &\subseteq \tquoc{\Set{ \tcons{\pC}{m_1}[\vec x_1]{t} \given \tquoc{t_2} \kleeneq T_2' \land t \in (\tquoc{t_1'}) \tint T_2' }} \cup {}\\
      &\phantom{{}={}} \tquoc{\Set{ \tcons{s_2}{m_2}[\vec x_2]{t} \given \tquoc{(\immred{t_1}{\obsc{m_2}{\vec x_2}})} \kleeneq T_1' \land t \in (\tquoc{t_2'}) \tint T_1' }}\\
      &= \Set{ t \given \tquoc{t_2} \kleeneq T_2' \land t \in (\tquoc{t_1'}) \tint T_2' } \cup {}\\
      &\phantom{{}={}} \tquoc{\Set{ \tcons{s_2}{m_2}[\vec x_2]{t} \given \tquoc{(\immred{t_1}{\obsc{m_2}{\vec x_2}})} \kleeneq T_1' \land t \in (\tquoc{t_2'}) \tint T_1' }}\\
      &= \Set{ t \given t \in (\tquoc{t_1'}) \tint (\tquoc{t_2}) } \cup {}\\
      &\phantom{{}={}} \tquoc{\Set{ \tcons{s_2}{m_2}[\vec x_2]{t} \given \tquoc{(\immred{t_1}{\obsc{m_2}{\vec x_2}})} \kleeneq T_1' \land t \in (\tquoc{t_2'}) \tint T_1' }}\\
      &= (\tquoc{t_1'}) \tint (\tquoc{t_2}) \cup {}\\
      &\phantom{{}={}} \tquoc{\Set{ \tcons{s_2}{m_2}[\vec x_2]{t} \given \tquoc{(\immred{t_1}{\obsc{m_2}{\vec x_2}})} \kleeneq T_1' \land t \in (\tquoc{t_2'}) \tint T_1' }}\\
      &= (\tquoc{t_1'}) \tint (\tquoc{t_2}) \cup {}\\
      &\phantom{{}={}} \Set{ \tcons{s_2}{m_2}[\vec x_2]{t} \given \tquoc{(\immred{t_1}{\obsc{m_2}{\vec x_2}})} \kleeneq T_1' \land t \in (\tquoc{t_2'}) \tint T_1' },\\
    \shortintertext{which by \cref{lemma:trace-operations:2}:}
      &= (\tquoc{t_1'}) \tint (\tquoc{t_2}) \cup {}\\
      &\phantom{{}={}} \Set{ \tcons{s_2}{m_2}[\vec x_2]{t} \given \immred{(\tquoc{t_1})}{\obsc{m_2}{\vec x_2}} \kleeneq T_1' \land t \in (\tquoc{t_2'}) \tint T_1' }\\
      &= (\tquoc{t_1'}) \tint (\tquoc{t_2}) \cup {}\\
      &\phantom{{}={}} \Set{ \tcons{s_2}{m_2}[\vec x_2]{t} \given \immred{(\tquoc{t_1'})}{\obsc{m_2}{\vec x_2}} \kleeneq T_1' \land t \in (\tquoc{t_2'}) \tint T_1' }\\
    \shortintertext{At this point, observe that if \(T_1'\) is defined, then the carrier of the head of \(\tquoc{t_1'}\), if it exists, cannot be \(\carrcn(m_2)\).
    Indeed, the head of \(\tquoc{t_1'}\) has sign \(\pI\), \(\pO\) or \(\pS\), and if its carrier were \(\carrcn(m_2)\), then the reduction by \(\obsc{m_2}{\vec x_2}\) would be undefined.
    By \cref{lemma:trace-operations:2}, this last union is:}
      &= (\tquoc{t_1'}) \tint (\tquoc{t_2}) \cup {}\\
      &\phantom{{}={}} \tquoc{\Set{ \tcons{s_2}{m_2}[\vec x_2]{t} \given \immred{(\tquoc{t_1'})}{\obsc{m_2}{\vec x_2}} \kleeneq T_1' \land t \in (\tquoc{t_2'}) \tint T_1' }}\\
    \shortintertext{By the just-above observation, the following interleaving must be given by a commuting case:}
      &\subseteq (\tquoc{t_1'}) \tint (\tquoc{t_2}) \cup {}\\
      &\phantom{{}={}} \tquoc{t_1'} \tint (\tcons{s_2}{m_2}[\vec x_2]{(\tquoc{t_2'})})\\
      &= (\tquoc{t_1'}) \tint (\tquoc{t_2})\\
      &= (\tquoc{t_1}) \tint (\tquoc{t_2}).
  \end{align*}
  This case analysis is exhaustive and completes the proof.
\end{proofE}

\subsection{Type-theoretic properties}

\Cref{sec:type-reduction} gave a trace semantics for process specifications.
We relate the behaviour of trace operators to this trace assignment.
Recall that a process specification \(\jwfps{\Pi}{\mc{G}}\) is well-formed only if every channel appearing free in a type in \(\Pi, \mc{G}\) is typed by some other type assignment in the specification.

\Cref{prop:session-types:13} semantically characterizes hiding internal channels in a specification.
The well-formedness hypothesis is required to ensure that no types become ill-scoped.
For example, it precludes deleting \(a : \Tu\) from \(\jwfps{\Pi}{\proci{\Delta}{\Iota, a : \Tu}{c : \TCaseU{a}{C}}}\).

\begin{propositionE}
  \label{prop:session-types:13}
  If \(\jtt{t}{\jwfps{\Pi}{\mc{G} \scons \proci{\Delta}{\Iota,\Gamma}{a : A}}}\) and \(\jwfps{\Pi}{\mc{G} \scons \proci{\Delta}{\Iota}{a : A}}\) is well-formed, then \(\jtt{(\tdel{t}{\dom(\Gamma)})}{\jwfps{\Pi}{\mc{G} \scons \proci{\Delta}{\Iota}{a : A}}}\).
\end{propositionE}

\begin{proofE}
  Recall that \(\jtt{t}{\jwfps{\Pi}{\mc{G} \scons \proci{\Delta}{\Iota,\Gamma}{a : A}}}\) is inductively defined in \cref{sec:type-reduction} using a subset relationship.
  We proceed by induction on this derivation.

  The base case, induced by \( \sembr{ \jwfts{\Pi} { \cdot } } \supseteq \Set{ \tempt } \) holds vacuously.

  For the inductive steps, we show several representative cases.

  Assume that \(\jtt{t}{ \jwfps{\Pi}{\mc{G} \scons (\jwfpi { \Delta } { \Iota, \Gamma } { a : \Tplus_{l \in L} A_l }) }}\) because \(t = \tcons{\pO}{\mLabel{a}{k}}{t'}\) with \(\jtt{t'}{ \jwfts { \immred{\Pi}{\pi} } { \immred{\mc{G}}{\pi} \scons (\jwfpi { \immred{\Delta}{\pi} } { \immred{(\Iota, \Gamma)}{\pi} } { a : A_k })} }\),  where \(\pi = \pLabel{a}{k}\).
  By the induction hypothesis, \(\jtt{(\tdel{t'}{\dom(\Gamma)})}{ \jwfts { \immred{\Pi}{\pi} } { \immred{\mc{G}}{\pi} \scons (\jwfpi { \immred{\Delta}{\pi} } { \immred{\Iota}{\pi} } { a : A_k })} }\).
  Then by the same clause, \(\jtt{(\tdel{t}{\dom(\Gamma)}}{ \jwfps{\Pi}{\mc{G} \scons (\jwfpi { \Delta } { \Iota } { a : \Tplus_{l \in L} A_l }) }}\).

  Assume that \(\jtt{t}{ \jwfts { \Pi } { \mc{G} \scons (\jwfpi { \Delta } { \Iota, \Gamma, a : \Tplus_{l \in L} A_l } { c : C })} }\) because \(t = \tcons{\pS}{\mLabel{a}{k}}{t'}\) with \(\jtt{t'}{ \jwfts { \immred{\Pi}{\pi} } { \immred{\mc{G}}{\pi} \scons (\jwfpi { \immred{\Delta}{\pi} } { \immred{(\Iota, \Gamma)}{\pi}, a : A_k } { c : \immred{C}{\pi} })} }\), where \(\pi = \pLabel{a}{k}\).
  This case involves two subcases: the first where \(a\) is deleted with \(\Gamma\), and where \(a\) is not deleted.
  The second subcase is analogous to the previous case, so assume that \(a\) is deleted.
  We must show that
  \[
    \jtt{\tdel{t}{\dom(\Gamma, a : \Tplus_{l \in L} A_l )}}{ \jwfts { \Pi } { \mc{G} \scons (\jwfpi { \Delta } { \Iota } { c : C })} }.
  \]
  Observe that the specifications
  \[
    \jwfts { \immred{\Pi}{\pi} } { \immred{\mc{G}}{\pi} \scons (\jwfpi { \immred{\Delta}{\pi} } { \immred{(\Iota, \Gamma)}{\pi}, a : A_k } { c : \immred{C}{\pi} })}
    \quad
    \text{and}
    \quad
    \jwfts { \Pi } { \mc{G} \scons (\jwfpi { \Delta } { \Iota, \Gamma, a : A_k } { c : C })}
  \]
  are syntactically equal by the well-formedness assumption (reduction by an observed communication behaves as the identity on types that do not mention the communication's carrier channel).
  By the induction hypothesis,
  \[
    \jtt{\tdel{t'}{\dom(\Gamma, a : A_k )}}{ \jwfts { \immred{\Pi}{\pi} } { \immred{\mc{G}}{\pi} \scons (\jwfpi { \immred{\Delta}{\pi} } { \immred{\Iota}{\pi} } { c : \immred{C}{\pi} })} }
  \]
  By the above syntactic observation,
  \[
    \jtt{\tdel{t'}{\dom(\Gamma, a : A_k )}}{     \jwfts { \Pi } { \mc{G} \scons (\jwfpi { \Delta } { \Iota } { c : C })} }.
  \]
  By definition of the deletion operation,
  \[
    \jtt{\tdel{t}{\dom(\Gamma, a : \Tplus_{l \in L} A_l )}}{     \jwfts { \Pi } { \mc{G} \scons (\jwfpi { \Delta } { \Iota } { c : C })} }.
  \]

  The remaining cases are analogous.
\end{proofE}

The following proposition characterizes independent parallel executions.
To concretize it, consider specifications \(\jwfps{\Pi, \Delta_2, \Iota_2, a_2 : A_2}{\proci{\Delta_1}{\Iota_1}{a_1 : A_1}}\) and \(\jwfps{\Pi, \Delta_1, \Iota_1, a_1 : A_1}{\proci{\Delta_2}{\Iota_2}{a_2 : A_2}}\).
They specify two processes whose types can observe each other's channels.
We can compose these two specifications in a single specification: \(\jwfps{\Pi}{\proci{\Delta_1}{\Iota_1}{a_1 : A_1} \scons \proci{\Delta_2}{\Iota_2}{a_2 : A_2}}\).
The proposition states that interleaving the traces of each individual specification results in traces allowed by the composed specification.
This proposition is useful when characterizing traces involving channel transmissions.
More generally, where \(\overline{\mc{G}}\) denotes the typed channels in \(\mc{G}\):

\begin{propositionE}
  \label{conj:session-types:1}
  If \(\jtt{t_1}{\getrh{WCons}{1}}\) and \(\jtt{t_2}{\getrh{WCons}{2}}\), then \(t_1 \tint t_2 \subseteq \jtts{\getrc{WCons}}\).
\end{propositionE}

\begin{proofE}
  By lexicographic induction on the lengths of \(t_1\) and \(t_2\).
  The base cases where either trace is \(\tempt\) are obvious.
  Assume next that \(t_i = \tcons{s_i}{m_i}[\vec x_i]{\tau_i}\) for \(i = 1,2\), and set \(\pi_i = \obsc{m_i}{\vec x_i}\).
  We proceed by case analysis on which clause defines \(t_1 \tint t_2\).

  The first synchronizing case, where \(m_1 = m_2\) and \(\vec x_1 = \vec x_2\) and, without loss of generality, \(s_1 = \pO\) and \(s_2 = \pI\), is impossible.
  Indeed, it would require \(\mc{G}_1\) and \(\mc{G}_2\) to type the same channel, but this would render the specification \(\getrc{WCons}\) ill-formed.

  The second synchronizing case, where \(m_1 = m_2\) and \(\vec x_1 = \vec x_2\), and without loss of generality, \(s_2 = \pC\) is possible.
  We proceed by case analysis on \(s_1\) and \(m_1\).
  If \(s_1 = \pI\) and \(m_1 = \mLabel{k}{a}\), then
  \begin{enumerate}
  \item \(t_1 = \tcons{\pO}{\mLabel{k}{a}}{\tau_1}\)
  \item \(t_2 = \tcons{\pC}{\mLabel{k}{a}}{\tau_2}\)
  \end{enumerate}
  Inversion on the derivations \(\jtt{t_1}{\getrh{WCons}{1}}\) and \(\jtt{t_2}{\getrh{WCons}{2}}\) gives:
  \begin{enumerate}[resume]
  \item \(\mc{G}_1 = \mc{G}_1' \scons (\jwfpi { \Delta_1 } { \Iota_1 } { a : \Tplus_{l \in L} A_l })\)
  \item \(\jtt{t_1}{\jwfts { \Pi, \overline{\mc{G}_2} } { \mc{G}_1' \scons (\jwfpi { \Delta_1 } { \Iota_1 } { a : \Tplus_{l \in L} A_l })}}\)
  \item \(\jtt{t_2}{ \jwfts { \Pi, \overline{\mc{G}_1', \Delta_1, \Iota_1},  a : \Tplus_{l \in L} A_l } { \mc{G}_2 } }\)
  \item \(\jtt{\tau_1}{ \jwfts { \immred{(\Pi, \overline{\mc{G}_2})}{\pi_1} } { \immred{\mc{G}_1'}{\pi_1} \scons (\jwfpi { \immred{\Delta_1}{\pi_1} } { \immred{\Iota_1}{\pi_1} } { a : A_k })} } \)
  \item \(\jtt{\tau_2}{ \jwfts { \immred{(\Pi, \overline{\mc{G}_1', \Delta_1, \Iota_1})}{\pi_1},  a : A_k } { \immred{ \mc{G}_2 }{\pi_1} }}\)
  \end{enumerate}
  By the induction hypothesis:
  \begin{enumerate}[resume]
  \item \({\tau_1 \tint \tau_2} \subseteq \jtts{ \jwfts{ \immred{\Pi}{\pi_1} }{ \immred{\mc{G}_1'}{\pi_1} \scons (\jwfpi { \immred{\Delta_1}{\pi_1} } { \immred{\Iota_1}{\pi_1} } { a : A_k }) \scons \immred{\mc{G}_2}{\pi_1}}} \)
  \end{enumerate}
  Then by definition of trace sets for specifications,
  \begin{enumerate}[resume]
  \item \(\tcons{\pO}{\mLabel{k}{a}}{(\tau_1 \tint \tau_2)} \subseteq \jtts{ \jwfts{ \Pi }{ \mc{G}_1' \scons (\jwfpi { \Delta_1 } { {\Iota_1} } { a :  \Tplus_{l \in L} A_l }) \scons \mc{G}_2}}\)
  \end{enumerate}
  But this is exactly the desired inclusion \(t_1 \tint t_2 \subseteq \jtts{\getrc{WCons}}\).
  The analysis is analogous for the other subcases \(s_1 \in \Set{ \pO, \pS, \pC }\).
  It is also analogous for other values of \(m_1\).

  Finally, we consider the commuting cases where \(\carrcn(m_1) \neq \carrcn(m_2)\).
  In this case,
  \begin{align*}
    &\tcons{s_1}{m_1}[\vec x_1]{t_1} \tint \tcons{s_2}{m_2}[\vec x_2]{t_2}\\
    &= \Set{ \tcons{s_1}{m_1}[\vec x_1]{t} \given \immred{t_2}{\obsc{m_1}{\vec x_1}} \kleeneq T_2' \land t \in t_1 \tint T_2' } \cup {}\\
    &\cup \Set{ \tcons{s_2}{m_2}[\vec x_2]{t} \given \immred{t_1}{\obsc{m_2}{\vec x_2}} \kleeneq T_1' \land t \in T_1' \tint t_2 }
  \end{align*}
  We must show that each of these two subsets is contained in \(\jtts{\getrc{WCons}}\).
  We do so for the first; the second inclusion will follow by symmetry.
  If \(T_2'\) is undefined, then the inclusion is immediate (the set is empty).
  If \(T_2'\) is defined, then we consider two subcases: whether \(s_1 = \pC\) or \(s_1 \neq \pC\).

  If \(s_1 = \pC\), then by inversion on \(\jtt{t_1}{\getrh{WCons}{1}}\):
  \begin{enumerate}
  \item \(\jtt{t_1}{\jwfts{ \Pi, \overline{\mc{G}_2} }{\mc{G}_1}}\),
  \item \(\jtt{t_1'}{\jwfts{\Pi', \overline{\mc{G}_2'}}{\immred{\mc{G}_1}{\pi_1}}}\) for some \(\Pi', \overline{\mc{G}_2'}\) such that
  \item \(\Pi, \overline{\mc{G}_2} \psred{\pi_1} \Pi', \overline{\mc{G}_2'}\).
  \end{enumerate}
  That \(T_2' = \immred{t_2}{\pi_1}\) is defined implies that \(\carrcn(\pi_1) \notin \dom(\mc{G}_2)\), so \(\carrcn(\pi_1) \in \dom(\Pi)\).
  It follows that \(\Pi, \overline{\mc{G}_2} \psred{\pi_1} \Pi', \overline{\mc{G}_2'}\) because:
  \begin{enumerate}[resume]
  \item \(\Pi \psred{\pi_1} \Pi'\) and
  \item \(\overline{\mc{G}_2'} = \immred{\mc{G}_2}{\pi_1}\).
  \end{enumerate}
  These facts jointly imply
  \begin{enumerate}[resume]
  \item \(\jwfps{\Pi, \overline{\mc{G}_2}}{\mc{G}_1} \psred{(\pC, \pi_1)} \jwfps{\Pi', \overline{\immred{\mc{G}_2}{\pi_1}}}{\immred{\mc{G}_1}{\pi_1}}\).
  \end{enumerate}
  By \cref{prop:session-types:36}, \(\jtt{T_2'}{\jwfps{\Pi', \overline{\immred{\mc{G}_2}{\pi_1}}}{\immred{\mc{G}_1}{\pi_1}}}\).
  By the induction hypothesis,
  \[
    t_1' \tint T_2' \subseteq \jtts{\jwfps{\Pi'}{\immred{\mc{G}_1}{\pi_1} \scons \immred{\mc{G}_2}{\pi_1}}}.
  \]
  By the same clause giving \(\jtt{t_1}{\getrh{WCons}{1}}\), we deduce
  \[
    \tcons{\pC}{m_1}[\vec x_1]{(t_1' \tint T_2')} \subseteq \jtts{\jwfps{\Pi'}{\immred{\mc{G}_1}{\pi_1} \scons \immred{\mc{G}_2}{\pi_1}}},
  \]
  which is what we wanted to show.

  The case where \(s_1 \neq \pC\) is analogous.
  This completes the case analyses so we conclude the result.
\end{proofE}

An alternate form of specification composition results in synchronization.
Indeed, consider processes specifications \(\jwfps{\Pi, \Delta_2, \Iota_2, c : C}{\proci{\Delta_1}{\Iota_1}{a : A}}\) and \(\jwfps{\Pi, \Delta_1, \Iota_1}{\proci{\Delta_2, a : A}{\Iota_2}{{c : C}}}\).
The parallel composition operation composes them along the channel \(a\) so that they synchronize on this channel.
Subject to side conditions, their composition satisfies \(\jwfps{\Pi}{\proci{\Delta_1, \Delta_2}{\Iota_1, a : A, \Iota_2}{c : C}}\) where \(a\) becomes an inner synchronization channel.
This synchronizing parallel composition is also characterized by synchronized interleaving:

\begin{propositionE}
  \label{cor:session-types:4}
  If \(\jtt{t_1}{\jwfps{\Pi, \Delta_2, \Iota_2, c : C}{\proci{\Delta_1}{\Iota_1}{a : A}}}\) and\\
  \(\jtt{t_2}{\jwfps{\Pi, \Delta_1, \Iota_1}{\proci{\Delta_2, a : A}{\Iota_2}{{c : C}}}}\), then \(t_1 \tint t_2 \subseteq \jtts{\jwfps{\Pi}{\proci{\Delta_1, \Delta_2}{\Iota_1, a : A, \Iota_2}{c : C}}}\).
\end{propositionE}

\begin{proofE}
  By induction on definition of \(t_1 \tint t_2\).

  The base cases where traces \(t_1\) or \(t_2\) are \(\tempt\) are impossible by inversion on the derivations of \({\jtt{t_1}{\jwfps{\Pi, \Delta_2, \Iota_2, c : C}{\proci{\Delta_1}{\Iota_1}{a : A}}}}\) and \(\jtt{t_2}{\jwfps{\Pi, \Delta_1, \Iota_1}{\proci{\Delta_2, a : A}{\Iota_2}{{c : C}}}}\).

  In the inductive cases, \(t_i = \tcons{s_i}{m_i}[\vec x_i]{t_i'}\).

  Assume first that we are in the principal synchronizing case: \(\carrcn(m_1) = \carrcn(m_2)\) and \(\vec x_1 = \vec x_2\), and without loss of generality, \(s_1 = \pO\) and \(s_2 = \pI\).
  We proceed by case analysis on \(m_1\).
  If \(m_1 = \mLabel{k}{a}\), then
  \begin{enumerate}
  \item \(t_1 = \tcons{\pO}{\mLabel{k}{a}}{\tau_1}\)
  \item \(t_2 = \tcons{\pI}{\mLabel{k}{a}}{\tau_2}\)
  \item \(A = \Tplus_{l \in L} A_l\) with \(k \in L\).
  \end{enumerate}
  Set \(\pi = \obsc{\mLabel{k}{a}}{\emptyset}\).
  Inversion on the derivations \(\jtt{t_1}{\jwfps{\Pi, \Delta_2, \Iota_2, c : C}{\proci{\Delta_1}{\Iota_1}{a : A}}}\) and
  \(\jtt{t_2}{\jwfps{\Pi, \Delta_1, \Iota_1}{\proci{\Delta_2, a : A}{\Iota_2}{{c : C}}}}\) gives:
  \begin{enumerate}[resume]
  \item \(\jtt{\tau_1}{    \jwfts
    { \immred{(\Pi, \Delta_2, \Iota_2, c : C)}{\pi} }
    {
    (\jwfpi
    { \immred{\Delta_1}{\pi} }
    { \immred{\Iota_1}{\pi} }
    { a : A_k })}
  }\)
\item \(\jtt{\tau_2}{
    \jwfts
    { \immred{(\Pi, \Delta_1, \Iota_1)}{\pi} }
    {
      (\jwfpi
    { \immred{\Delta_2}{\pi}, a : A_k }
    { \immred{\Iota_2}{\pi} }
    { c : \immred{C}{\pi} })}
    }\)
  \end{enumerate}
  By the induction hypothesis, \( \tau_1 \tint \tau_2 \subseteq \jtts{ \jwfts{ \immred{\Pi}{\pi} } { \jwfpi{\immred{(\Delta_1, \Delta_2)}{\pi} } { \immred{(\Iota_1, \Iota_2)}{\pi}, a : A_k } { c : \immred{C}{\pi} } }}\).
  It then follows that \( \tcons{\pS}{\mLabel{k}{a}}{(\tau_1 \tint \tau_2)} \subseteq \jtts{\jwfps{\Pi}{\proci{\Delta_1, \Delta_2}{\Iota_1, a : A, \Iota_2}{c : C}}}\).
  But this is exactly what we wanted to show, for \(t_1 \tint t_2 = \tcons{\pS}{\mLabel{k}{a}}{(\tau_1 \tint \tau_2)}\).
  The case analyses for other choices of \(m_1\) are analogous.

  Assume next that it is the synchronizing case dealing with constraints.
  In particular, \(m_1 = m_2\), \(\vec x_1 = \vec x_2\), and assume that \(s_2 = \pC\) (this is not without loss of generality, but the case where \(s_1 = \pC\) is analogous).
  Set \(\pi = \obsc{m_1}{\vec x_1}\).
  By inversion on the elementhood relation for \(t_2\), we know that \(\carrcn(m_2) \neq a\).
  We proceed by case analysis on \(s_1\) and \(m_1\).
  Assume first that \(s_1 = \pI\) and \(m_1 = \mLabel{b}{k}\).
  Then by inversion on elementhood for \(t_1\), we have \(\Delta_1 = \Delta, b : \Tplus_{l \in L} B_l\) with \(k \in L\), and by inversion on elementhood for both traces:
  \begin{enumerate}
  \item \(\jtt{t_1}{\jwfps{\Pi, \Delta_2, \Iota_2, c : C}{\proci{\Delta_1}{\Iota_1}{a : A}}}\)
  \item \(\jtt{t_1'}{\jwfps{\immred{(\Pi, \Delta_2, \Iota_2, c : C)}{\pi}}{\proci{\immred{\Delta}{\pi}, b : B_k}{\immred{\Iota_1}{\pi}}{a : \immred{A}{\pi}}}}\)
  \item \(\jtt{t_2}{\jwfps{\Pi, \Delta_1, \Iota_1}{\proci{\Delta_2, a : A}{\Iota_2}{{c : C}}}}\)
  \item \(\jtt{t_2'}{\jwfps{\immred{(\Pi, \Iota_1, \Delta)}{\pi}, b : B_k}{\immred{\proci{\Delta_2, a : A}{\Iota_2}{{c : C}}}{\pi}}}\)
  \end{enumerate}
  By the induction hypothesis,
  \[
    t_1' \tint t_2' \subseteq \jtts{\jwfps{\immred{\Pi}{\pi}}{\proci{\immred{(\Delta, \Delta_2)}{\pi}, b : B_k}{\immred{(\Iota_1, \Iota_2, a : A)}{\pi}}{c : \immred{C}{\pi}}}}.
  \]
  By the same clause giving elementhood for \(t_1\),
  \[
    t_1 \tint t_2 = \tcons{s_1}{m_1}[\vec x_1]{(t_1' \tint t_2')} \subseteq \jtts{\jwfps{\Pi}{\proci{\Delta_1, \Delta, b : \Tplus_{l \in L} B_l}{\Iota_1, a : A, \Iota_2}{c : C}}}.
  \]
  This is exactly what we wanted to show.
  The case analyses for other choices of \(s_1\) and \(m_1\) are equally mundane.

  Finally, assume that \(t_1 \tint t_2\) is given by a commuting case, where \(\carrcn(m_1) \neq \carrcn(m_2)\).
  In this case,
  \begin{align*}
    &\tcons{s_1}{m_1}[\vec x_1]{t_1'} \tint \tcons{s_2}{m_2}[\vec x_2]{t_2'}\\
    &= \Set{ \tcons{s_1}{m_1}[\vec x_1]{t} \given \immred{t_2}{\obsc{m_1}{\vec x_1}} \kleeneq T_2' \land t \in t_1' \tint T_2' } \cup {}\\
    &\cup \Set{ \tcons{s_2}{m_2}[\vec x_2]{t} \given \immred{t_1}{\obsc{m_2}{\vec x_2}} \kleeneq T_1' \land t \in T_1' \tint t_2' }
  \end{align*}
  We must show that each of these two subsets is contained in \(\jtts{\jwfps{\Pi}{\proci{\Delta_1, \Delta_2}{\Iota_1, a : A, \Iota_2}{c : C}}}\).
  We do so for the first subset; the second inclusion is analogous.
  If \(T_2'\) is undefined, then the inclusion is immediate (the set is empty).
  Assume next that \(T_2'\) is defined; this implies that \(\carrcn(m_1) \neq a\).
  Again, we proceed by case analysis on \(s_1\) and \(m_1\), and give an illustrative case.
  Assume first that \(s_1 = \pS\) and \(m_1 = \mLabel{b}{k}\).
  By inversion on elementhood for \(t_1\),
  \begin{enumerate}
  \item \(\Iota_1 = \Iota_1', b : \odot_{l \in L} B_l\) with \(k \in L\) and \({\odot} \in \Set{{\Tplus}, {\Tamp}}\);
  \item \(\jtt{t_1'}{\jwfts{ \immred{(\Pi, \Delta_2, \Iota_2, c : C)}{\pi} }{\proci{\immred{\Delta_1}{\pi}}{\immred{\Iota_1'}{\pi}, b : B_k}{a : \immred{A}{\pi}}}}\).
  \end{enumerate}
  By \cref{prop:session-types:36},
  \begin{enumerate}[resume]
  \item \(\jtt{T_2'}{\jwfps{ \immred{(\Pi, \Delta_1, \Iota_1')}{\pi}, b : B_k}{\immred{\proci{\Delta_2, a : A}{\Iota_2}{c : C}}{\pi}} }\).
  \end{enumerate}
  By the induction hypothesis,
  \[
    t_1' \tint T_2' \subseteq \jtts{\jwfps{\immred{\Pi}{\pi}}{\proci{\immred{(\Delta_1, \Delta_2)}{\pi}}{\immred{(\Iota_1', \Iota_2, a : A)}{\pi}, b : B_k}{c : \immred{C}{\pi}}}}.
  \]
  By the same clause giving the elementhood relationship for \(t_1\), we deduce
  \[
    \tcons{s_1}{m_1}[\vec x_1]{(t_1' \tint T_2')} \subseteq \jtts{\jwfps{\Pi}{\proci{\Delta_1, \Delta_2}{\Iota_1',  b : \odot_{l \in L} B_l, \Iota_2, a : A}{c : C}}},
  \]
  which is what we wanted to show.
  The remaining cases are analogous.
  This completes the case analyses so we conclude the result.
\end{proofE}

\section{Typechecking Processes}
\label{sec:typech-proc}

We give a typechecking algorithm to statically check that a process satisfies its specification, \ie, that each execution of the process is permitted by its specification.
It is driven by a collection of inference rules and operates in two alternating phases.

The first phase decomposes process compositions into individual processes.
It then determines the next channel \(a\) that a given process will use to communicate, and kicks off the second phase.
The second phase focusses on the type of \(a\).
It first tries to reduce the focussed type to weak head normal form, and then proceeds to typecheck the given process with respect to that weak head normal type.
Because our system is compositional, types may depend on ambient channels whose meaning will eventually be given by composition, and type reduction may cause constraints to be imposed on ambient communications.
To capture these, our algorithm generates a collection constraints that must be satisfied by the environment in order for the process to be well-typed.

\subsection{Phase one:  uniform process typing}

\begin{figure}
  \begin{gatherrules}
    \judgmentbox
    {\jtproc{\Pi}{\Iota}{P}{\Delta}{a : A}{\tfnt{T}}}
    {Process \(P\) satisfies specification \(\jwfps{\Pi}{\proci{\Delta}{\Iota}{a : A}}\) subject to constraints \(\tfnt{T}\)}
    \\
    \getrule*{New}
    \\
    \adjustbox{max width=\linewidth}{\getrule*{Par}}
    \\
    \getrule*{STFoc-R}
    \\
    \getrule*{STFoc-L}
  \end{gatherrules}
  \caption{Uniform Process Typing Judgment}
  \label{fig:gagillionth-iteration:9}
\end{figure}

The first phase, which we call uniform typing, is driven by a judgment \(\jtproc{\Pi}{\Iota}{P}{\Delta}{a : A}{\tfnt{T}}\) inductively defined by the rules of \cref{fig:gagillionth-iteration:9}.
We use colours to indicate the modes of each parameter in the judgment, where \inm{red} indicates inputs to the judgment and \outm{blue} indicates outputs.
Here, we attempt to check process \(P\) against the well-formed specification \(\jwfps{\Pi}{\proci{\Delta}{\Iota}{a : A}}\).
The judgment outputs a set \(\tfnt{T}\) of traces that consist of the communications of \(P\) interleaved with constraints on channels in \(\Pi\).
By our soundness result (\cref{theorem:typing-processes:1}), restricting each trace in \(\tfnt{T}\) to its observable actions (\ie, those with sign \(\pI\), \(\pO\), or \(\pS\)) will result in the set \(\ptraces{P}\) of traces of \(P\).
We implicitly assume that process specifications are well-formed, \ie, that all names are in scope.

The rule \getrn{New} introduces internal channels that can be used to type process \(P\).
The constraint set for its conclusion is obtained by deleting the new channels from the constraint set for \(P\).
This is analogous to the trace semantics of hiding in \cref{sec:trace-semantics}.
Because we assume that all specifications are well-formed, we know that no types in \(\jwfps{\Pi}{\proci{\Delta}{\Iota}{c : C}}\) attempt to observe \(a_1, \dotsc, a_n\).

The rule \getrn{Par} ensures that two processes can be composed only if they impose compatible constraints.
This is ensured by premise \(\getrh{Par}{3}\) that holds if and only if:
\begin{equation}
  \label{eq:typing-processes:1}
  \forall t_1 \in \tfnt{T}_1\;.\; \forall t_2 \in \tfnt{T}_2\;.\; \left(\tquoc{t_1} \pint \tquoc{t_2}\right) = \tquoc{\left(t_1 \tint t_2\right)}.
\end{equation}
Intuitively, it specifies that if traces \(t_1\) and \(t_2\) corresponding to \(P\) and \(Q\) can be composed as process executions, then each synchronized interleaving of \(t_1\) and \(t_2\) qua process traces can be obtained by synchronizing \(t_1\) and \(t_2\) as constraint traces.
In other words, we can interleave the constraints in \(t_1\) and \(t_2\) such that all of the constraints are consistent with the process actions in \(t_1\) and \(t_2\), and recover all valid process interleavings.
The key point of this definition is that we can recover all valid process interleavings despite the presence of constraints, \ie, that \(\left(\tquoc{t_1} \pint \tquoc{t_2}\right) \subseteq \tquoc{\left(t_1 \tint t_2\right)}\); we already know by \cref{prop:session-types:24} that adding constraints cannot produce new process interleavings, \ie, that \(\left(\tquoc{t_1} \pint \tquoc{t_2}\right) \supseteq \tquoc{\left(t_1 \tint t_2\right)}\).

\begin{example}
  \label{ex:typing-processes:1}
  We illustrate how constraint satisfaction rules out incompatible compositions by attempting to compose the following two processes (respectively, \(P\) and \(Q\)) in parallel:
  \begin{align*}
    &\jtproc{\cdot}{\cdot}{\tRecvL{a}{\Set{ l \Rightarrow \tSendL{b}{\mt{0}}{\tWait{a}{\tClose{b}}}}_{l \in \bit}}}{a : \Tlist{\bit}{1}}{b : \Tlist{\bit}{1}}{\tfnt{T}_1}\\
    &\jtproc{a : \Tlist{\bit}{1}}{\cdot}{\tRecvL{a}{\Set{ l \Rightarrow \tSendL{b}{l}{\tWait{a}{\tClose{b}}}}_{l \in \bit}}}{b : \Tlist{\bit}{1}}{c : \Tidlist{a}{1}}{\tfnt{T}_2}.
  \end{align*}
  Their composition should not satisfy \( \jtproc{\cdot}{b : \Tlist{\bit}{1}}{\tPar{a}{P}{Q}}{a : \Tlist{\bit}{1}}{c : \Tidlist{a}{1}}{\tfnt{T}} \) because an input of \(\mt{1}\) on \(a\) does not result in the output \(\mt{1}\) on \(c\) specified by \(\Tidlist{a}{1}\).
  The following two traces violate \eqref{eq:typing-processes:1} (they have no interleavings) and keep \getrn{Par} from being applied:
  \begin{align*}
    &\tcons{\pI}{\mLabel{a}{\mt{1}}}{\tcons{\pO}{\mLabel{b}{\mt{0}}}{\tcons{\pI}{\mClose{a}}{\tcons{\pO}{\mClose{b}}{\tempt}}}} \in \tfnt{T}_1\\
    &\tcons{\pI}{\mLabel{b}{\mt{0}}}{\tcons{\pC}{\mLabel{a}{\mt{0}}}{\tcons{\pI}{\mLabel{c}{\mt{0}}}{\cdots}}} \in \tfnt{T}_2
  \end{align*}
\end{example}

The rules \getrn{STFoc-R} and \getrn{STFoc-L} uniformly identify the next channel \(a\) on which a process will communicate, and kick off the second phase that reduces the type of \(a\).
This channel \(a\) is called the \defin{principal channel} of the process, and is given by \(\princ(P) = a\) for
\[
  P \coloneqq \tClose{a} \mid \tWait{a}{Q} \mid \tSendL{a}{k}{Q} \mid
  \tRecvL{a}{\Set{l \Rightarrow Q_l}_{l \in L}} \mid \tSendC{a}{b}{Q_1}{Q_2} \mid \tRecvC{b}{a}{Q}.
\]
We remark that there is exactly one rule in the first phase for each process forming construct.

\subsection{Phase two: focussed type reduction}
\label{sec:phase-two:-focussed}

\begin{figure}
  \begin{gatherrules}
    \judgmentbox
    {\jtred{\Pi}{\Iota}{P}{\Delta}{\focus{a : A}}{\tfnt{T}}}
    {Focussed reduction of \(a : A\) on the right}
    \\
    \getrule*{RTu-R}
    \\
    \getrule*{RTplus-R}
    \\
    \getrule*{RTamp-R}
    \\
    \adjustbox{max width=\linewidth}{
      \infer[\getrn*{RTot-R}]{
        \getrc{RTot-R}
      }{
        \begin{array}[b]{c}
          \getrh{RTot-R}{1}\\
          \getrh{RTot-R}{2}
        \end{array}
        &
        \getrh{RTot-R}{3}
        &
        \getrh{RTot-R}{4}
      }
    }
    \\
    \getrule*{RTlolly-R}
  \end{gatherrules}
  \caption{Right-Focussed Process Typing --- Focussed Type in Weak Head
    Normal Form}
  \label{fig:process-language:2}
\end{figure}

\begin{figure}[h]
  \centering
  \begin{gatherrules}
    \getrule*{Red-RTu}
    \\
    \adjustbox{max width=\linewidth}{\getrule*{Red-RTLbl}}
    \\
    \adjustbox{max width=\linewidth}{\getrule*{Red-RTChan}}
  \end{gatherrules}
  \caption{Right-Focussed Process Typing --- Reducing the Type in Focus}
  \label{fig:process-language:3}
\end{figure}

The second phase is driven by a pair of type-reduction judgments
\(\jtred{\Pi}{\Iota}{P}{\Delta}{\focus{a : A}}{\tfnt{T}}\)
and
\({\jtred{\Pi}{\Iota}{P}{\Delta, \focus{a : A}}{c : C}{\tfnt{T}}}\)
that reduce a focussed type \(\focus{a : A}\) to weak head normal form, and checks that if the principal channel of \(P\) is \(a\), then \(P\) communicates on \(a\) according to \(A\).
The right-focussed judgment is inductively defined by the rules of \cref{fig:process-language:2} (weak head normal cases) and \cref{fig:process-language:3} (reduction cases); the analogous left-focussed judgment exchanges sending and receiving.

For example, the rule \getrn{RTu-R} of \cref{fig:process-language:2} specifies that the process \(\tClose{a}\) communicates according to the weak head normal type \(a : \Tu\), provided it has empty client and internal contexts.
The client context must be empty to ensure that we do not accidentally discard any clients.
The internal context must be empty to ensure that we use all internal channels in the specification.
We must do so, for the current process may have been spawned by a larger process (typed by channels in \(\Pi\)) that observes channels in \(\Iota\), and we must not allow these channels to be discarded.
Because no constraints are imposed on the environment, the output constraint set is exactly \(\ptraces{\tClose{a}}\).

The rule \getrn{RTplus-R} specifies that a process \(\tSendL{a}{k}{P}\) is a server for a channel \(a : \Tplus_{l \in L \cup \Set{k}} A_l\) provided \(P\) is well-typed when \(a : A_k\) and the remainder of the specification has been reduced by \({\pLabel{a}{k}}\).
This reduction ensures that all types that depend on \(a\) observe the transmitted label \(k\).
The resulting constraint set is given by prefixing constraint set \(\tfnt{T}\) for \(P\) by the output action \((\pO, \mLabel{a}{k})\).

Channel transmission \(\tSendC{a}{b}{P}{Q}\) spawns a process \(P\) that provides a fresh channel \(b\), sends \(b\) over \(a\), and continues as \(Q\).
Processes \(P\) and \(Q\) cannot share any channels.
This is ensured by \getrn{RTot-R}: it treats all channels used or provided by \(P\) as ambient in \(Q\)'s typing judgment and vice-versa.
The rule also ensures that their specifications impose compatible constraints, analogously~to~\getrn{Par}.

Sometimes, the focussed channel is not yet weak head normal.
In these cases, we must reduce it to a weak head normal type (see \cref{fig:process-language:3}).
The rule \getrn{Red-RTu} reduces a focussed channel of type \(\TCaseU{c}{A}\).
This reduction is justified so long as a close message is observed on ambient channel \(c\).
This constraint is captured by prefixing the constraint \((\pC, \mClose{c})\) on the constraint set \(\tfnt{T}\) generated by the premise.
The rule \getrn{Red-RTLbl} can generate constraints for a subset of labels in a choice to support composition with processes that only send a subset of the permitted labels.
The remaining reduction rules are analogous.
We remark that the rules of \cref{fig:process-language:3} only reduce types that observe ambient channels.
A focussed type that observes a used or provided channel cannot be reduced because we cannot reduce types based on future communications.
Indeed, the observation required to reduce the type can only be provided by the process being checked, but the process will communicate on the focussed principal channel before it will communicate on the channel being observed.
Types that observe local channels are instead immediately reduced by the rules of \cref{fig:process-language:2}.

\begin{example}
  \label{ex:typing-processes:2}
  Set \(C = \TCaseU{a}{\Tu}\) and \(P_{x,y} = (\tWait{x}{\tClose{y}})\).
  We illustrate the mechanics of our typechecking algorithm by attempting to typecheck \(\tNew{b}{\Tu}{(\tPar{b}{P_{a,b}}{P_{b,c}})}\) against the specification \(\jwfps{\cdot}{\proci{a : 1}{\cdot}{c : C}}\).
  Typechecking succeeds if we can build a derivation of \( \jtproc { \cdot } { \cdot } { \tNew{b}{\Tu}{(\tPar{b}{P_{a,b}}{P_{b,c}})} } { a : \Tu } { c : C } { \mf{T} }\) for some \(\mf{T}\).
  First, we build a candidate derivation by considering only the inputs to the judgment (the red parts of the tree).
  The only possible bottom rule in our case is \getrn{New}, and we use proof search to complete the derivation.
  The focussed nature of our system ensures that search amounts to inversion: in each case, at most one rule can be used to extend the derivation.
  Eventually each branch either gets stuck or reaches a leaf (an axiom).
  The following candidate derivation shows that we can successfully build the red portion of the tree:
  \begingroup\small%
  \[
    \adjustbox{max width=\linewidth}{
      \infer[\getrn{New}]{
        \jtproc
        { \cdot }
        { \cdot }
        { \tNew{b}{\Tu}{(\tPar{b}{P_{a,b}}{P_{b,c}})} }
        { a : \Tu }
        { c : C }
        { \tdel{(\mc{L}_2 \pint \mc{R}_3)}{\Set{b}} }
      }{
        \infer[\getrn{Par}]{
          \jtproc
          { \cdot }
          { b : \Tu }
          { \tPar{b}{P_{a,b}}{P_{b,c}} }
          { a : \Tu }
          { c : C }
          { \mc{L}_2 \pint \mc{R}_3 }
        }{
          \infer[\getrn{STFoc-L}]{
            \jtproc
            { c : C }
            { \cdot }
            { P_{a,b} }
            { a : \Tu }
            { b : \Tu }
            { \mc{L}_2 }
          }{
            \infer[\getrn{RTu-L}]{
              \jtred
              { c : C }
              { \cdot }
              { P_{a,b} }
              { \focus{a : \Tu} }
              { b : \Tu }
              { \mc{L}_2 }
            }{
              \infer[\getrn{STFoc-R}]{
                \jtred
                { c : \Tu }
                { \cdot }
                { \tClose{b} }
                { \cdot }
                { b : \Tu }
                { \mc{L}_1 }
              }{
                \infer[\getrn{RTu-R}]{
                  \jtred
                  { c : \Tu }
                  { \cdot }
                  { \tClose{b} }
                  { \cdot }
                  { \focus{b : \Tu} }
                  { \mc{L}_1 }
                }{
                }
              }
            }
          }
          &
          \infer[\getrn{STFoc-L}]{
            \jtproc
            { a : \Tu }
            { \cdot }
            { P_{b,c} }
            { b : \Tu }
            { c : C }
            { \mc{R}_3 }
          }{
            \infer[\getrn{RTu-L}]{
              \jtred
              { a : \Tu }
              { \cdot }
              { P_{b,c} }
              { \focus{b : \Tu} }
              { c : C }
              { \mc{R}_3 }
            }{
              \infer[\getrn{STFoc-R}]{
                \jtproc
                { a : \Tu }
                { \cdot }
                { \tClose{c} }
                { \cdot }
                { c : C }
                { \mc{R}_2 }
              }{
                \infer[\getrn{Red-RTu}]{
                  \jtproc
                  { a : \Tu }
                  { \cdot }
                  { \tClose{c} }
                  { \cdot }
                  { \focus{c : C} }
                  { \mc{R}_2 }
                }{
                  \infer[\getrn{RTu-R}]{
                    \jtproc
                    { \cdot }
                    { \cdot }
                    { \tClose{c} }
                    { \cdot }
                    { \focus{c : \Tu} }
                    { \mc{R}_1 }
                  }{
                  }
                }
              }
            }
          }
        }
      }
    }
  \]\endgroup%
  Assuming every branch reaches a leaf, the axioms specify the leaf's judgment's output (the trace set in blue).
  We thread these trace sets back down through the derivation tree, using the operations specified in each rule to build that rules output trace set.
  In this case:
  \begingroup\small%
  \begin{align*}
    \mc{L}_1 &= \Set{ \tcons{\pO}{\mClose{b}}{\tempt} } & \mc{L}_2 &= \tcons{\pI}{\mClose{a}}{\mc{L}_1}  & & \\
    \mc{R}_1 &= \Set{ \tcons{\pO}{\mClose{c}}{\tempt} } & \mc{R}_2 &= \tcons{\pC}{\mClose{a}}{\mc{R}_1} & \mc{R}_3 &= \tcons{\pI}{\mClose{b}}{\mc{R}_2}\\
    \mc{L}_2 \pint \mc{R}_3 &= \mathrlap{\Set{ \tcons{\pI}{\mClose{a}}{\tcons{\pS}{\mClose{b}}{\tcons{\pO}{\mClose{c}}{\tempt}}} }} & & & &\\
    \tdel{(\mc{L}_2 \pint \mc{R}_3)}{\Set{b}} &= \mathrlap{\Set{ \tcons{\pI}{\mClose{a}}{\tcons{\pO}{\mClose{c}}{\tempt}}}} &&&&
  \end{align*}%
  \endgroup%
  Typechecking~succeeds if the resulting derivation is valid (including the side condition \eqref{eq:typing-processes:1}~for~\getrn{Par}).
\end{example}

Our rules determine a terminating typechecking algorithm when all choice types are indexed by finite label sets.
Indeed, all constraint sets and their traces are finite, so trace operations terminate, and the alternating phases reduce the complexity of processes and their specifications.
Most of the typechecking complexity lies in computing operations on traces.
We conjecture that they could efficiently be implemented by representing traces as tries or radix trees.

\section{Safety Properties}
\label{sec:safety-properties}

Our typechecking algorithm is sound: if we can typecheck a process against a process specification, then all executions of that process are permitted by its specification.
Put differently, our typechecking algorithm ensures that well-typed processes communicate safely:

\begin{theoremE}[Soundness]
  \label{theorem:typing-processes:1}
  The typechecking algorithm is sound: if any of
  \begin{enumerate}
  \item \(\jtproc{\Pi}{\Iota}{P}{\Delta}{a : A}{\tfnt{T}}\),
  \item \(\jtred{\Pi}{\Iota}{P}{\Delta}{\focus{a : A}}{\tfnt{T}}\), or
  \item \(\jtred{\Pi}{\Iota}{P}{\Delta, \focus{a : A}}{c : C}{\tfnt{T}}\),
  \end{enumerate}
  then \(\tquoc{\tfnt{T}} = \ptraces{P}\) (the set \(\tfnt{T}\) extents the executions of \(P\) with constraints) and \(\tfnt{T} \subseteq \jtts{\jwfts{\Pi}{\jwfpi{\Delta}{\Iota}{a : A}}}\) (the constraint set \(\tfnt{T}\) satisfies the specification \(\jwfts{\Pi}{\jwfpi{\Delta}{\Iota}{a : A}}\)).
\end{theoremE}

\begin{proofE}
  By induction on the derivation.
  \begin{proofcases}
  \item[\getrn{New}]
    The result is immediate by the induction hypothesis, \cref{prop:session-types:13}, and \cref{eq:process-language:8}.

  \item[\getrn{Par}]
    Recall
    \[
      \adjustbox{max width=\linewidth}{
        \getrule{Par}
      }
    \]
    By the induction hypothesis, \(\tquoc{\tfnt{T}_1} = \ptraces{P}\) and \(\tquoc{\tfnt{T}_2} = \ptraces{Q}\).
    By definition of \(\tfnt{T}_1 \therefore \tfnt{T}_2\),
    \[
      \ptraces{\tPar{a}{P}{Q}} = \ptraces{P} \pint \ptraces{Q} = (\tquoc{\tfnt{T}_1}) \pint (\tquoc{\tfnt{T}_2}) = \tquoc{(\tfnt{T}_1 \tint \tfnt{T}_2)}.
    \]
    The inclusion \(\tquoc{\tfnt{T}_1} \tint \tquoc{\tfnt{T}_2} \subseteq \sembr{\jwfps{\Pi}{\proci{\Delta_1, \Delta_2}{\Iota_1, a : A, \Iota_2}{c : C}}}\) is given by \cref{cor:session-types:4}.

  \item[\getrn{STFoc-R}] The result is immediate by the induction hypothesis.

  \item[\getrn{STFoc-L}] The result is immediate by the induction hypothesis.

  \item[\getrn{RTu-R}]
    Recall
    \[
      \getrule{RTu-R}
    \]
    The inclusion \(\tfnt{T} = \Set{ \tcons{\pO}{\mClose{a}}{\tempt} } \subseteq \jtts{\jwfts{\Pi}{\jwfpi{\cdot}{\cdot}{a : \Tu}}}\) is easily checked.
    Moreover, \(\tdel{\tfnt{T}}{\freecn(\Pi)} = \tfnt{T} = \ptraces{\tClose{a}}\) by \cref{eq:process-language:1}, so we are done.

  \item[\getrn{RTplus-R}]
    The equality is by the induction hypothesis and \cref{eq:process-language:3}.
    The inclusion is by the induction hypothesis and \cref{psred:Tplus-R}.

  \item[\getrn{RTamp-R}]
        The equality is by the induction hypothesis and \cref{eq:process-language:4}.
    The inclusion is by the induction hypothesis and \cref{psred:Tamp-R}.

  \item[\getrn{RTplus-L}] Analogous to \getrn{RTamp-R}

  \item[\getrn{RTamp-L}] Analogous to \getrn{RTplus-R}.

  \item[\getrn{RTot-R}]
    By the induction hypothesis, \(\tquoc{\tfnt{T}_1} = \ptraces{P}\) and \(\tquoc{\tfnt{T}_2} = \ptraces{Q}\).
    By \cref{eq:process-language:5},
    \begin{align*}
      &\ptraces{\tSendC{a}{b}{P}{Q}}\\
      &= \Set{ \tcons{\pO}{\mChan{a}}[b]{t} \given t_P \in \ptraces{P}, t_Q \in \ptraces{Q}, t \in t_P \pint t_Q }\\
      &= \Set{ \tcons{\pO}{\mChan{a}}[b]{t} \given t_P \in \tquoc{\tfnt{T}_1}, t_Q \in \tquoc{\tfnt{T}_2}, t \in t_P \pint t_Q }\\
      &= \Set{ \tcons{\pO}{\mChan{a}}[b]{t} \given t \in \tquoc{\tfnt{T}_1} \pint \tquoc{\tfnt{T}_2} }\\
      \shortintertext{which by \(\getrh{RTot-R}{4}\),}
      &= \Set{ \tcons{\pO}{\mChan{a}}[b]{t} \given t \in \tquoc{(\tfnt{T}_1 \tint \tfnt{T}_2)} }\\
      &= \tcons{\pO}{\mChan{a}}[b]{(\tquoc{(\tfnt{T}_1 \tint \tfnt{T}_2)})}\\
      &= \tquoc{(\tcons{\pO}{\mChan{a}}[b]{(\tfnt{T}_1 \tint \tfnt{T}_2)})}
    \end{align*}
    The desired inclusion is given by the induction hypothesis, \cref{conj:session-types:1}, and \cref{psred:Tot-R}.

  \item[\getrn{RTlolly-R}]
    Recall
    \[
      \getrule{RTlolly-R}
    \]
    The equality is by the induction hypothesis and \cref{eq:process-language:6}.
    The inclusion is by the induction hypothesis and \cref{psred:Tlolly-R}.

  \item[\getrn{RTot-L}] Analogous to \getrn{RTlolly-R}.

  \item[\getrn{RTlolly-L}] Analogous to \getrn{RTot-R}.

  \item[\getrn{RTu-L}]
    The desired inclusion is given by the induction hypothesis and \cref{psred:Tu-L}.
    The desired equality is given by the induction hypothesis and \cref{eq:process-language:2}.

  \item[\getrn{Red-RTu}, \getrn{Red-LTu}, \getrn{Red-RTLbl}, \getrn{Red-LTLbl}, \getrn{Red-RTChan}, or \getrn{Red-LTChan}]
    The inclusion is given by the induction hypothesis and the appropriate choice of \cref{eq:session-types:1,eq:session-types:2,eq:session-types:3}.
    The equality is immediate by the induction hypothesis and the definition of \(\tquoc{({-})}\).\qedhere
  \end{proofcases}
\end{proofE}

\begin{lemmaE}[][all end]
  \label{lemma:safety:1}
  If \(t \in \jtts{\jwfps{\cdot}{\mc{G}}}\), then \(\tquoc{t} = t\).
\end{lemmaE}

\begin{proofE}
  By induction on the derivation \(t \in \jtts{\jwfps{\cdot}{\mc{G}}}\).
\end{proofE}

\begin{corollaryE}
  \label{cor:typing-processes:1}
  If \(\jtproc{\cdot}{\Iota}{P}{\Delta}{a : A}{\tfnt{T}}\), then \(\tfnt{T} = \ptraces{P}\) and \(\ptraces{P} \subseteq \ptraces{\jwfps{\cdot}{\proci{\Delta}{\Iota}{a : A}}}\).
\end{corollaryE}

\begin{proofE}
  Immediate by \cref{lemma:safety:1,theorem:typing-processes:1}.
\end{proofE}

A key feature of \chase is its modular approach to process specification.
Processes whose specifications impose no constraints on their environment can always be composed:

\begin{propositionE}[Compositionality]%
  \label{prop:safety:1}
  If \(\jtproc{\cdot}{\Iota_1}{P}{\Delta_1}{a : A}{\tfnt{T}_1}\) and\\
  \({\jtproc{\cdot}{\Iota_2}{Q}{\Delta_2, a : A}{c : C}{\tfnt{T}_2}}\),
  then \(\jtproc{\cdot}{\Iota_1, a : A, \Iota_2}{\tPar{a}{P}{Q}}{\Delta_1, \Delta_2}{c : C}{\tfnt{T}_1 \pint \tfnt{T}_2}\).
\end{propositionE}

\begin{proofE}
  Assume
  \begin{itemize}
  \item \(\jtproc{\cdot}{\Iota_1}{P}{\Delta_1}{a : A}{\tfnt{T}_1}\) and
  \item \(\jtproc{\cdot}{\Iota_2}{Q}{\Delta_2, a : A}{c : C}{\tfnt{T}_2}\).
  \end{itemize}
  By induction on the two derivations, we can weaken them to
  \begin{itemize}
  \item \(\jtproc{\Delta_2, \Iota_2, c : C}{\Iota_1}{P}{\Delta_1}{a : A}{\tfnt{T}_1}\) and
  \item \(\jtproc{\Delta_1, \Iota_1}{\Iota_2}{Q}{\Delta_2, a : A}{c : C}{\tfnt{T}_2}\).
  \end{itemize}
  To show \(\jtproc{\cdot}{\Iota_1, a : A, \Iota_2}{\tPar{a}{P}{Q}}{\Delta_1, \Delta_2}{c : C}{\tfnt{T}_1 \pint \tfnt{T}_2}\), it is sufficient by \cref{cor:typing-processes:1} and \getrn{Par} to show that \(\tfnt{T}_1 \therefore \tfnt{T}_2\), \ie, that
  \[
    \forall t_1 \in \tfnt{T}_1\;.\; \forall t_2 \in \tfnt{T}_2\;.\; \left(\tquoc{t_1} \pint \tquoc{t_2}\right) = \tquoc{\left(t_1 \tint t_2\right)}.
  \]
  Indeed, if  \(\tfnt{T}_1 \therefore \tfnt{T}_2\), then \(\jtproc{\cdot}{\Iota_1, a : A, \Iota_2}{\tPar{a}{P}{Q}}{\Delta_1, \Delta_2}{c : C}{\tfnt{T}_1 \tint \tfnt{T}_2}\) by \getrn{Par}.
  We show that \(\tfnt{T}_1 \therefore \tfnt{T}_2\).
  By \cref{cor:typing-processes:1,lemma:safety:1}, \(\tquoc{t_i} = t_i\) for all \(t_i \in \tfnt{T}_i\) with \(i = 1, 2\).
  We deduce
  \[
    \left(\tquoc{t_1} \tint \tquoc{t_2}\right) = t_1 \tint t_2 = \tquoc{\left(t_1 \tint t_2\right)},
  \]
  where the last equality stems from the fact that \(\tquoc{\cdot}\) is the identity on traces with no constraints, and interleaving two traces without constraints does not introduce any constraints.
  This is what we wanted to show.
\end{proofE}

\section{Extending \chase}
\label{sec:extending-chase}

\chase provides a foundation for the development of session-typed languages with rich specifications.
In this section, we conjecturally sketch extensions along several axes that allow \chase to capture richer computational phenomena: selection, asynchrony, value dependency, and recursion.

\subsection{Selection}
\label{sec:selection}

Go~\cite{project_2024:_go_progr_languag_specif} includes a \emph{select} statement that, given a collection of communication actions to perform, randomly performs one from the set of actions that are possible.
For example, if processes \(P\) and \(Q\) attempt to communicate with a process \(R\), then process \(R\) could use a select statement to communicate with the first of the two processes that is ready.
We show how to extend \chase with a linear variant of a select statement building on ideas from differential linear logic~\cite{ehrhard_2018:_introd_to_differ_linear_logic} and its applications to concurrency~\cite{rocha_caires_2023:_safe_session_based, rocha_caires_2021:_propos_as_types_shared_state}.
Differential linear logic offers a means of \emph{adding} proofs of the same sequent.
In the intuitionistic case, if we ascribe the sequents of intuitionistic linear logic with proof terms (processes), we get the rule
\begingroup\small
\[
  \infer{
    \Gamma \vdash P + Q :: a : A
  }{
    \Gamma \vdash P :: a : A
    &
    \Gamma \vdash Q :: a : A
  }
\]
\endgroup%
The possible behaviours of \(P + Q\) are all those of \(P\) or \(Q\), so the sum denotes \(\sembr{P + Q} = \sembr{P} \cup \sembr{Q}\).
This clause satisfies the identities described by \citeauthor{rocha_caires_2021:_propos_as_types_shared_state}, namely, that process composition distributes over sums and that all processes are idempotent with regard to the sum, \ie, that \(P + P \equiv P\) for all \(P\).
Our discussion suggests the following typing rule for sums of \chase processes:
\begingroup\small
\[
  \infer[\rn{Sum}]{
    \jtproc{\Pi}{\Iota}{P_1 + P_2}{\Delta}{a : A}{\mc{T}_1 \cup \mc{T}_2}
  }{
    \jtproc{\Pi}{\Iota}{P_1}{\Delta}{a : A}{\mc{T}_1}
    &
    \jtproc{\Pi}{\Iota}{P_2}{\Delta}{a : A}{\mc{T}_2}
  }
\]
\endgroup%

Returning to our motivating example, suppose that we want to receive channels \(a\) and \(c\) over \(b\) and \(d\), respectively, but that we do not want to impose a particular order on which of \(b\) or \(d\) is used first.
The following sum of processes receives on whichever of \(b\) or \(d\) is ready first, non-deterministically breaking any ties, and communicates over the other before continuing as \(R'\):
\begingroup\small%
\[
  (\tRecvC{a}{b}{\tRecvC{c}{d}{R'}}) + (\tRecvC{c}{d}{\tRecvC{a}{b}{R'}}).
\]
\endgroup%

\subsection{Asynchrony}
\label{sec:asynchrony}

Adapting \chase to use an asynchronous communication semantics is straightforward, but it comes at the cost of technical complexity.
The intended semantics treats channels as unbounded FIFO buffers, where processes can always send (add a message to a buffer), and they can receive a message if one is buffered.
We implement FIFO buffers using \emph{continuation channels}, where each channel carries exactly one message, and each message consists of the datum plus the name of the (continuation) channel that will carry the next message.
To preclude interference between channels, continuation channel names are globally fresh.
To illustrate, consider the trace that describes sending on \(a\) the label \(l\), receiving label \(k\), and closing \(a\).
In the synchronous setting, we simply send the three messages on \(a\), giving a trace \( \tcons{\pO}{\mLabel{a}{l}}{\tcons{\pI}{\mLabel{a}{k}}{\tcons{\pO}{\mClose{a}}{\tempt}}} \).
In the asynchronous setting, the label \(l\) is paired with a fresh name \(a_1\) when sent on \(a\); label \(k\) is paired with a fresh name \(a_2\) and sent on \(a_1\); and the close message is sent on \(a_2\).
To enforce global freshness, ensure that trace sets remain finite, and preserve compositionality, we model continuation channels in traces by bound channel names.
The corresponding trace is \( \tcons{\pO}{\mLabel{a}{l}}[a_1]{\tcons{\pI}{\mLabel{a_1}{k}}}[a_2]{\tcons{\pO}{\mClose{a_2}}{\tempt}} \), where \(a_1\) and \(a_2\) are bound.
Modifying the semantic clauses to capture asynchronous communication is then a matter of adapting the trace operations of \cref{sec:operation-traces} to handle continuation channels, and of threading bindings for continuation channels throughout the semantic clauses.
For example, \cref{eq:process-language:3} becomes \(\ptraces{\tSendL{a}{k}{P}} = \tcons{\pO}{\mLabel{a}{k}}[a]{ \ptraces{P} }\), where \(a\) is bound in the traces in \(\ptraces{P}\).
To type processes in the asynchronous setting, we must extend the syntax of observed communications \(\pi\) to include the name of the continuation channel, and extend type reductions \(\immred{A}{\pi}\) to substitute the continuation channel for the carrier of \(\pi\) in \(A\).
Because type reduction is localized to the operation \(\immred{A}{\pi}\) and all reductions occur in lockstep in the typing rules, we never need to worry about a continuation channel name going out of scope, and no other changes are required.

\subsection{Value-dependent session types}
\label{sec:value-dependency}

\newcommand{\tSendV}[3]{\ms{send}\ #2\ \ms{on}\ #1;\ #3}
\newcommand{\tRecvV}[3]{\ms{let}\ #2 = \ms{rcv}\ #1\ \ms{in}\ #3}
\newcommand{\Tforall}[2]{\forall #1\mskip\thinmuskip .\mskip\thinmuskip#2}
\newcommand{\Texists}[2]{\exists #1\mskip\thinmuskip .\mskip\thinmuskip#2}
\newcommand{\TB}[1]{[\,#1\,]}
\newcommand{\tZ}{\ms{z}}
\newcommand{\tS}[1]{\ms{s}(#1)}
\newcommand{\TCaseV}[3]{\ms{CASE}\ #1\ \{\,\ms{val}\ #2 \Rightarrow #3\,\}}
\newcommand{\Tnat}{\ms{nat}}
\newcommand{\mVal}[2]{\ms{val}\ #2\ \ms{on}\ #1}
\newcommand{\jtypef}[3]{#1 \vdash_{\ms{f}} #2 : #3}
\newcommand{\jtredF}[7]{\inm{#3} \Rightarrow \inm{\jwfps{\inm{#7};\ \inm{#1}}{\proci{\inm{#4}}{\inm{#2}}{\inm{#5}}}} \mathbin{/\!\!/} \outm{#6}}
\newcommand{\jtprocF}[7]{\inm{#3} \Vdash \inm{\jwfps{\inm{#7};\ \inm{#1}}{\proci{\inm{#4}}{\inm{#2}}{\inm{#5}}}} \mathbin{/\!\!/} \outm{#6}}
\newcommand{\valof}[1]{\mb{val}[#1]}
\newcommand{\expof}[1]{\mb{exp}[#1]}

\chase can be extended to support value-dependent session types~\cite{toninho_2011:_depen_session_types}, where session types can depend on values drawn from an underlying dependent functional language.
Value-dependent type formers \(\Tforall{x : \tau}{A}\) and \(\Texists{x : \tau}{A}\) respectively specify servers that receive or send values \(v\) of type \(\tau\) and then communicate according to \(\subst{v}{x}{A}\).
We first extend our syntax:
\begingroup\small%
\begin{alignat*}{5}
  \text{Processes}\quad P &\Coloneqq \cdots&&\grammid \tSendV{a}{M}{P} &\quad&\text{Evaluate term \(M\) and send its value on \(a\)}\\
                                      &&&\grammid \tRecvV{a}{x}{P} &&\text{Bind received value to \(x\) in \(P\)}\\
  \text{Session types}\quad A &\Coloneqq \cdots &&\grammid \Tforall{x : \tau}{A} &&\text{Dependent value receiving}\\
                                      &&&\grammid \Texists{x : \tau}{A}&&\text{Dependent value sending}\\
                                      &&&\grammid \TCaseV{a}{x}{A}&&\text{Value observation}
\end{alignat*}\endgroup%

Value observation \(\TCaseV{a}{x}{A}\) observes a transmitted value \(v\) on a channel \(a\) and reduces to the type \(\subst{v}{x}{A}\).
Its reduction is defined analogously to the other immediate reductions:
\begingroup\small%
\[
  \immred{(\TCaseV{c}{x}{A})}{\pi} =
  \begin{cases}
    \subst{M}{x}{A} &\text{if }\pi = \mVal{c}{M}\\
    \TCaseV{c}{x}{\immred{A}{\pi}} &\text{if }\carrcn(\pi) \neq c
  \end{cases}
\]\endgroup

Write \(\valof{\tau}\) for the sets of closed values of type \(\tau\).
We give meaning to value transmission by allowing terms to be embedded in traces:\footnote{Extending our semantics to capture evaluation in the underlying dependent type theory is beyond the scope of this sketch.}
\begingroup\small%
\[
  \ptraces{\tSendV{a}{M}{P}} = \tcons{\pO}{\mVal{a}{M}}{ \sembr{P} },
  \qquad
  \ptraces{\tRecvV{a}{x}{P}} = \bigcup\nolimits_{v \in \valof{\tau}} \tcons{\pI}{\mVal{a}{v}}{ \sembr{P} }.
\]\endgroup
We then extend interleaving to use the underlying definitional equality on terms so that whenever \(M\) and \(v\) are definitionally equal, \((\tcons{\pO}{\mVal{a}{M}}{t_1}) \pint (\tcons{\pI}{\mVal{a}{v}}{t_2}) = \tcons{\pS}{\mVal{a}{v}}{(t_1 \pint t_2)}\)

To type processes that use values, we assume a typing judgment \(\jtypef{\Gamma}{M}{\tau}\) for the functional language and extend the process typing judgment to include \(\Gamma\).
The right process typing rules follow a similar pattern as for label transmission, and the other rules follow analogously:%
\begingroup\small%
\allowdisplaybreaks
\begin{gatherrules}
  \infer[\rn{\({\Rightarrow}{\forall}\)R}]{
    \jtredF{\Pi}{\Iota}{\tRecvV{a}{x}{P}}{\Delta}{\focus{a : \Tforall{x : \tau}{A}}}{\bigcup\nolimits_{v \in \valof{\tau}} \tcons{\pI}{\mVal{a}{v}}{\subst{v}{x}{\mc{T}}}}{\Gamma}
  }{
    \jtprocF{\immred{\Pi}{\pi}}{\immred{\Iota}{\pi}}{P}{\immred{\Delta}{\pi}}{a : A}{\mc{T}}{\Gamma, x : \tau}
    &
    \pi = \mVal{a}{x}
  }
  \\
  \infer[\rn{\({\Rightarrow}{\exists}\)R}]{
    \jtredF{\Pi}{\Iota}{\tSendV{a}{M}{P}}{\Delta}{\focus{a : \Texists{x : \tau}{A}}}{\tcons{\pO}{\mVal{a}{M}}\mc{T}}{\Gamma}
  }{
    \jtprocF{\immred{\Pi}{\pi}}{\immred{\Iota}{\pi}}{P}{\immred{\Delta}{\pi}}{a : \subst{M}{x}{A}}{\mc{T}}{\Gamma}
    &
    \jtypef{\Gamma}{M}{\tau}
    &
    \pi = \mVal{a}{M}
  }
  \\
  \adjustbox{max width=\linewidth}{
    \infer[\rn{\({\Rightarrow}{\forall\exists}\)[R]}]{
      \jtredF{\Pi, c : (\sharp x : \tau .C)}{\Iota}{P}{\Delta}{\focus{a : \TCaseV{c}{x}{A}}}{\bigcup\nolimits_{v \in \valof{\tau}} \tcons{\pC}{\mVal{a}{v}}{\subst{v}{x}{\mc{T}}}}{\Gamma}
    }{
      \jtprocF{\immred{\Pi}{\pi}}{\immred{\Iota}{\pi}}{P}{\immred{\Delta}{\pi}}{a : A}{\mc{T}}{\Gamma, x : \tau}
      &
      \pi = \mVal{a}{x}
      &
      {\sharp} \in \Set{{\forall}, {\exists}}
    }
  }
\end{gatherrules}
\endgroup

\subsection{Recursion}
\label{sec:recursion}

\newcommand{\tRec}[5]{\ms{nrec}\ #1\ \{ \mt{z} \Rightarrow #2 \mid \mt{s}(#3)\ \mt{with}\ #4 \Rightarrow #5\} }

\newcommand{\Trec}[4]{\ms{NREC}\ #1\ \{ \mt{z} \Rightarrow #2 \mid \mt{s}(\uscore)\ \mt{with}\  #3 \Rightarrow #4\} }
\newcommand{\TS}[1]{\mb{S}(#1)}
\newcommand{\updmap}[2]{[#1 \mid #2]}
\newcommand{\timp}{\mathsf{impossible}}

\newcommand{\Tcstr}[2]{#1 \Rrightarrow #2}
\newcommand{\tCall}[1]{\ms{call}\ #1}
\newcommand{\tDecl}[2]{#1 = #2}
\newcommand{\TDecl}[2]{#1 = #2}
\newcommand{\tProg}[2]{\tCall{#1}\ \ms{where}\ #2}

\chase can be extended to support mutually recursive types and processes using coinductively defined typing judgments.
We implement mutually recursive processes using process declarations \(\tDecl{f(\vec x)}{P_f}\) gathered in an implicit global context \(\Sigma_P\).
These declarations specify that the process named \(f\) is given by its implementation \(P_f\) involving channels \(\vec x\).
The syntax \(\tCall{f(\vec a)}\) calls process \(f\) with channel names \(\vec a\), and it corresponds to running \(\subst{\vec a}{\vec x}{P_f}\).
Equirecursive types are given by type declarations \(\TDecl{X[\vec x]}{A_X}\) gathered in an implicit global context \(\Sigma_A\).
Channel name variables \(\vec x\) specify the channels that \(A_X\) can observe, and \(X[\vec a]\) is definitionally equal to \(\subst{\vec a}{\vec x}{A_X}\).
A complete program \(\mathbb{P}\) is a called process name paired with explicit declarations for each called name.
\begingroup\small%
\begin{alignat*}{5}
  \text{Processes}&\quad& P &\Coloneqq \cdots \grammid \tCall{f(\vec a)}  &\quad&\text{Call named process \(f\) with channels \(\vec a\)}\\
  \text{Session types}&\quad& A &\Coloneqq \cdots \grammid X[\vec x]&&\text{Equirecursive type names}\\
  \text{Type declarations}&\quad& \Sigma_A &\Coloneqq \cdot \grammid \Sigma_A, \TDecl{X[\vec x]}{A_X}&&\text{\(X[\vec x]\) is definitionally equal to \(A_X\)}\\
  \text{Process declarations}&\quad& \Sigma_P &\Coloneqq \cdot \grammid \Sigma_P, \tDecl{f(\vec x)}{P_f}&&\text{Process named \(f\) is implemented by \(P_f\)}\\
  \text{Programs}&\quad& \mathbb{P} &\Coloneqq \tProg{f(\vec a)}{\Sigma_P}&&\text{Complete or top-level program}
\end{alignat*}\endgroup%
We assume that recursive declarations are contractive~\cite[300]{pierce_2002:_types_progr_languag}.
This standard assumption simplifies our theory and allows us to interpret equirecursive types as regular trees.
To ensure type reduction \(\immred{A}{\pi}\) remains well defined, we interpret its definition coinductively instead of inductively.

\begin{example}
  \label{ex:extensions:3}
  For each fixed type \(A\), the type \(\ms{list}_A = \Tplus \Set{ \ms{nil} : \Tu, \ms{cons} : A \otimes \ms{list}_A }\) specifies lists of channels of type \(A\).
  The type \( \ms{stream}_A= \Tamp \Set{ \ms{head} : A, \ms{tail} : \ms{stream}_A } \) specifies a stream of channels of type \(A\).
  Given a process declaration \(\tDecl{\ms{u}(x)}{(\tRecvL{x}{\Set{\ms{head} \Rightarrow \tClose{x} \mid \ms{tail} \Rightarrow \tCall{\ms{u}(x)}}})}\), the process ``\(\tCall{\ms{u}(a)}\)'' is a closed process providing \(a : \ms{stream}_\Tu\).
\end{example}

\begin{example}
  \label{ex:extensions:2}
  The types \(\ms{Bits}\) and \(\ms{IdBitSeq}\ a\) and the process \(\ms{id}\) from \cref{sec:client-serv-arch,sec:spec-part-ident} are encoded by the following declarations, where we use the label \(\$\) and type \(\Tu\) to specify finite~sequences:
  \begingroup\small%
  \begin{align*}
    \ms{Bits} &= \Tplus \Set{ \$ : \Tu, \mt{0} : \ms{Bits}, \mt{1} : \ms{Bits} }\\
    \ms{IdBitSeq}[x] &= \TCaseL{x}{\{\,\$ \Rightarrow \Tplus \Set{\$: \TCaseU{x}{\Tu}}\\
              &\qquad\qquad\;\mid \mt{0} \Rightarrow  \Tplus \Set{ \mt{0} : \ms{IdBitSeq}[x] } \mid \mt{1} \Rightarrow \Tplus \Set{ \mt{1} : \ms{IdBitSeq}[x] } \,\}}\\
    \tDecl{\ms{id}(x,y) &}{\tRecvL{x}{\Set{\$ \Rightarrow \tWait{x}{\tClose{y}} \mid \mt{0} \Rightarrow \tSendL{y}{\mt{0}}{\tCall{\ms{id}(x,y)}} \mid \mt{1} \Rightarrow \tSendL{y}{\mt{1}}{\tCall{\ms{id}(x,y)}} }}}.
  \end{align*}
  \endgroup
\end{example}

We extend our semantics to support calling named processes.
Let \(\mc{P}\) be the set of declared calling interfaces \(f(\vec x)\).
A process environment is a map \(\rho : \mc{P} \to \mb{TSets}\) giving the meaning of declared processes, where \(\mb{TSets}\) is the lattice of sets of traces ordered by inclusion.
We generalize \(\sembr{P}\) from an element of \(\mb{TSets}\) to a continuous map \(\sembr{P} : (\mc{P} \to \mb{TSets}) \to \mb{TSets}\).
Calling \(f(\vec x)\) with channels \(\vec a\) means retrieving its meaning from the environment and instantiating it with the names~\(\vec a\):
\begingroup\small%
\[
  \ptraces{\tCall{f(\vec a)}}\rho = \Set{ \subst{\vec a}{\vec x}{t} \given t \in \rho(f(\vec x)) }.
\]
\endgroup%
We adapt the existing clauses by threading through the environments, \eg: \( \ptraces{\tPar{a}{P}{Q}}\rho = \sembr{P}\rho \pint \sembr{Q}\rho \).
To capture parallel composition in the presence of potentially infinite traces, we must extend \(\pint\) to be a fair merge operator.
This ensures that each element from each input traces appears at a finite depth in the interleavings.
We omit the details due to space constraints.

Complete programs are processes executed in an environment where every called process name has a corresponding declaration.
Their semantics are given by
\begingroup\small%
\begin{align*}
  \sembr{\tProg{f(\vec a)}{\Sigma_P}} &= \sembr{\tCall{f(\vec a)}}\sembr{\Sigma_P}, \text{ where}\\
  \sembr{\tDecl{f_1(\vec x_1)}{P_1}, \dotsc, \tDecl{f_n(\vec x_n)}{P_n}} &= \ms{FIX}(\lambda \rho \in (\mc{P} \to \mb{TSets}).
  \left[ \rho \mid f_1(\vec x_1) \mapsto \sembr{P_1}\rho \mid \cdots \mid f_n(\vec x_n) \mapsto\sembr{P_n}\rho \right]),
\end{align*}%
\endgroup%
\(\ms{FIX}\) is the continuous fixed-point operator on \(\mc{P} \to \mb{TSets}\), and \(\left[g \mid y_1 \mapsto v_1 \mid \cdots \mid y_n \mapsto v_n\right](x)\) is \(v_i\) if \(x = y_i\) and \(g(x)\) otherwise.
Intuitively, \(\sembr{\Sigma_P}\) maps each \(f_i(\vec x_i)\) to the meaning of \(\sembr{P_i}\) in the fixed point giving meaning to the mutually recursive declarations.

\begin{example}
  \label{ex:extensions:4}
  If \(\Sigma_P = (\tDecl{f(x)}{\tSendL{x}{\mt{1}}{\tCall{g(x)}}}),\;\; (\tDecl{g(x)}{\tSendL{x}{\mt{0}}{\tCall{f(x)}}})\), then
  \begingroup\small%
  \begin{align*}
    \sembr{\Sigma_P} &= \ms{FIX}\left(\lambda \rho . \left[ \rho \mid f \mapsto \tcons{\pO}{\mLabel{x}{\mt{1}}}{\rho(g)} \mid g \mapsto \tcons{\pO}{\mLabel{x}{\mt{0}}}{\rho(f)} \right]\right)\\
    \sembr{\tProg{f(a)}{\Sigma_P}} &= \Set{ \tcons{\pO}{\mLabel{a}{\mt{1}}}{\tcons{\pO}{\mLabel{a}{\mt{0}}}{\tcons{\pO}{\mLabel{a}{\mt{1}}}{\cdots}}} }.
  \end{align*}
  \endgroup
\end{example}

Process specifications still denote sets \(\sembr{\jwfps{\Pi}{\mc{G}}}\) of traces satisfying the specification.
However, to capture recursive behaviour, we adapt its definition to use a coinductively defined elementhood relation.
Elementhood is given by a coinductive reading of the rules defining \(t \in \sembr{\jwfps{\Pi}{\mc{G}}}\) generated by the obvious analogs of \cref{sec:type-reduction}, \eg:
\begingroup\small%
\[
  \infer{
    \tcons{\pO}{\mClose{a}}{t} \in
    \sembr{
      \jwfts{\Pi}
      { \mc{G}
        \scons
        (\jwfpi
        { \cdot }
        { \cdot }
        { a : \Tu })}
    }}{
    t \in     \sembr{
      \jwfts{\immred{\Pi}{\pi}}{ \immred{\mc{G}}{\pi} }
    }
    &
    \pi = \mClose{a}
  }
  \quad
  \infer{
    \tcons{\pO}{\mLabel{a}{k}}{t} \in   \sembr{
      \jwfts
      { \Pi }
      { \mc{G}
        \scons
        (\jwfpi
        { \Delta }
        { \Iota }
        { a : \Tplus_{l \in L} A_l })}
    }
  }{
    t \in  \sembr{
      \jwfts
      { \immred{\Pi}{\pi} }
      { \immred{\mc{G}}{\pi}
        \scons
        (\jwfpi
        { \immred{\Delta}{\pi} }
        { \immred{\Iota}{\pi} }
        { a : A_k })}
    }
    &
    \pi = \mLabel{a}{k}
  }
\]\endgroup%

\newcommand{\jwfe}[3]{#1;\ #2 \vdash #3}
\newcommand{\jent}[3]{#1;\ #2 \vdash #3}
\newcommand{\jtprocR}[6]{\inm{#3} \Vdash \jwfps{\inm{#1}}{\proci{\inm{#4}}{\inm{#2}}{\inm{#5}}} \mathbin{/\!\!/} \inm{#6}}

In order to type mutually recursive processes, we give our typing rules a coinductive reading and treat each component as an \inm{input} to the judgment.
Then, we extend our rules with:
\begingroup\small%
\[
  \infer[\rn{Call}]{
    \jtprocR{\Pi}{\Iota}{\tCall{f}(\vec a)}{\Delta}{c : C}{\mc{T}}
  }{
    \jtprocR{\Pi}{\Iota}{\subst{\vec a}{\vec x}P_f}{\Delta}{c : C}{\mc{T}}
    &
    (\tDecl{f(\vec x)}{P_f}) \in \Sigma_P
  }
\]
\endgroup%
This system can be given a finitary presentation by moving from arbitrary coinductive derivations to circular derivations (finite derivations with loops)~\cite{fortier_santocanale_2013:_cuts_circul_proof,derakhshan_pfenning_2022:_circul_proof_as}.
We refer the reader to \cite{somayyajula_pfenning_2022:_type_based_termin_futur} for a discussion of how to do so to achieve typechecking in finite time, and also a discussion of how to encode inductive and coinductive types using value dependency and (general) equirecursive types.

\section{Related Work}
\label{sec:related-future-work}

\chase builds on linear-logical foundations of session types pioneered by \textcite{caires_pfenning_2010:_session_types_intuit_linear_propos}.
Indeed, the simply typed fragment of our system is based on a proofs-as-processes interpretation of multiplicative-additive intuitionistic linear logic (MAILL), where propositions correspond to binary session types that specify communications on communication channels.
A key difference with our work is that such interpretations typically combine parallel composition and hiding into a single operation.
This is because process composition corresponds to the cut rule of linear logic.
In contrast, we have separate parallel composition and hiding operators, and we track internal channels.
This separation ensures that composition does not hide channels observed by specifications, thereby ensuring that specifications remain well-defined after process composition.
We can recover cut-style process composition with a derived rule.
Our uniform process typing system can then be seen as a conservative extension of typical interpretations of identity-free MAILL.
Another key difference is our use of focussing.
Processes can be specified with observing types, but we can only check specific communication actions (\eg, sending a label)) against weak head normal types (\eg, an internal choice).
This means that to type check a process, we must reduce the type of its principal channel to weak head normal form, potentially generating constraints on ambient communications along the way.
Our focussing approach ensures that we only reduce the types of principal channels.
This avoids the premature or spurious generation of constraints on ambient channels that could affect constraint compatibility in process composition, and it ensures that type checking is~deterministic.

There are various dependent type systems for binary session types.
\Textcite{toninho_2011:_depen_session_types, toninho_yoshida_2018:_depen_session_typed_proces} interpret logical quantifiers as value-dependent session types in languages with functional programming features.
As discussed in \cref{sec:value-dependency}, value-dependent session types depend on transmitted functional values.
Variations include label-dependent session types~\cite{thiemann_vasconcelos_2019:_label_depen_session_types}, which reify labels as a first-class type, and arithmetic refinements~\cite{das_pfenning_2020:_session_types_with_arith_refin}, which offer type-level arithmetic using transmitted natural numbers.
Like other binary session types, value-dependent session types take a channel-local approach: a type only considers communication on a single channel.
In contrast, types in \chase specify protocols on individual channels while potentially depending on ambient communications.

Multiparty session types~\cite{honda_2016:_multip_async_session_types} are an alternative approach to specifying systems of communicating processes.
Instead of specifying communications on individual channels, multiparty session types specify interactions between multiple parties using a global type, and then for each party project a local type specifying its interactions with the other parties.
Though multiparty session types can specify complex interactions, they are not compositional and typically require that entire systems be specified at once.
To achieve compositionality, \textcite{stolze_2023:_compos_partial_multip} introduced \emph{partial multiparty sessions types}, which specify multiparty sessions where some parties may be missing.
Determining when partial sessions can be composed and the resulting composition is technically complex.
In contrast, \chase is designed to ensure compositional process specifications, and the composition of process specifications is determined by a handful of inference rules.

The dynamics of processes in proofs-as-processes interpretations of intuitionistic linear logic is often given by multiset rewriting semantics~\cite{cervesato_scedrov_2009:_relat_state_based}.
We instead describe the behaviour of processes using a trace semantics.
We do so to simplify relating the meaning of processes (a set of traces) with the meaning of process specifications (a set of allowed traces).
It also provides a means of ensuring that processes and specifications are mutually compatible when typechecking their composition.
We conjecture that our trace semantics captures the same observable communication behaviours~\cite{atkey_2017:_obser_commun_seman_class_proces, kavanagh_2022:_fairn_commun_based} as multiset-based semantics.
It is challenging to define trace semantics or ordered semantics for process calculi with name generation or \(\alpha\)-conversion.
Some approaches, \eg, typed event structures for \(\pi\)-calculi~\cite{varacca_yoshida_2010:_typed_event_struc_linear}, specify in advance all names that will be created; others use complex semantic composition operators involving renamings~\cite{crafa_2007:_compos_event_struc}.
In contrast, we use binding to range over all possible fresh names in a manner reminiscent of higher-order abstract syntax.
This avoids unintended name clashes when interleaving traces during process composition.
Our compact presentation also minimizes the number of traces considered by our typechecking algorithm.
Traces with binding are similar to nominal sequences~\cite{gabbay_ghica_2012:_game_seman_nomin_model}, which use a coabstraction operator to bind atoms appearing later in a sequence.
In contrast, our traces bind not atoms, but names found in later messages.
As a result, our traces more closely resemble abstract binding trees.

\section{Conclusion}

In this work we describe \chase, a message-observing binary session type system.
It uses type-level processes to specify protocols that vary based on ambient communications. This allows us to express more precise safety guarantees that cannot be expressed by prior work.

Our main goal in designing \chase is to ensure compositionality. We achieve this using a novel trace-based semantics that uses traces with bindings to compactly handle higher-order communications. Our main technical results are type safety and compositionality for well-typed processes.

In the future, we plan to extend \chase with shared channels.
This would allow \chase to capture shared data structures like shared queues, and shared services like databases.

\nopagebreak

\begin{acks}                            %
  This work was funded in part by a \grantsponsor{GS501100000038}{Natural Sciences and Engineering Research Council of Canada}{http://dx.doi.org/10.13039/501100000038} \grantnum{GS501100000038}{Postdoctoral Fellowship} (awarded to the first author) and Discovery Grant (grant number \grantnum{GS501100000038}{206263}).

\end{acks}

\ifbool{appendix}{%
\appendix

\section{Omitted Rules}
\label{sec:ommited-rules}

The following rules are those of \cref{fig:process-language:2,fig:process-language:3}, plus any rules omitted from those figures:
\begingroup\allowdisplaybreaks
\begin{gatherrules}
  \judgmentbox
  {\begin{array}{l}
     \jtred{\Pi}{\Iota}{P}{\Delta}{\focus{a : A}}{\tfnt{T}}
     \\
     \jtred{\Pi}{\Iota}{P}{\Delta, \focus{a : A}}{c : C}{\tfnt{T}}
   \end{array}}
 {Focussed reduction of \(a : A\)}
 \\
 \getrule{RTu-R}
 \\
 \getrule*{RTu-L}
 \\
 \getrule{RTplus-R}
 \\
 \getrule*{RTplus-L}
 \\
 \getrule{RTamp-R}
 \\
 \getrule*{RTamp-L}
 \\
 \adjustbox{max width=\linewidth}{
   \infer[\getrn{RTot-R}]{
     \getrc{RTot-R}
   }{
     \begin{array}[b]{c}
       \getrh{RTot-R}{1}\\
       \getrh{RTot-R}{2}
     \end{array}
     &
     \getrh{RTot-R}{3}
     &
     \getrh{RTot-R}{4}
   }
 }
 \\
 \getrule*{RTot-L}
 \\
 \getrule{RTlolly-R}
 \\
  \adjustbox{max width=\linewidth}{
   \infer[\getrn{RTlolly-L}]{
     \getrc{RTlolly-L}
   }{
     \begin{array}[b]{c}
       \getrh{RTlolly-L}{1}\\
       \getrh{RTlolly-L}{2}
     \end{array}
     &
     \getrh{RTlolly-L}{3}
     &
     \getrh{RTlolly-L}{4}
   }
 }
 \\
 \getrule{Red-RTu}
 \\
 \getrule*{Red-LTu}
 \\
 \adjustbox{max width=\linewidth}{\getrule{Red-RTLbl}}
 \\
 \adjustbox{max width=\linewidth}{\getrule*{Red-LTLbl}}
 \\
 \adjustbox{max width=\linewidth}{\getrule{Red-RTChan}}
 \\
 \adjustbox{max width=\linewidth}{\getrule*{Red-LTChan}}
\end{gatherrules}
\endgroup

\section{Omitted semantic clauses}

The following clauses were omitted from \cref{sec:type-reduction}.

Where \(\pi = \pLabel{a}{k}\), \(m = \mLabel{a}{k}\), and \(k \in L\):
\begin{align}
  \sembr{
  \jwfts
  { \Pi }
  { \mc{G}
  \scons
  (\jwfpi
  { \Delta }
  { \Iota }
    { a : \Tamp_{l \in L} A_l })}
    }
  &\supseteq
    \tcons{\pI}{m}{
    \sembr{
    \jwfts
    { \immred{\Pi}{\pi} }
    { \immred{\mc{G}}{\pi}
    \scons
    (\jwfpi
    { \immred{\Delta}{\pi} }
    { \immred{\Iota}{\pi} }
    { a : A_k })}
    }}
    \label{psred:Tamp-R}
      \\
  \sembr{
  \jwfts
  { \Pi }
  { \mc{G}
  \scons
  (\jwfpi
  { \Delta, a : \Tamp_{l \in L} A_l }
  { \Iota }
  { c : C })}
  }
  &\supseteq
      \scalemath{0.99}{
    \tcons{\pO}{m}{
    \sembr{
    \jwfts
    { \immred{\Pi}{\pi} }
    { \immred{\mc{G}}{\pi}
    \scons
    (\jwfpi
    { \immred{\Delta}{\pi}, a : A_k }
    { \immred{\Iota}{\pi} }
    { c : \immred{C}{\pi} })}
    }}}
    \label{psred:Tamp-L}
  \\
  \sembr{
  \jwfts
  { \Pi }
  { \mc{G}
  \scons
  (\jwfpi
  { \Delta }
  { \Iota, a : \Tamp_{l \in L} A_l }
  { c : C })}
  }
  &\supseteq
    \scalemath{0.99}{
    \tcons{\pS}{m}{
    \sembr{
    \jwfts
    { \immred{\Pi}{\pi} }
    { \immred{\mc{G}}{\pi}
    \scons
    (\jwfpi
    { \immred{\Delta}{\pi} }
    { \immred{\Iota}{\pi}, a : A_k }
    { c : \immred{C}{\pi} })}
    }}}
\end{align}

Where \(\pi = (\pChan{a}{b})\) with \(b\) fresh, and \(m = \mChan{a}\):
\begin{align}
  &\sembr{\jwfts
    { \Pi }
    { \mc{G}
    \scons
    (\jwfpi
    { \Delta }
    { \Iota, a : (b : B) \Tot (a : A) }
    { c : C })}}
    \nonumber\\
  &\supseteq
    \tcons{\pS}{\pi}[b]{
    \sembr{
    \jwfts
    { \immred{\Pi}{\pi} }
    { \immred{\mc{G}}{\pi}
    \scons
    (\jwfpi
    { \immred{\Delta}{\pi} }
    { \immred{\Iota}{\pi}, b : B, a : A }
    { c : \immred{C}{\pi} })}
    }}
    \label{psred:Tot-S}
  \\
  &\sembr{\jwfts
    { \Pi }
    { \mc{G}
    \scons
    (\jwfpi
    { \Delta }
    { \Iota }
    { a : (b : B) \Tlolly (a : A) })}}
    \nonumber\\
  &\supseteq
    \tcons{\pI}{m}[b]{
    \sembr{
    \jwfts
    { \immred{\Pi}{\pi} }
    { \immred{\mc{G}}{\pi}
    \scons
    (\jwfpi
    { \immred{\Delta}{\pi}, b : B }
    { \immred{\Iota}{\pi} }
    { a : A })
    }}}
    \label{psred:Tlolly-R}
  \\
  &\sembr{\jwfts
    { \Pi }
    { \mc{G}
    \scons
    (\jwfpi
    { \Delta }
    { \Iota, a : (b : B) \Tlolly (a : A) }
    { c : C })}}
    \nonumber\\
  &\supseteq
    \tcons{\pS}{\pi}[b]{
    \sembr{
    \jwfts
    { \immred{\Pi}{\pi} }
    { \immred{\mc{G}}{\pi}
    \scons
    (\jwfpi
    { \immred{\Delta}{\pi} }
    { \immred{\Iota}{\pi}, b : B, a : A }
    { c : \immred{C}{\pi} })}
    }}
    \label{psred:Tlolly-S}
  \\
  &\sembr{\jwfts
    { \Pi }
    { \mc{G}
    \scons
    (\jwfpi
    { \Delta_1, \Delta_2, a : (b : B) \Tlolly (a : A) }
    { \Iota_1, \Iota_2 }
    { c : C })}}
    \nonumber\\
  &\supseteq
    \tcons{\pO}{m}[b]{
    \sembr{
    \jwfts
    { \immred{\Pi}{\pi} }
    { \immred{\mc{G}}{\pi}
    \scons
    (\jwfpi
    { \immred{\Delta_1}{\pi} }
    { \immred{\Iota_1}{\pi} }
    { b : B })
    \scons
    (\jwfpi
    { \immred{\Delta_2}{\pi}, a : A }
    { \immred{\Iota_2}{\pi} }
    { c : \immred{C}{\pi} })
    }}}
    \label{psred:Tlolly-L}
\end{align}

\section{Proofs}

\printProofs
}{} %

\printbibliography

\end{document}